\documentstyle[12pt,epsfig]{article} 
\def\baselinestretch{1.3}
\newcommand{\ba}{\begin{array}}
\newcommand{\ea}{\end{array}}
\newcommand{\bd}{\begin{displaymath}}
\newcommand{\ed}{\end{displaymath}}
\newcommand{\be}{\begin{equation}}
\newcommand{\ee}{\end{equation}}
\newcommand{\bea}{\begin{eqnarray}}
\newcommand{\eea}{\end{eqnarray}}

\def\a{\alpha}

\def\b{\beta}

\def\l{\lambda}

\def\q2 {q^2}

\def\bt{\begin{table}}
\def\et{\end{table}}

\catcode`@=11 % This allows us to modify PLAIN macros.
\def \gsim{\mathrel{\mathpalette\@versim>}}
\def \lsim{\mathrel{\mathpalette\@versim<}}
\def \@versim#1#2{\lower0.4ex\vbox{\baselineskip\z@skip\lineskip\z@skip
     \lineskiplimit\z@\ialign{$\m@th#1\hfil##\hfil$%
     \crcr#2\crcr\sim\crcr}}}
\catcode`@=12 % at signs are no longer letters

\begin{document}

%THE TEXT STARTS HERE
%\begin{flushright}
%{\large  HRI-P-07-01-001}
%\end{flushright}

\begin{center}
{\Large\bf Non-universal gaugino masses: a signal-based analysis
for the Large Hadron Collider}\\[15mm]
Subhaditya Bhattacharya\footnote{E-mail: subha@mri.ernet.in},
AseshKrishna Datta\footnote{E-mail: asesh@mri.ernet.in} 
and Biswarup Mukhopadhyaya\footnote{E-mail: biswarup@mri.ernet.in}\\
{\em Regional Centre for Accelerator-based Particle Physics \\
     Harish-Chandra Research Institute\\
Chhatnag Road, Jhusi, Allahabad - 211 019, India}
\\[20mm] 
\end{center}

\begin{abstract}
We discuss the signals at the Large Hadron Collider (LHC) for scenarios
with non-universal gaugino masses in supersymmetric (SUSY) theories. 
We perform a multichannel analysis, and 
consider the ratios of event rates in different channels
such as $jets~+~{E}_T\!\!\!\!\!\!/~$  , $same$ - and $opposite$-$sign~ 
dileptons$ 
$+jets+ {E}_T\!\!\!\!\!\!/~$ , as well as $single-lepton$ and $trilepton$
 final states together with $jets~+~{E}_T\!\!\!\!\!\!/~$  . 
Low-energy SUSY spectra corresponding
to high-scale gaugino non-universality arising from different breaking
schemes of SU(5) as well as SO(10) Grand Unified (GUT) SUSY models 
are considered, with both degenerate low-energy sfermion masses and 
those arising from a supergravity scenario. We present the numerical 
predictions over a wide range of the parameter space using the event 
generator {\tt Pythia}, 
specifying the event selection criteria and pointing out
regions where signals are likely to be beset with backgrounds. Certain
broad features emerge from the study, which may be useful in identifying
the signatures of different GUT breaking schemes and distinguishing them 
from a situation with a universal gaugino mass at high scale.
The absolute values of the predicted event rates for different scenarios
are presented together with the various event ratios, so that
these can also be used whenever necessary.
\end{abstract}

\vskip 1 true cm

\newpage
\setcounter{footnote}{0}

\def\baselinestretch{1.5}

\section{Introduction} 

In its bid to unravel new laws of physics around the TeV scale,
the Large Hadron Collider (LHC) experiment will place considerable
emphasis on the search for supersymmetry (SUSY) \cite{Book,Sally,Gl}. 
Apart from stabilizing the electroweak symmetry breaking sector and
providing a rather tantalizing hint of Grand Unification, SUSY (in the
$R$-parity conserving version) also provides a cold dark matter candidate
in the form of the stable lightest supersymmetric particle (LSP) \cite{Martin}.
With this in view, the SUSY signals that are most frequently talked about 
are those where a large amount of missing (transverse) momentum is carried 
away by a 
pair of LSP's resulting from decay chains of superparticles produced 
in the initial hard scattering process \cite{Sally,Martin}. The missing energy 
is accompanied
by hard jets and/or leptons, and their relative numbers as well as signs
in  the observed final states are expected to direct us to specific regions of
the `signature space', indicating, in turn,  where one stands in the parameter
space of the overseeing SUSY theory 
\cite{Bourjaily:2005ja,CDF,Ash,Barger,Matchev,Baer,ATLAS}.

Locating oneself correctly in the signature space also helps one in knowing
whether some of the usual or simplifying assumptions made about SUSY are 
actually tenable.  For example, signals can be qualitatively different if 
$R$-parity (with $R = (-1)^{3B+L+2S}$) is violated \cite{Barbier},  
or the LSP is,
contrary to common expectations, not the lightest neutralino ($\chi^0_1$) 
\cite{BM}. Very heavy scalars can also warrant a different analysis of SUSY
signals \cite {Nima,Guidice,Djouadi2}. Similarly,  signals may also be quite 
different if, instead of the supergravity
(SUGRA) scheme  controlling SUSY breaking,  gauge mediated SUSY
breaking (GMSB) \cite{Martin,GMSB}or anomaly mediated SUSY breaking (AMSB)
\cite{Martin,AMSB} is 
operative.  A more difficult problem is, however,  posed if the signals are
not qualitatively new but are found to differ from usual expectations  only 
on detailed quantitative scrutiny. Here we undertake an analysis of one 
such situation, where, contrary to the most popular outcome of SUSY
embedded in a Grand Unified Theory (GUT) framework, the gaugino masses at 
the high scale  are  {\em not} unified. 
\cite{Ellis,Drees}. 

In the simplest SUGRA models, all  low-scale parameters are derived from
a universal gaugino mass ($M_{1/2}$),  a universal scalar mass ($m_0$),
the trilinear soft SUSY-breaking parameter ($A_0$) and the sign of
the Higgsino mass parameter (sgn($\mu$)) for each value of
$\tan\beta$, the ratio of the two Higgs vacuum expectation values 
\cite{Martin}.  A universal gaugino mass occurs in the simplest form 
of a SUSY GUT.  
Its immediate consequence is that
the  three low-energy gaugino masses corresponding to SU(3), SU(2)
and U(1) are in the ratio of the corresponding fine-structure constants:
${\frac{M_3}{\alpha_3}}~=~  {\frac{M_2}{\alpha_2}}~=~  {\frac{M_1}{\alpha_1}}$
 \cite{Martin}.
This relation governs the low-energy chargino and neutralino masses vis-a-vis
the gluino mass. It has  profound implications on the strengths of different
types of signals,  since  gluinos are liable to be copiously produced at the
LHC, and the cascades initiated by them involve the charginos and neutralinos
at various stages \cite{Gl,Ash}.  Therefore, if  gaugino mass universality 
at high scale
does not hold,  it means that both the spectrum and the compositions of
the charginos and neutralinos are subject to marked variations, 
so that the final states 
have different rates compared to the universal case both through 
kinematics and dynamics.

While departure from universality may well indicate that one is
not facing a  SUSY GUT scenario,  it may, interestingly, still be
the consequence of a GUT framework. The gaugino masses arise from
the gauge kinetic function whose trivial nature, as we shall see 
in the next section,  implies a universal gaugino mass when SUSY 
is broken at high scale. This is possible if the combination
of hidden sector fields involved in the function is a singlet 
under the  GUT group. However, it is always possible to generate mass 
terms via higher GUT representations,
which in turn create inequality among $M_1$, $M_2$ and $M_3$ 
at the high scale itself.
It is also possible to have more than one GUT representations involved 
in SUSY breaking, in which case the non-universality arises from a 
linear combination of the effects mentioned above \cite{Ellis,Drees,Chamoun,
Pran,Cremmer}.

Identifying departure from universality in SUSY signals is important 
at more than one levels \cite{CP1,fine-tunning,mu-egamma,noscale,CP2}. 
First, one would like to know whether 
or not the gaugino
mass relation corresponding to a particular  
GUT representation is involved.
The absence of any such obvious relation, however, still keeps 
SUSY GUT's alive, if the analysis of signals reveals that a linear 
combination of GUT multiplets is involved.
It is only the decisive failure of such a finer analysis that 
can rule out a framework
based on GUT. Therefore, if SUSY signals in some channel(s) are indeed seen at 
the LHC, the exercise of tracing them back to some underlying GUT framework,
 be it with gaugino mass universality or not, is of utmost importance.

Testing gaugino non-universality at the LHC, however, is not easy,
especially if the ambitious task of looking for higher GUT representations
has to be undertaken. There has been some detailed analysis  of events 
and kinematics for non-universal gaugino masses in the context of the Tevatron 
\cite {Baer1,Huitu,Sourav}, with reference to SU(5).  
Some phenomenological studies have been performed 
on different types of signals at the LHC, 
too \cite {Choi:2007ka,Bt1,Bt2},  but the systematic investigation
that is required to link the departure from universality to GUT 
representations has not so far been carried out in detail.

In our study, different representations of SUSY SU(5) and SO(10) 
are considered.
No specific SUGRA origin of  scalar masses is assumed in the general 
analyses, and we deliberately
(and perhaps artificially) adhere to degenerate squark and slepton 
masses at low energy in each case. However, 
we also present side by side the  consequences of a SUGRA scenario with
universal scalar masses at high scale. 
In each case, we consider a comprehensive set of SUSY signals, 
such as  $jets$ + ${E}_T\!\!\!\!\!\!/~$,  $same-sign$ as well as 
$opposite-sign 
dileptons$, $one~ isolated~ lepton$ and  $trileptons$ alongwith 
$jets$ + ${E}_T\!\!\!\!\!\!/~$ (so called multichannel analysis 
\cite {Datta:1999uh,baer11}). After subjecting the 
calculated event rates for these different final states and for 
different parameter values to 
such cuts as to suppress the standard model (SM) backgrounds, we look at their
various ratios. This reduces uncertainties due to jet energy resolution, 
jet energy scale, parton distribution functions and so on. It 
also ensures that the departure from gaugino 
universality, rather than  the overall scale of superparticle 
masses, is the decisive factor.  Thereafter, we compare these ratios 
with the corresponding cases with a universal gaugino mass. 
The squark and gluino
masses are kept at the same values during this comparison,  since the most 
important cascades are dictated by them, and their masses can be 
approximately found out from
the LHC data from ${E}_T\!\!\!\!\!\!/~$ and effective mass distributions. 
Although we confine
ourselves to a relatively rudimentary analysis, 
it is expected that more elaborate ones
can be built on it following the same strategy. 
It is our belief that such an approach
will mean full utilization of  the LHC data in following up on any 
signature of SUSY, an exercise that is eminently appropriate 
at the present juncture \cite{Nima2}.

In section 2,  we briefly  review the process by which non-universality arises
at the GUT scale, and summarise the high-scale mass relations of gauginos in
different GUT representations responsible for the non-universality. 
The strategy adopted in selecting the relevant SUSY parameters, and the 
event selection criteria for LHC, are outlined in section 3. The 
analysis of predicted signals for SU(5)  and SO(10) are presented in sections
4.1 and 4.2, respectively.  We summarise and conclude in section 5.
Appendix A contains the various chargino and neutralino masses for
different scenarios, while the absolute values of event rates in different
channels (which has been found to be necessary supplements to the various
ratios presented in the main text) are listed in Appendix B.

\section{Non-universal SUSY GUT and gaugino mass ratios}

In this section we review the issues that govern non-universality of 
supersymmetry breaking gaugino masses, arising under the influence of
various GUT representations responsible for the SUSY breaking terms.

We adhere to a scenario where all soft SUSY breaking effects arise
via hidden sector interactions in an underlying supergravity (SUGRA) framework.
Specifically, we are considering supersymmetric SU(5) and SO(10) gauge theories
with an arbitrary chiral matter superfield content coupled to N=1 supergravity.
The essential theoretical principles governing high-scale non-universality in
gaugino masses as well as in gauge couplings have been discussed in a number
of earlier works in the context of both SU(5) \cite{Ellis,Drees} 
and SO(10) \cite{Chamoun} gauge groups respectively. 
Later works that addressed the 
related phenomenology (mostly in the context of SU(5) ) are by and large 
based on these principles \cite{Pran,Cremmer,Baer1,Huitu}.

All gauge and matter terms including gaugino masses in the N=1 supergravity
lagrangian depend crucially on two fundamental functions of chiral
superfields. One of them is the gauge kinetic function 
$f_{\alpha \beta}(\Phi)$ which is an
analytic function of the left-chiral superfields $\Phi_{i}$. It 
transforms as a symmetric product of the adjoint representation as gauge
superfields belong to the adjoint representation of the underlying gauge
group ($\alpha$, $\beta $ being the gauge generator indices). The other is the 
real function $G(\Phi_{i},\Phi^*_{i})$  with $G = K + ln|W|$ where $K$ is the 
K$\ddot a$hler potential and $W$ is the superpotential. $G$ is a real function
of the chiral superfields $\Phi_{i}$ and is a gauge singlet. 
However, $f_{\alpha\beta}$ in general has a non-trivial gauge transformation 
property. Based on whether its functional 
dependence on the chiral superfields involves singlet 
or non-singlet irreducible
representations of the underlying gauge group, one has universal or
non-universal gaugino masses at the GUT scale, when SUSY is broken.

In the component field notation, the part of the N=1 supergravity lagrangian
containing kinetic energy and mass terms for gauginos and gauge bosons 
(including only terms containing the real part of $f(\Phi)$) reads 
\cite{Cremmer}

\bea
e^{-1} {\mathcal{L}}= -\frac{1}{4} Re f_{\alpha \beta}(\phi)(-1/2
\bar{\lambda}^
{\alpha} D\!\!\!\!/ \lambda^{\beta})-\frac{1}{4} Re f_{\a \b}(\phi)F_{\mu
  \nu}^{\a}F^{\beta \mu \nu} \nonumber \\
 +\frac{1}{4} e^{-G/2}
G^{i}((G^{-1})^{j}_{i})[\partial f^*_{\a \b}(\phi^*)/\partial
{\phi^{*j}}]\l^{\a}\l^{\b} + h.c
\eea  
\noindent
where $G^i = \partial{G}/\partial{\phi_{i}}$ and $(G^{-1})^{i}_j$ is the
inverse matrix of 
${G^j}_i\equiv \partial{G} / {\partial{\phi^{*i}}\partial{\phi_j}}$,
$\l^{\a}$ is the gaugino field, and 
 $\phi$ is the scalar component of the chiral superfield $\Phi$. 
The $F$-component of $\Phi$ enters the last term to generate gaugino masses. 
Thus, following equation (1), the lagrangian can be expressed as \cite{Huitu}

\bea
e^{-1} {\mathcal{L}}= -\frac{1}{4} Re f_{\alpha \beta}(\phi)(-1/2
\bar{\lambda}^
{\alpha} D\!\!\!\!/ \lambda^{\beta})-\frac{1}{4} Re f_{\a \b}(\phi)F_{\mu
  \nu}^{\a}F^{\beta \mu \nu} \nonumber\\
+\frac{F^{j}_{\acute{\a} \acute{\b}}}{2}  [\partial{f^{*}_{\a
  \b}(\phi^{*j})}/\partial{{\phi^{*j}}_{\acute{\a} \acute{\b}}}]\l^{\a}\l^{\b}
 + h.c
\eea
\noindent
where \bea
F^{j}_{\acute{\a} \acute{\b}}=\frac{1}{2} e^{-G/2}
[G^{i}((G^{-1})^{j}_{i})]_{\acute{\a} \acute{\b}}
\eea 

The $\Phi^{j}$ s can be classified into two categories: a set of
GUT singlet supermultiplets $\Phi^{S}$, and a set of
non-singlet ones $\Phi^N$. The non-trivial gauge kinetic function $f_{\a
  \b}(\Phi^{j}) $can be expanded in terms of the non-singlet
components in the following way \cite{Ellis,Drees,Huitu}:

\bea
f_{\a \b}(\Phi^{j})= f_{0}(\Phi^{S})\delta_{\a
  \b}+\sum_{N}\xi_{N}(\Phi^s)
\frac{{\Phi^{N}}_{\a \b}}{M}+ {\mathcal{O}}(\frac{\Phi^N}{M})^2
\eea
\noindent
where $f_0$ and $\xi^N$ are functions of chiral singlet superfields, and
$M$ is the reduced Planck mass$=M_{Pl}/\sqrt{8\pi}$.

In principle, the gauge kinetic function $f_{\a \b}$ is a function of all
chiral superfields $\Phi^{j}$. However, those which contribute significantly 
at the minimum of the potential by acquiring large vacuum expectation 
values (vev) are (i) gauge singlet fields which are part of the hidden 
sector (i.e. the fields $\Phi^{S}$), and 
(ii) fields associated with the spontaneous breakdown of the
GUT  group to $SU(3)\times SU(2)\times U(1)$ 
(i.e. the fields $\Phi^{N}$) \cite{Ellis,Drees}. 
In equation (3), the contribution to the gauge kinetic function
from  $\Phi^{N}$ has to come through symmetric products of the 
adjoint representation of
associated GUT group, since $f_{\a \b}$ on the left side of (3) has such
transformation property. Thus  $f_{\a \b}$ can have the `non-trivial' 
contribution of the second type of terms only if one has chiral superfields
belonging to representations which can arise from the symmetric products
of two adjoint representations \cite{Pran}. For SU(5), for example,
one can have contributions to $f_{\a \b}$ from all possible 
non-singlet irreducible representations to which $\Phi^{N}$
can belong : 
$$
(24\times 24)_{symm}=1+24+75+200
$$
\noindent
For SO(10), the possible representations are :
$$
(45\times 45)_{symm}=1+54+210+770\\ 
$$
\noindent
The contribution to  $f_{\a \b}$ can also come from any linear combination 
of the singlet and possible non-singlet representations (as shown above) 
in case of both SU(5) and SO(10).
It is now almost clear from (2)  
that these non-singlet representations can be responsible for 
non-universal gaugino mass terms at the GUT scale.\\
In order to obtain the low energy effective theory, we replace the 
fields ${\Phi^{S}}$  and ${\Phi^{N}}$
in the gauge kinetic term (3) by their vev's and get 
$\langle f_{\a \b} \rangle$.
The value of $\langle f_{\a \b} \rangle$ which determines the gaugino mass
matrix crucially depends on the specific
representation (or their linear combinations) responsible for 
the process \cite{Ellis,Drees}. 
It is important to note here that, in this analysis, the breakdown 
of the symmetry from
SU(5) to the SM gauge group  has been assumed to take place at 
the GUT scale ($M_{X}$)
itself. When there is an intermediate gauge group $H$ 
(as is possible for SO(10)), 
the vev of the gauge kinetic function depends not only on the chosen 
non-singlet representation but also crucially on the
intermediate group $H$ in the breaking chain \cite{Chamoun}. In
addition, the presence of intermediate scale can also affect the vev of the 
gauge kinetic function and hence gaugino mass ratios at the GUT scale 
\cite{Chamoun}.

Next,  the kinetic energy terms are restored to the canonical form
by rescaling the gauge superfields, by defining

\be 
{F^{\a}}_{\mu \nu} \rightarrow {\hat{F}^{\a}}_{\mu \nu}={\langle Re f_{\a
    \b} \rangle}^{\frac{1}{2}}{F^{\b}}_{\mu \nu} 
\ee
and    
\be
\l^{\a}  \rightarrow {\hat{\l}}^{\a}={\langle Re f_{\a
    \b} \rangle}^{\frac{1}{2}}\l^{\b}
\ee  
Simultaneously, the gauge couplings are also rescaled (as a result of (4)):
\be
g_{\a}(M_{X}){\langle Re f_{\a \b} \rangle}^{\frac{1}{2}}\delta_{\a \b}= 
g_{c}(M_{X}) 
\ee
where $g_{c}$ is the universal coupling constant at the GUT scale. This
shows clearly that the first consequence of a non-trivial
gauge kinetic function is non-universality of the gauge couplings $g_{\a}$ 
at the GUT scale, if $\langle f_{\a \b} \rangle$ carries a gauge index.

Once SUSY is broken by non-zero vev's of the $F$ components of 
hidden sector chiral superfields, the coefficient of 
the last term in equation(2) is replaced by \cite{Ellis,Drees,Huitu}
\be
{\langle { F_{\a \b}}^{i} \rangle}= {\mathcal{O}}(m_{\frac{3}{2}} M)
\ee
where $m_{\frac{3}{2}}= exp(-\frac{\langle G\rangle}{2})$ 
is the gravitino mass. Taking 
into account the rescaling of the gaugino fields 
(as stated earlier in equation (4)and (5)) in equation (6), 
the gaugino mass matrix can be written down as in \cite{Huitu} or 
\cite{Ellis,Pran} 
\be
M_{\a}(M_{X})\delta_{\a \b}=\sum_{i}\frac{{\langle F^{i}_{\acute{\a}
      \acute{\b}}\rangle}}{2} \frac{\langle \partial{f_{\a
    \b}(\phi^{*i})}/\partial{{\phi^{*i}}_{\acute{\a} \acute{\b}}} \rangle}
{\langle Re f_{\a \b} \rangle}
\ee
or 
\be
M_{\a}(M_{X})\delta_{\a \b}=\frac{1}{4} e^{-G/2}
G^{i}((G^{-1})^{j}_{i})\frac{\langle \partial f^*_{\a \b}(\phi^*)/\partial
{\phi^{*j}}\rangle}{{\langle Re f_{\a \b} \rangle}}
\ee
which demonstrates that the gaugino masses are non-universal at the GUT scale. 
The underlying reason for this is the fact that  
$\langle f_{\a \b} \rangle$ can 
be shown to acquire the form {\large $ f_{\a}\delta_{\a \b}$} 
\cite {Ellis,Drees},
thanks to the symmetric character of the representations.
Consequently, the derivatives on the right-hand side of the above equations
acquire such forms as to render $M_\alpha$ non-universal in the gauge indices.
On the contrary, if symmetry breaking occurs via gauge
singlet fields only,  one has {\large$f_{\a \b}=f_{0}\delta_{\a \b}$} 
from equation (4) and as a result, 
{\large $\langle f_{\a \b} \rangle=f_{0}$}. Thus
both gaugino masses and the gauge couplings are unified at the GUT scale, 
as can be seen from equations (7) and (10).\\

Following the approach in \cite{Ellis,Drees,Pran,Huitu}, 
we make a further simplification 
by neglecting the non-universal contributions to the gauge couplings
at the GUT scale. The gaugino mass ratios at high scale thus obtained 
\cite{Ellis,Chamoun} are shown in Tables 1 and 2. We also present the 
approximate values of the ratios 
at the Electroweak Symmetry Breaking scale (EWSB) in those tables. 
While the effects corresponding
to all symmetric representations of SU(5) have been shown, we have presented
the case for only the lowest representation of SO(10). This is because,
SO(10) being a rank-5 gauge group, the low-energy consequences of the
mass ratios depend on not only
the specific breaking chain adopted, but also  the presence (or otherwise)
and magnitudes of intermediate breaking scale. A proliferation of such
features affects the collider phenomenology in too complicated a manner to
be related easily to high scale physics. Therefore, we illustrate our points
by taking the lowest relevant representation, and using the mass ratios
corresponding to two breaking chains, assuming that the breakdown to
$SU(3)\times SU(2)\times U(1)$  takes place at the GUT scale itself 
in each case. 

\noindent

\begin{center}

\begin{tabular}{|c|c|c|}

\multicolumn{3}{c}{Table 1: High-scale and approximate low-scale gaugino mass 
ratios for SU(5). }\\
\hline
 Representation & $M_{3}:M_{2}:M_{1}$ at $M_{GUT}$& $M_{3}:M_{2}:M_{1}$ at 
$M_{EWSB}$ \\
\hline 
{\bf 1} & 1:1:1 & 6:2:1 \\
\hline
{\bf 24} & 2:(-3):(-1) & 12:(-6):(-1)\\
\hline
{\bf 75} & 1:3:(-5) & 6:6:(-5)\\
\hline
{\bf 200} & 1:2:10 & 6:4:10  \\
\hline
\end {tabular}\\

\noindent

\end {center}

\noindent

\begin{center}

\begin{tabular}{|c|c|c|}

\multicolumn{3}{c}{Table 2: High-scale and approximate 
low-scale gaugino mass ratios for SO(10). }\\
\hline
 Representation & $M_{3}:M_{2}:M_{1}$ at $M_{GUT}$& $M_{3}:M_{2}:M_{1}$ at 
$M_{EWSB}$ \\
\hline
{ 1} & 1:1:1 &  6:2:1\\
\hline
{\bf 54(i)}: {$H \rightarrow SU(2) \times SO(7)$} & 1:(-7/3):1 & 7:(-5):1\\
\hline
{\bf 54(ii)}: {$H \rightarrow SU(4) \times SU(2) \times SU(2)$} & 1:(-3/2):(-1)
 & 7:(-3):(-1)\\
\hline
\end {tabular}\\
\noindent

\end {center}

\vspace{0.4 cm}

\section{SUSY signals and backgrounds: strategy for analysis}

In this section we discuss and analyse the difference 
in the collider signature 
due to non-universal gaugino 
masses at the GUT scale for various non-singlet 
representations of SU(5) and SO(10) GUT group in 
the context of the LHC.

\subsection {Choice of SUSY parameters}

In our analysis we have confined ourselves to 
$R$-parity conserving supersymmetry 
where the lightest neutralino is the LSP. 
Thus all SUSY signals at the LHC are
characterized by a large amount of missing $E_T$ 
carried by the LSP, together
with jets and/or leptons of various multiplicity.

A large part of our analysis is done for a scenario where the gaugino masses
are obtained through one-loop running from the non-universal mass parameters
at the high scale, whereas the low-energy scalar masses are all treated as 
phenomenological  inputs. Furthermore, since we wish to examine the effects
of gaugino non-universality in isolation, 
we have taken all the squark and slepton 
masses to be degenerate. This not only avoids 
special situations arising from
SUSY cascade decays due to a spread in the sfermion masses, but also keeps
the scenario above board 
by suppressing flavour-changing neutral currents (FCNC) \cite{FCNC}.
The Higgsino mass parameter $\mu$, too, is a free parameter here.
The mass parameters
of the Higgs sector are determined once $\mu$, 
the neutral pseudoscalar mass ($m_A$)
and $\tan\beta$ (the ratio of the two Higgs vev's) are specified.

Side by side, we also present an analysis pertaining to a non-universal SUGRA 
scenario where the low energy supersymmetric spectrum is generated 
from a common 
scalar mass $m_{0}$, common trilinear coupling $A_{0}$ and $sgn(\mu)$,
with non-universal gaugino masses $M_{i}$ at high scale 
arising from various non-singlet representations of SU(5) and SO(10).
While this allows a spread in the low-energy sfermion masses, it also
gives one the opportunity to compare the predicted collider results with
those in the phenomenological scalar spectrum mentioned above.
It has been made sure that in both this case and the previous one,
the parameter choices are consistent with the LEP bounds,
as far as the neutral Higgs mass, the lighter chargino mass etc.
are concerned \cite {LEP}. 

The spectrum in the first case is generated by the option {\bf pMSSM} in
the code {\tt SuSpect} v2.3 \cite{SUSPECT}.  It should be 
remembered that our goal here is
to generate a phenomenological low-energy 
spectrum with degenerate scalar masses,
but with the three gaugino mass parameters 
related not by high-scale universality
but by the specific conditions answering to various non-singlet GUT 
representations. In order to implement this, we resort to 
a two-step process. The first step is to give as inputs
non-universal gaugino masses at the GUT scale, and evolve them down
to low scale through one-loop renormalization group equations (which do not
involve scalar masses). This yields a phenomenological gaugino
spectrum which, to a reasonable approximation, corresponds to the
specific non-singlet GUT representation under scrutiny. 
In the second step, we feed the thus obtained gaugino masses, together with
the degenerate scalar masses (and the free parameters in the Higgs sector)
at the electroweak symmetry breaking (EWSB) scale,
into  {\tt SuSpect} as low energy inputs in the {\bf pMSSM} option. 
The subsequent
running of {\tt SuSpect} yields a low-energy spectrum which is basically
phenomenological, but ensures gauge coupling unification at high scale (see
discussion in the previous section), and is nonetheless consistent with
laboratory constraints on a SUSY scenario. We have used the low-energy
value of ${\alpha_3 (M_{Z})}^{\overline{MS}}= 0.1172$ 
for this calculation which is default in {\tt SuSpect}. 
Throughout the analysis we have assumed the top quark mass to be 171.4 GeV.
Electroweak symmetry breaking at the `default scale'  
$\sqrt{m_{\tilde{t_{L}}}m_{\tilde{t_{R}}}}$
has been ensured in this procedure, together with the requirement
of no tachyonic modes for sfermions. 
No radiative correction to gaugino masses has been considered, which
does not affect the main flow of our analysis in any significant way.
Full one-loop and the dominant two-loop corrections 
to the Higgs masses are incorporated.
And finally, consistency with low-energy constraints 
$b\rightarrow s\gamma$
and muon anomalous magnetic moment 
are checked for every combination of parameters
used in the analysis.  Preferring to be strictly confined to
accelerator signals, we have not considered dark matter constraints
in our analysis. For studies in this direction, we refer the reader 
to  \cite{Pran,Roy,DM1,DM2} where the issues related to dark matter 
in non-universal gaugino scenarios have been discussed. It should also
be remembered that, although we shall henceforth refer to this case as
{\bf pMSSM} for convenience, the low-energy spectrum is not purely
`phenomenological', since the gaugino masses at low energy actually
correspond to specific high-scale GUT-breaking conditions.

We attempt a representative analysis of the above situation 
by taking all possible combinations of parameters, arising out 
of the following choices, for each type of GUT breaking
scheme:\\
$m_{\tilde g}$= [500 GeV, 1000 GeV, 1500 GeV] \\
$m_{\tilde f}$= [500 GeV, 1000 GeV] \\
$\mu$= [300 GeV, 1000 GeV] \\ 
$\tan {\beta}$= [5, 40]

\noindent
where by $m_{\tilde f}$ we denote all the degenerate squark and slepton 
masses. This gives us a total of 24 combinations 
which include the most important 
kinematics regions in terms of $m_{\tilde g}$ and $m_{\tilde q}$ namely, 
(i) $m_{\tilde g} \gg m_{\tilde f}$,
(ii) $m_{\tilde q} \gg m_{\tilde g}$ and 
(iii) $m_{\tilde q} \simeq m_{\tilde g}$
 which crucially controls the final state scenario at the collider. 
Also the variation in $\mu$ changes the chargino and neutralino compositions 
which affect the various  decay branching fractions involved in the cascades.
We have also taken two values of $\tan\beta$, one close to the limit coming
from $e^+~ e^-$ collider data, and the other on the high side, since they
also control the chargino-neutralino sector. For all these points we keep
all the trilinear coupling constants $A_{0}$= 0 and the pseudoscalar Higgs
mass $m_{A}$= 1000 GeV.

For studying the other scenario, namely, gaugino mass non-universality in a 
{\bf SUGRA} setting, the spectrum is generated 
with the help of {\tt ISASUGRA}  
v7.75 \cite{ISAJET}. As mentioned earlier, here one uses as 
the inputs a common 
scalar mass $m_{0}$, a common trilinear coupling $A_{0}$,
 $\tan \beta$ and $sgn(\mu)$, along with non-universal gaugino masses
 $m_{i}$ at the GUT scale (with ratios as appropriate for various
GUT-breaking representations) and run down to low scale via
two-loop renormalization group equations. The chargino and neutralino 
spectra are given in Table A9, Appendix A.
We select a smaller number of samples than in the case of {\bf pMSSM},
taking  $ A_{0}=0$, $sgn(\mu)$ as positive and 
$\tan {\beta}$= 5. 
We choose $m_{0}$ at the GUT scale such that, for $m_{\tilde g}$= 1000 GeV 
at the low scale, the first two generations of squark masses
are clubbed around 1000 GeV. We know that the scalar mass thus obtained 
at the electroweak symmetry breaking scale with a high scale input by 
renormalisation group equation (RGE) has almost 90 \% contribution from
gauginos due to the running \cite{Rammond}.  
This value  turns out be 506 GeV the GUT scale. 
As is done earlier, we tune the SU(3) gaugino mass $M_{3}$ at the high 
scale to get $m_{\tilde g}$= 500 GeV, 1000 GeV and 1500 GeV.
We stick to  $ m_{0}$= 506 GeV at the GUT scale for all these cases.
%It may be noticed that the various hierarchies 
%between gluino and squark masses 
%mentioned earlier are not represented here; however, not all of them
%are permitted in the usual versions of SUGRA. 
%The scenario allowed in SUGRA
%but not included in the present study is 
%one with $m_{\tilde q} \gg m_{\tilde g}$. 
The low-energy spectrum is consistent with radiative electroweak symmetry
breaking as well as all other phenomenological constraints \cite{constraints}.

\subsection {Collider simulation}

The spectra generated as described in the previous section  
are fed into the event generator {\tt Pythia} 6.405 \cite{PYTHIA} 
by {\tt SLHA} interface \cite{sLHA} for the simulation of $pp$ collision 
with centre of mass energy 14 TeV.

We have used {\tt CTEQ5L} \cite{CTEQ} parton distribution functions, 
the QCD renormalisation and factorisation scales being
both set at the subprocess centre-of-mass energy  $\sqrt{\hat{s}}$.
All possible SUSY processes and decay chains have been kept open. 
In the illustrative study presented here, we  have switched
off initial and final state radiation as well as 
multiple interactions. However, we take hadronisation 
into account using the fragmentation functions inbuilt
in {\tt Pythia}. We have checked our analysis code against 
earlier studies done at the parton level in the 
MSSM framework \cite {Ash}. We also checked our code in 
the context of Tevatron using \cite{AD}. We checked all the cross-sections with
{\tt CalcHEP} also \cite{calchep}.

The standard final states in connection with $R$-parity
conserving SUSY have been looked for.  All of these 
have been discussed in the literature in different contexts
\cite {Ash,ATLAS,Baer1,Baer2}. These are 

\begin{itemize}
 \item Opposite sign dilepton (OSD) :
 $(\ell^{\pm}\ell^{\mp})+ (\geq 2)~ jets~ + {E_{T}}\!\!\!\!/$ 
  
\item Same sign dilepton (SSD) : 
$(\ell^{\pm}\ell^{\pm})+ (\geq 2)~jets~ + {E_{T}}\!\!\!\!/$

\item Single lepton ($(1\ell+jets)$):  
$1\ell~ + (\geq 2)~ jets~ + {E_{T}}\!\!\!\!/$

\item Trilepton ($(3\ell+jets)$): 
$3\ell~ + (\geq 2) ~jets~ + {E_{T}}\!\!\!\!/$

\item Inclusive jet ($jets$): $(\geq 3) ~jets~ + {E_{T}}\!\!\!\!/$    
\end{itemize}

\noindent
where $\ell$ stands for electrons or muons. The cuts used are as follows:

 \begin{itemize}

 \item  Missing transverse momentum $E_{T}\!\!\!\!/$ $\geq ~100$ GeV.

\item ${p_{T}}_\ell ~\ge ~20$ GeV and $|{{\eta}}_{\ell}| ~\le ~2.5$ 
\cite {Ash}.
 
\item An isolated lepton should have lepton-lepton separation
 ${\bigtriangleup R}_{\ell\ell}~ \geq 0.2$, lepton-jet separation 
 ${\bigtriangleup R}_{{\ell}j}~ \geq 0.4$, the energy deposit 
due to jet activity around a lepton ${E_{T}}$ within 
$\bigtriangleup R~ \leq 0.2$ of the lepton axis should be $\leq 10$ GeV.

\item ${E_{T}}_{jet} ~\geq ~100$ GeV and $|{\eta}_{jet}| ~\le ~2.5$  \cite{Ash}.
 
\end {itemize}

\noindent
where  $\bigtriangleup R = \sqrt {{\bigtriangleup \eta}^2
+ {\bigtriangleup \phi}^2}$ 
is the separation in pseudorapidity and azimuthal angle plane. 

Jets are formed in {\tt Pythia} using {\tt PYCELL} jet formation criteria
with $|{\eta}_{jet}| ~\le ~5.0$ in the calorimeter, 
$N_{\eta_{bin}}=100$ and $N_{\phi_{bin}}=64$. 
For a partonic jet to be considered as a jet initiator $E_{T}> 2$ GeV 
is required while a cluster of partonic jets to be called a hadron-jet
 $\sum_{parton} E_{T_{jet}}$ is required to be more than 20 GeV.
For a formed jet the maximum $\bigtriangleup R$ from the jet initiator 
is 0.4.

\subsection {Backgrounds} 

We have generated all dominant standard model (SM) events  
in {\tt Pythia} for the same final states, using the same  
factorisation scale, parton distributions and cuts. 
It has been found that 
$t\bar t$ production gives the most serious backgrounds 
in all channels excepting in the trilepton channel,
for  which the electroweak backgrounds are rather effectively 
removed by our event selection criteria.

The signal and background events have been all calculated for
an integrated luminosity of 300 fb$^{-1}$. 
As has been already mentioned, the ratios of events
in the different final states have been presented, which
presumably reduces some uncertainties in prediction.
Cases where the number of signal events in any of the channels
used in the ratio(s) is less than three have been left out.
Also, in the histograms (to be discussed in the next section),
cases where any of the entries in the ratio has
$\sigma=S/\sqrt{B}~\le~ 2$ ($S$,$B$ being the number of 
signal and background events) have been specially marked with a '\#',
since our observations on them may become useful 
if statistics can be improved.

\section{Prediction for different GUT representations}

 \subsection {Non-universal SU(5)}

\begin{figure}[t]
\begin{center}.
%\vspace*{-2.2cm}
\centerline{\epsfig{file=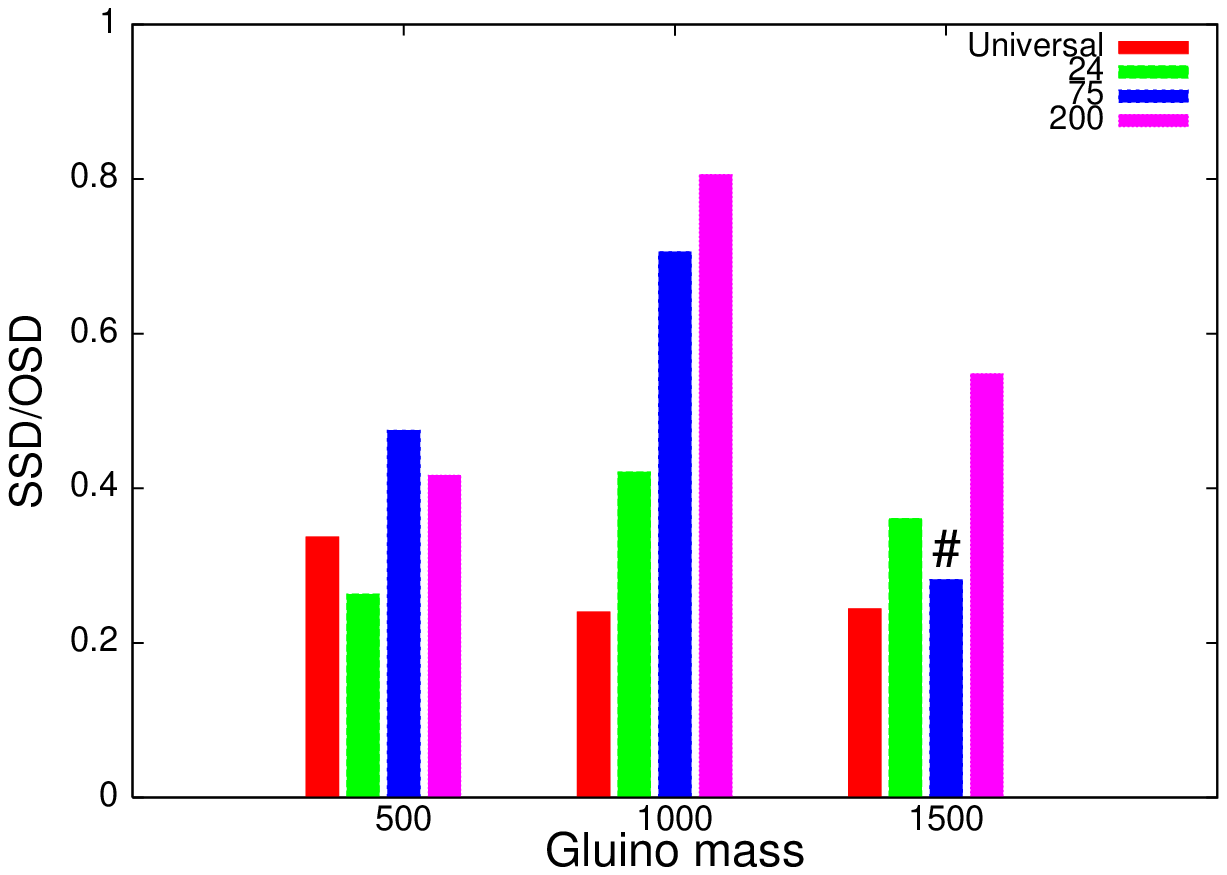,width=6.5 cm,height=5.5cm,angle=-0}
\hskip 20pt \epsfig{file=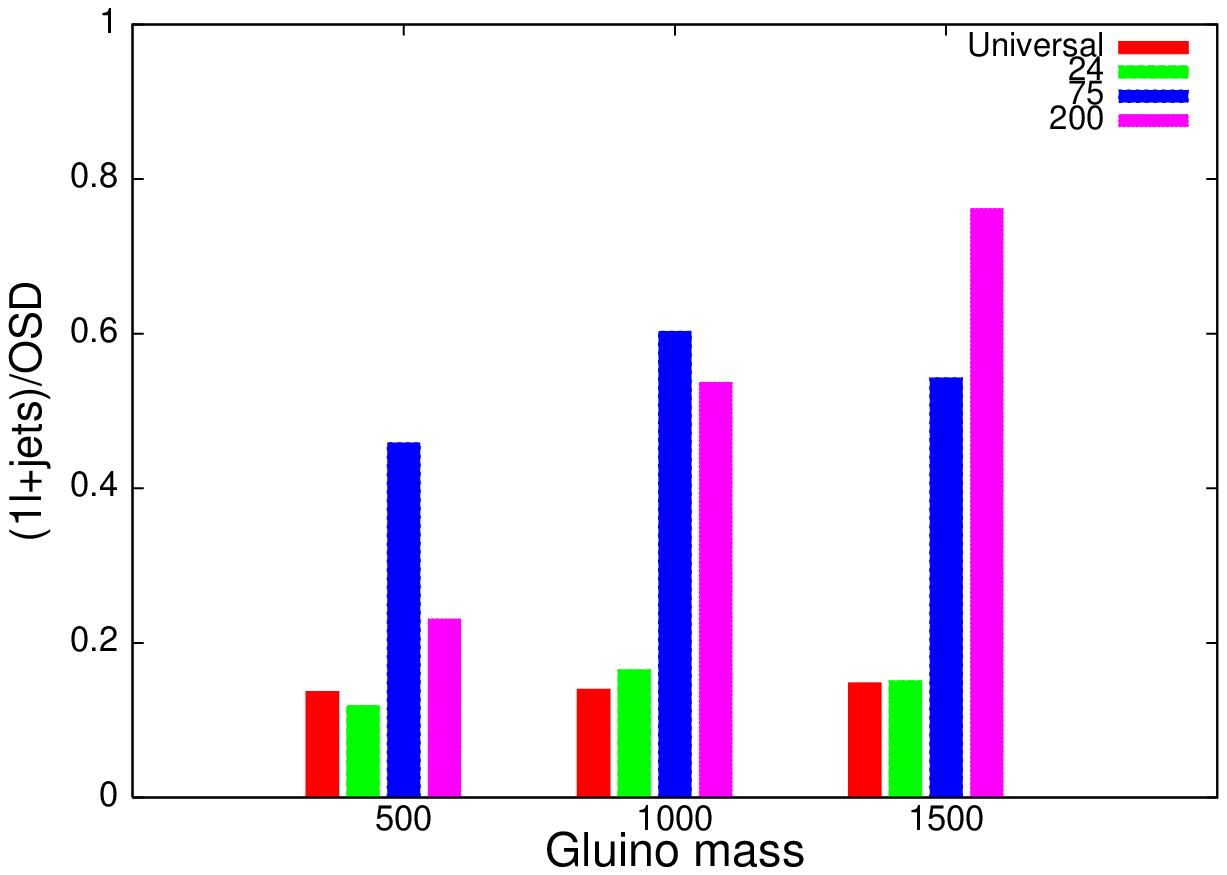,width=6.5cm,height=5.5cm,angle=-0}}
\vskip 10pt
{\epsfig{file=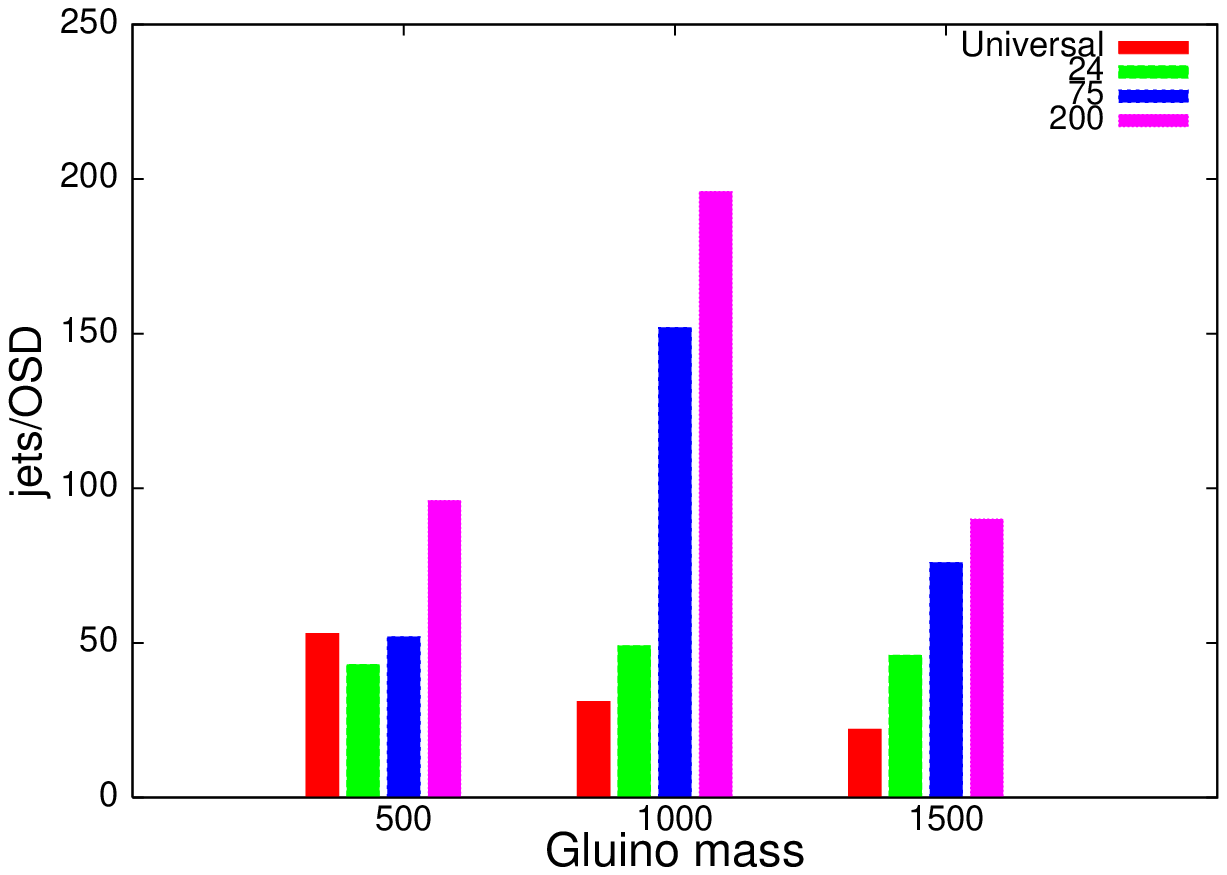,width=6.5 cm,height=5.50cm,angle=-0}}
%\hskip 20pt \epsfig{file=notrilep.eps,width=6.5cm,height=5.50cm,angle=-0}}
\caption{ Event ratios for {\bf pMSSM} in SU(5): $m_{\tilde f}=$500 GeV,
 $\mu=$300 GeV, $\tan{\beta}=5$} 
%distribution}
\end{center}
\end{figure}

 We discuss here the possibility of interpreting non-universality arising in 
various SU(5) representations, namely {\bf 24}, {\bf 75}, {\bf 200}, 
and  compare them with the
universal case.  For the {\bf pMSSM} kind of framework, and
adhering to the approach outlined already, we present in figures 
1 - 8 the ratios of the various types of signals for each of the above
schemes of non-universality. Figure 9 contains our prediction for SU(5) SUGRA.
We have taken the ratio of the number of
each type of signal event to the number of OSD events at the corresponding
point in the parameter space. Thus each panel shows four ratios, 
namely,  SSD/OSD, $(1\ell+jets)$/OSD, $(3\ell+jets)$/OSD and $jets$/OSD in 
the form of histograms.  
For reasons already mentioned,  
the ratio space is a rather reliable discriminator in the
signature space. However, as we shall see, there are regions where
all the ratios turn out to be of similar values for different GUT
representations. In order to address such cases and make the presentation 
complete, we also present the absolute values of the cross-sections
for each type of signal in Appendix B, while the chargino and neutralino
spectra in different cases are found in Appendix A. 

\begin{figure}[t]
\begin{center}
%\vspace*{-2.2cm}
\centerline{\epsfig{file=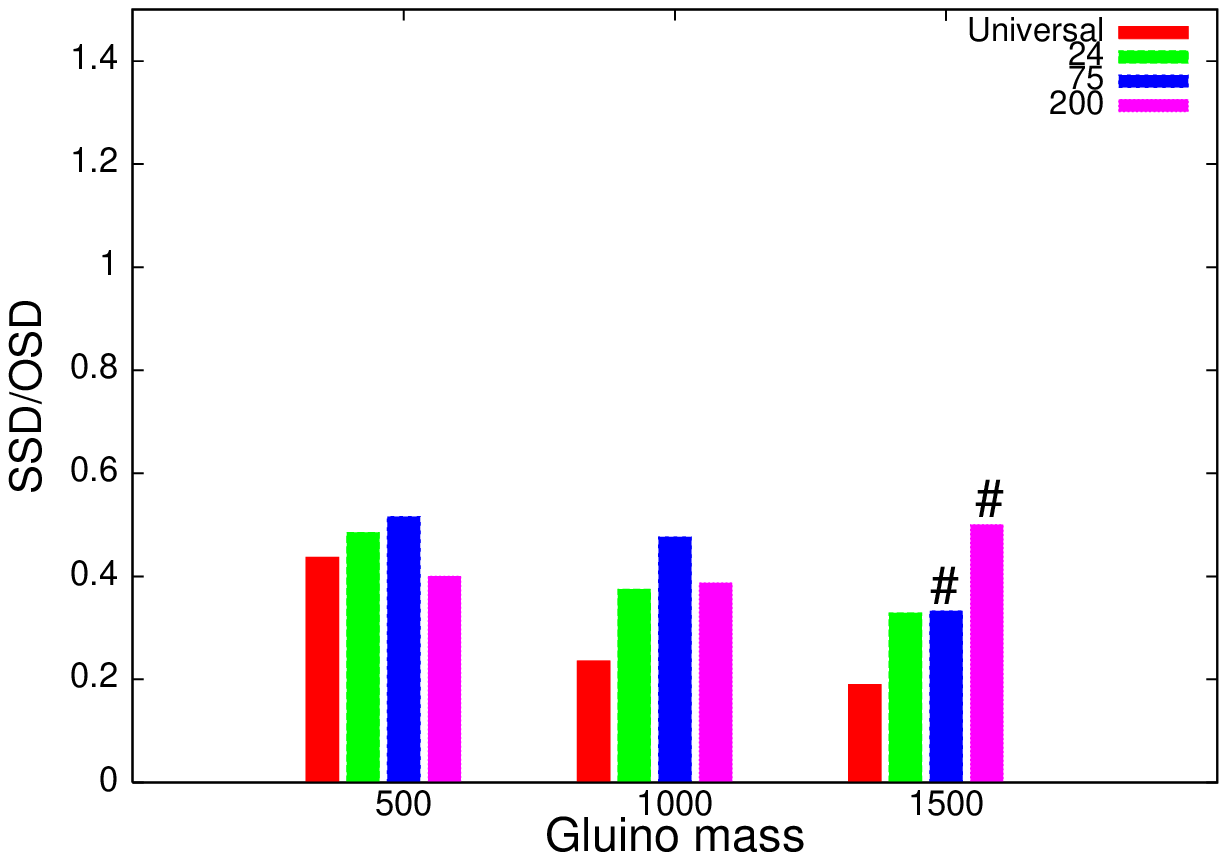,width=6.5 cm,height=5.50cm,angle=-0}
\hskip 20pt \epsfig{file=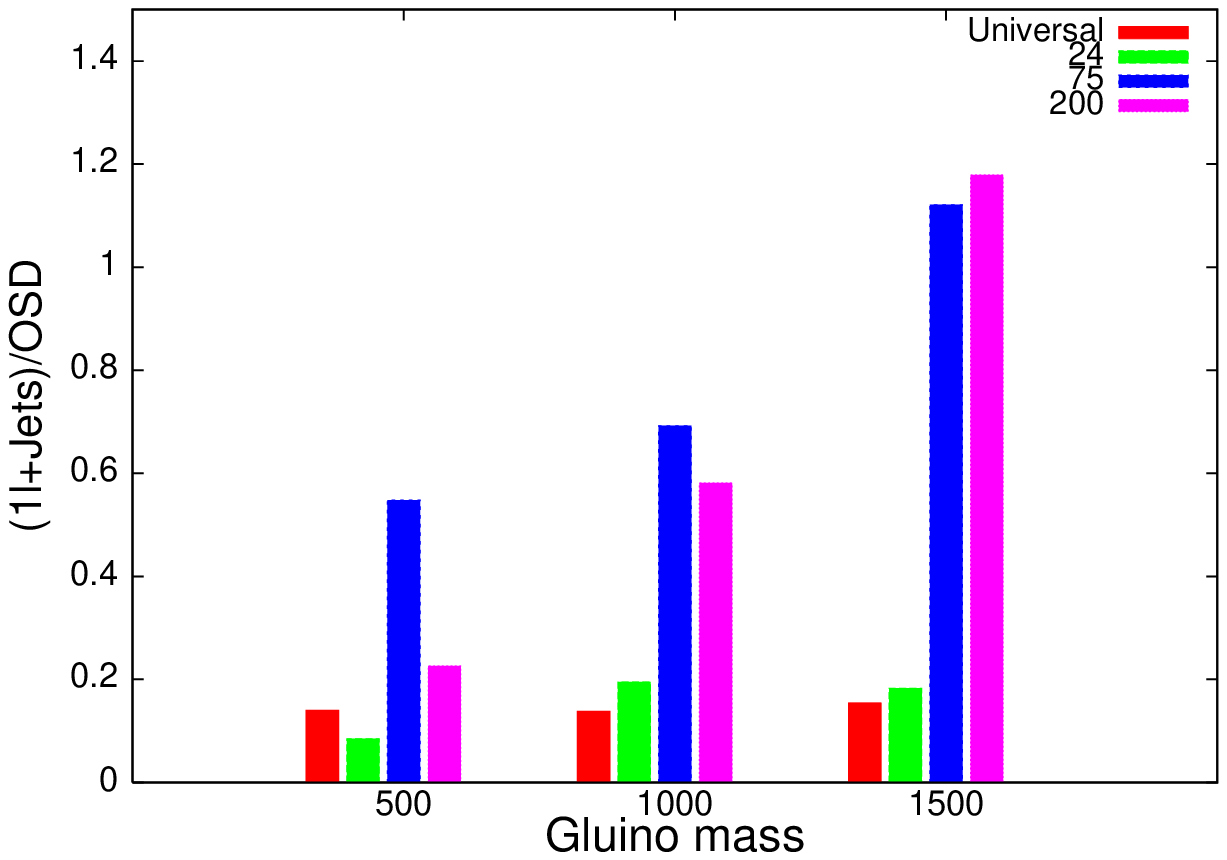,width=6.5cm,height=5.50cm,angle=-0}}
\vskip 10pt
\centerline{\epsfig{file=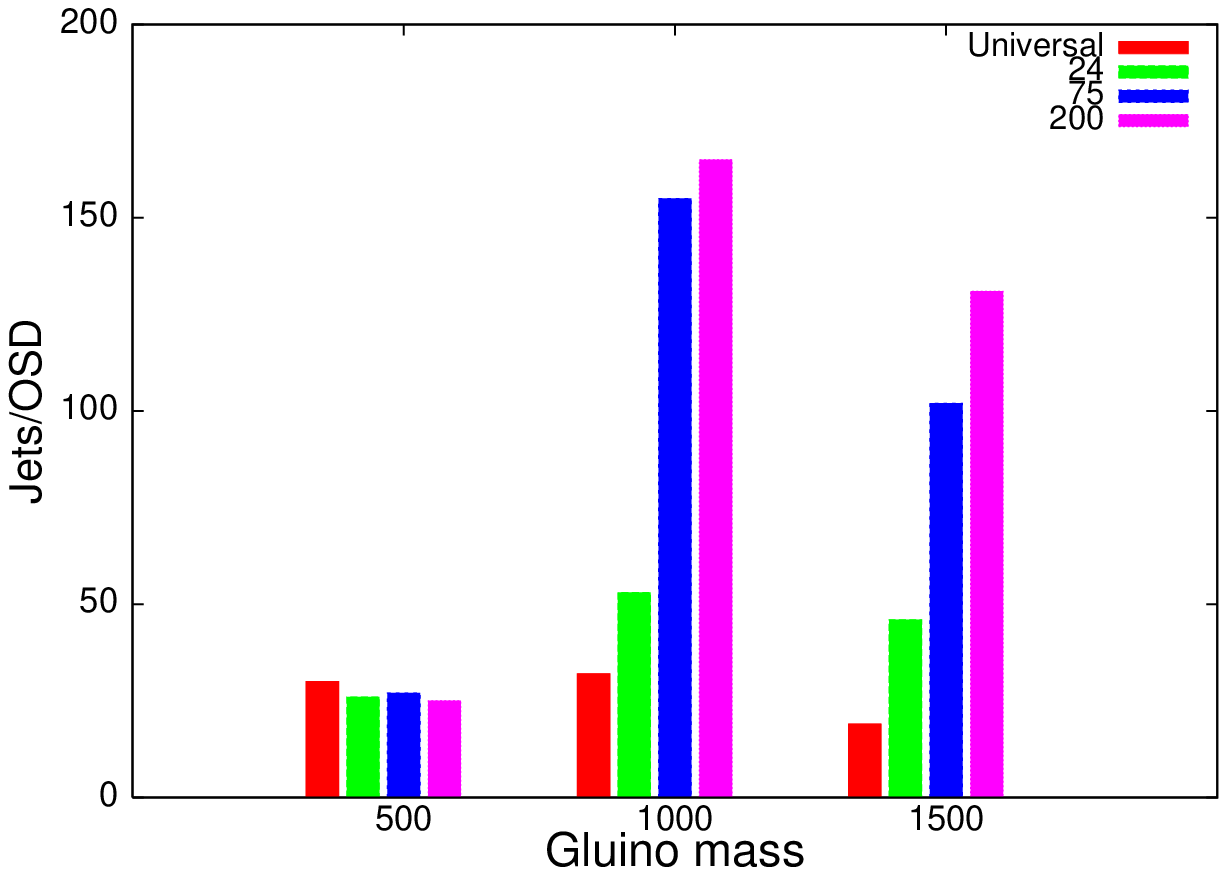,width=6.5 cm,height=5.50cm,angle=-0}
\hskip 20pt \epsfig{file=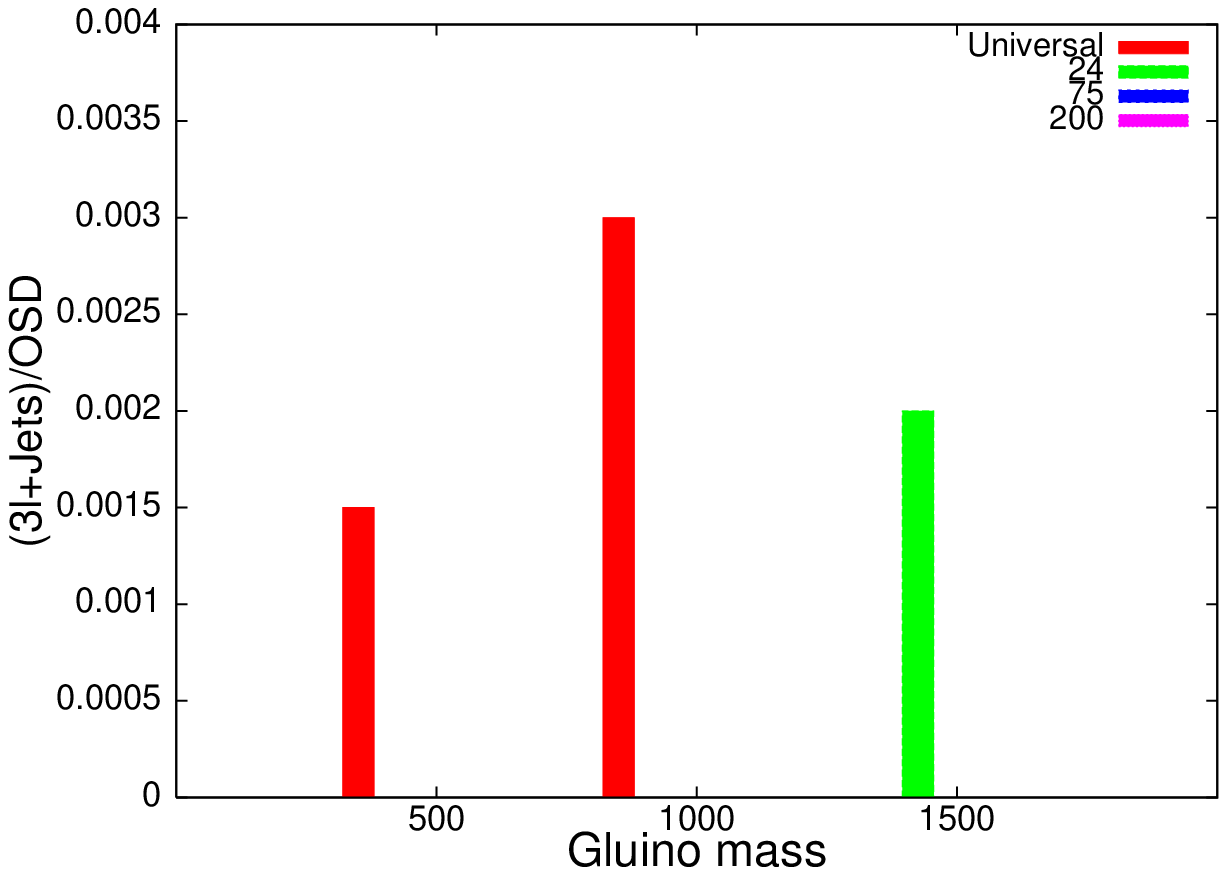,width=6.5cm,height=5.50cm,angle=-0}}
\caption{ Event ratios for {\bf pMSSM} in SU(5): $m_{\tilde f}=$500 GeV, 
$\mu=$300 GeV, $\tan{\beta}=40$} 
%distribution}
\end{center}

\end{figure}

 We plot a particular ratio (eg. SSD/OSD) along the y-axis 
for all non-singlet representations along with the universal one 
at three gluino masses 500 GeV, 1000 GeV and  1500 GeV in the x-axis 
with fixed sfermion mass $m_{\tilde f}$, $\mu$ and $\tan \beta$. 
We club all the different ratio plots in one pannel and discuss the outcome 
as a whole.

It can perhaps be assumed that, if SUSY signals are seen at the LHC,
their kinematic distributions in variables such as $p_{T}\!\!\!\!/$ or
effective mass will yield some useful information about the range of the gluino
and sfermion masses. Adding to this the information extracted from
the Higgs sector, one may be in a position to examine the 
aforementioned ratios, and compare them with our sample results.

In general, the wide multiplicity of parameters makes
the variation of different rates with GUT representations
far from transparent. However, a few features are 
broadly noticeable from figures 1 - 8, and we list them below,
before giving a brief account of each individual figure.

\begin{figure}[t]
\begin{center}
%\vspace*{-2.2cm}
\centerline{\epsfig{file=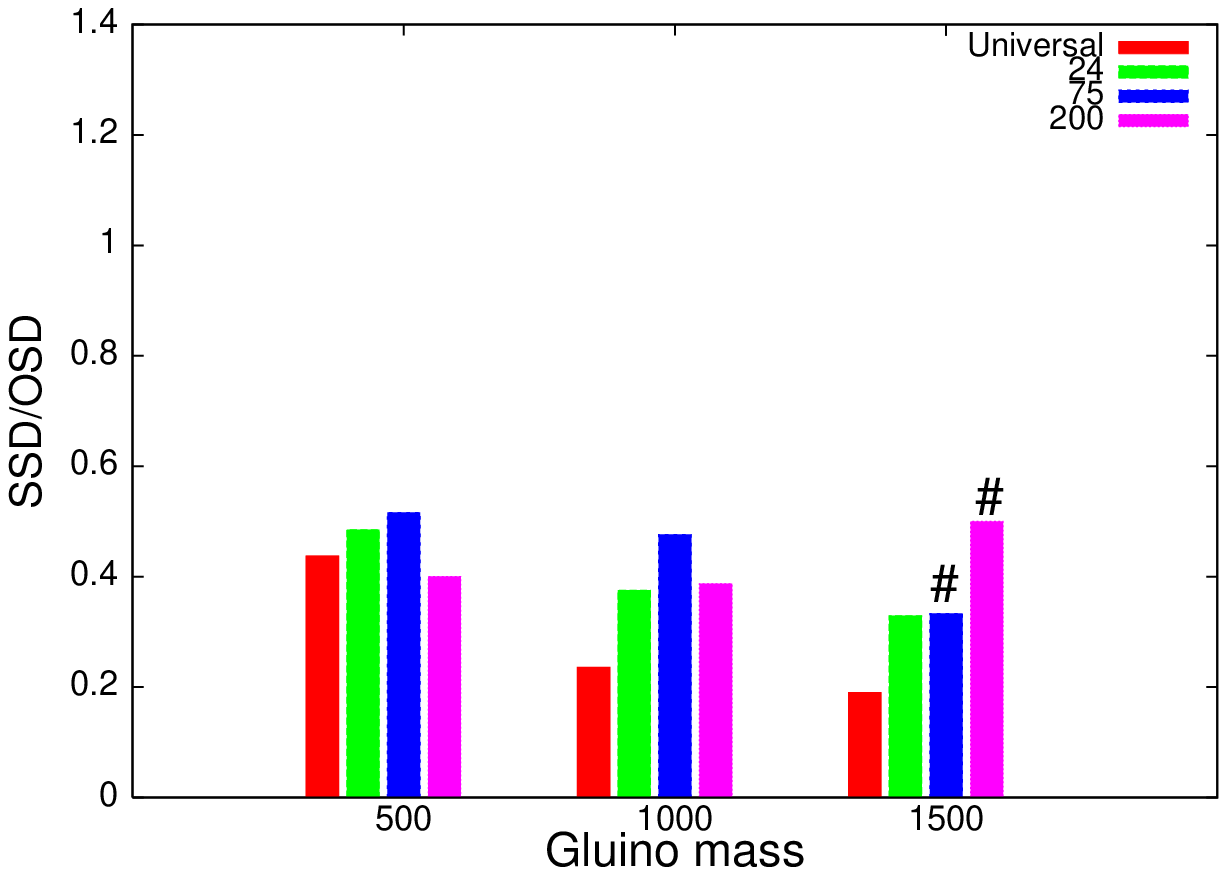,width=6.5 cm,height=5.50cm,angle=-0}
\hskip 20pt \epsfig{file=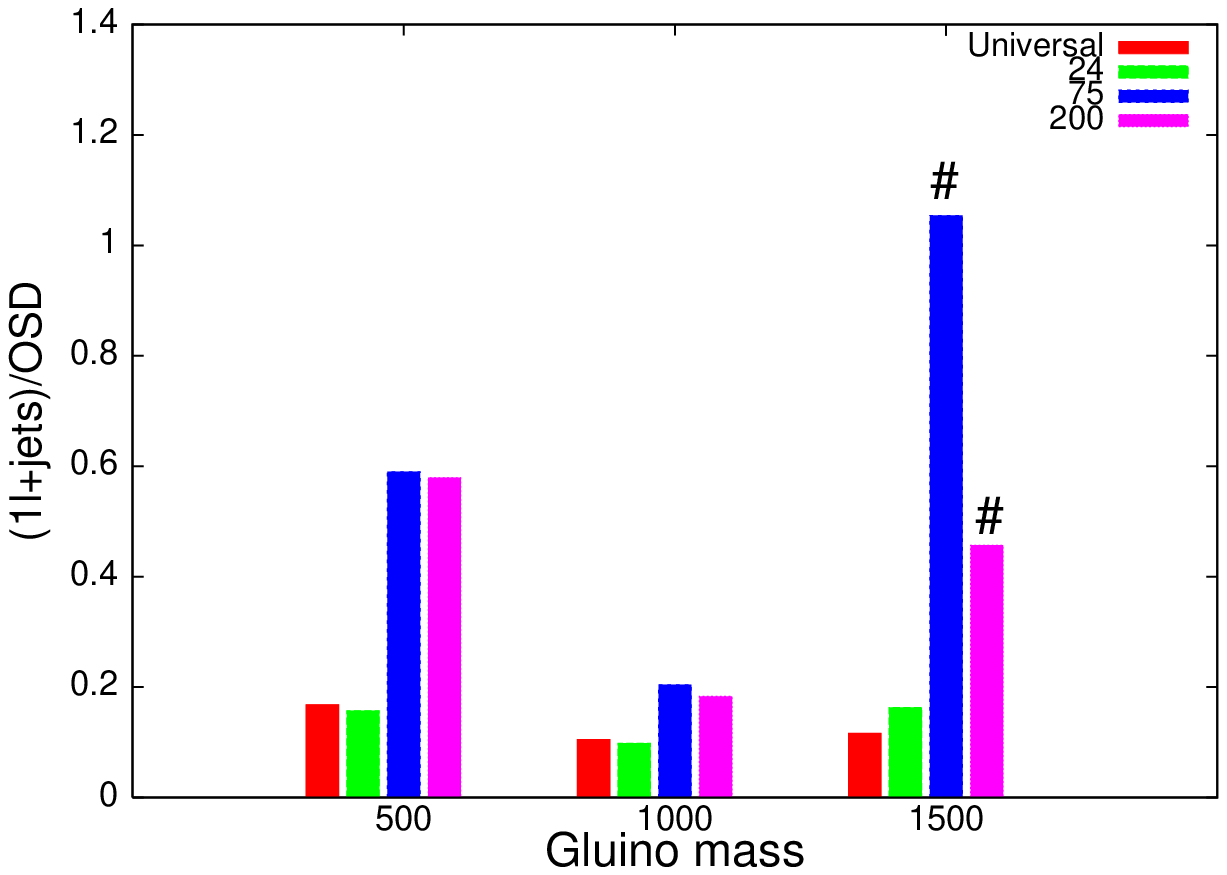,width=6.5cm,height=5.50cm,angle=-0}}
\vskip 10pt
\centerline{\epsfig{file=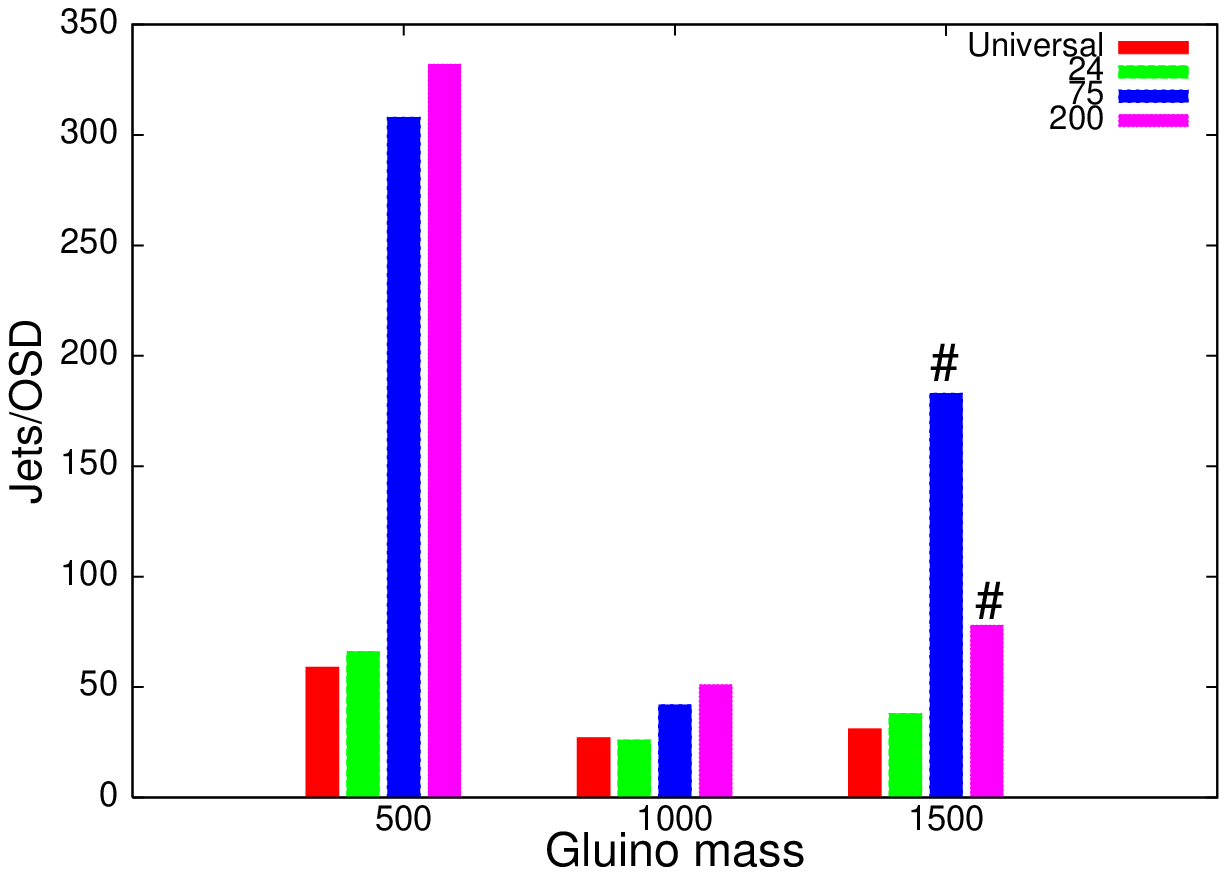,width=6.5 cm,height=5.50cm,angle=-0}
\hskip 20pt \epsfig{file=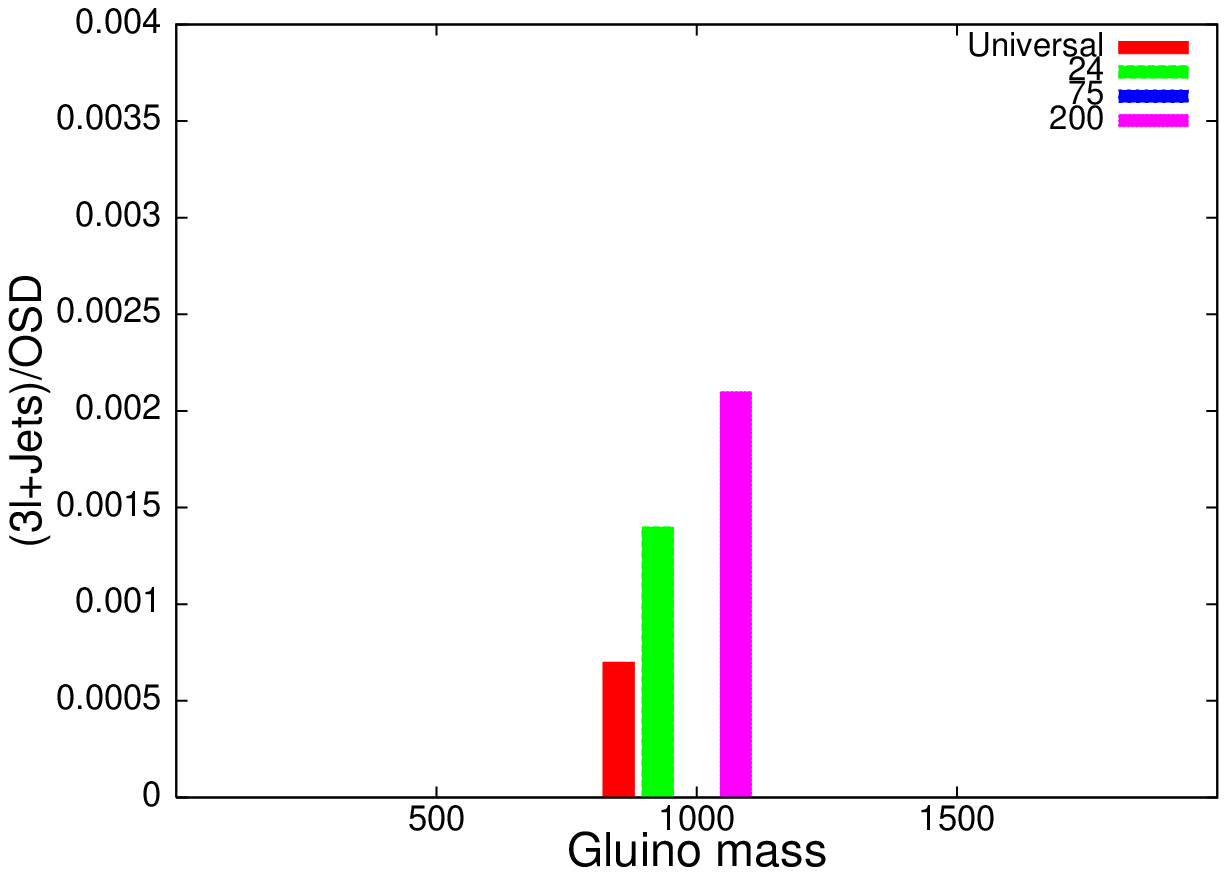,width=6.5cm,height=5.50cm,angle=-0}}
\caption{ Event ratios for {\bf pMSSM} in SU(5): $m_{\tilde f}=$1000 GeV, 
$\mu=$300 GeV, $\tan{\beta}=5$} 
%distribution}
\end{center}

\end{figure}

\begin{enumerate}

\item The event ratios for the representations {\bf 75} and  {\bf 200}
are mostly bigger than those for {\bf 24} and the universal case.
These correspond to the cases where the chargino and neutralino 
masses are relatively large compared to the gluino mass, which in turn
is an artifact of larger $M_{1}$ and $M_{2}$ compared to  $M_{3}$ at 
the GUT scale. The two worst sufferers due to this are the OSD and SSD 
events; of which the former suffers more. This is due to the different
masses and compositions of $\chi^0_2$ and $\chi^{\pm}_1$ (see next para),
which are principally responsible for the OSD and SSD events respectively.
The ratios for {\bf 200} are also separable from the 
others in at least one channel for a large number of cases.
In contrast, {\bf 24} and the universal case often 
behave similarly in the SSD/OSD, $(1\ell+jets)$/OSD and $jets$/OSD ratios.
While this indicates a partially available handle for discrimination 
over a substantial region of the parameter space, 
distinction between {\bf 24} and the universal case is 
possible relatively easily through absolute values 
of the event rates. However, in cases where distinguishing 
{\bf 75} and  {\bf 200} 
from the ratios are difficult, distinction from absolute number of
events are more challenging, because of the rather low rates of
events in such cases.

\begin{figure}[t]
\begin{center}
%\vspace*{-2.2cm}
\centerline{\epsfig{file=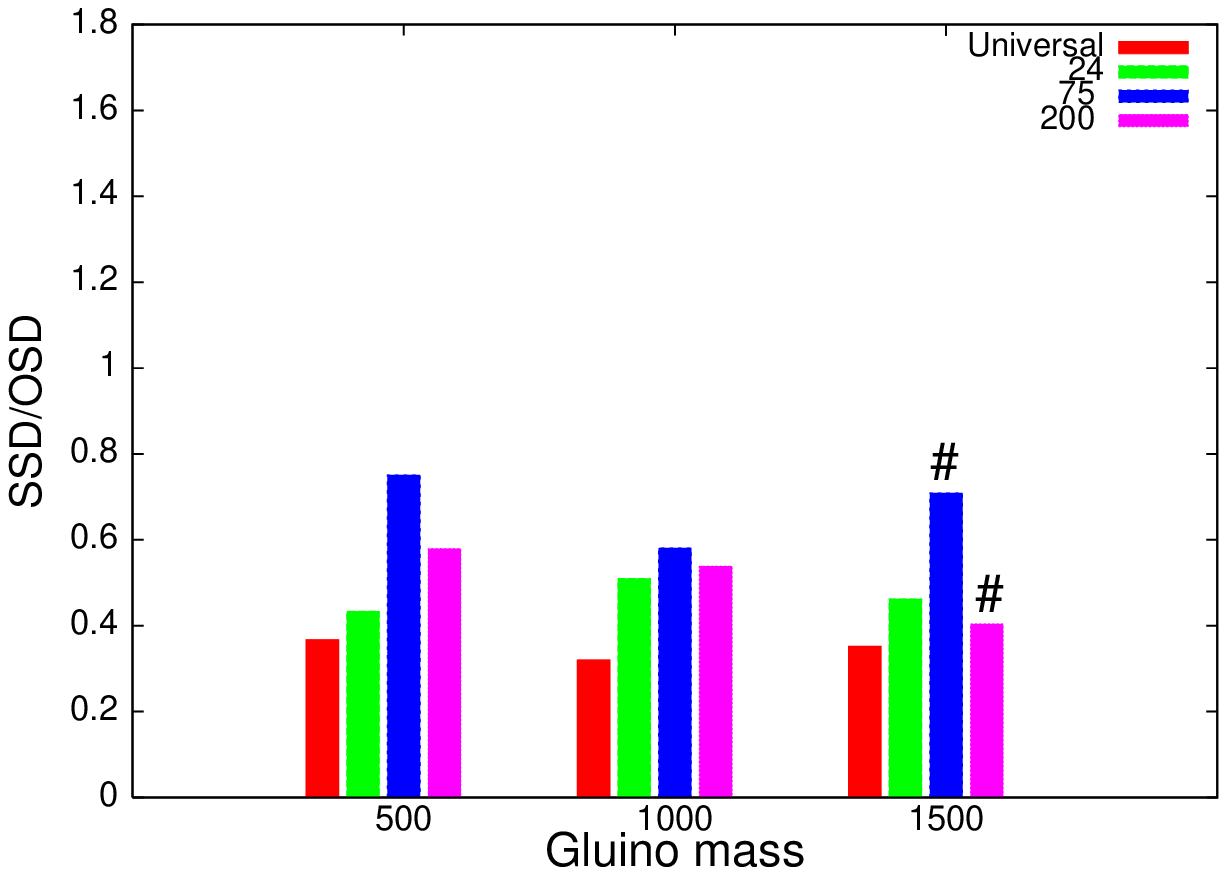,width=6.5 cm,height=5.50cm,angle=-0}
\hskip 20pt \epsfig{file=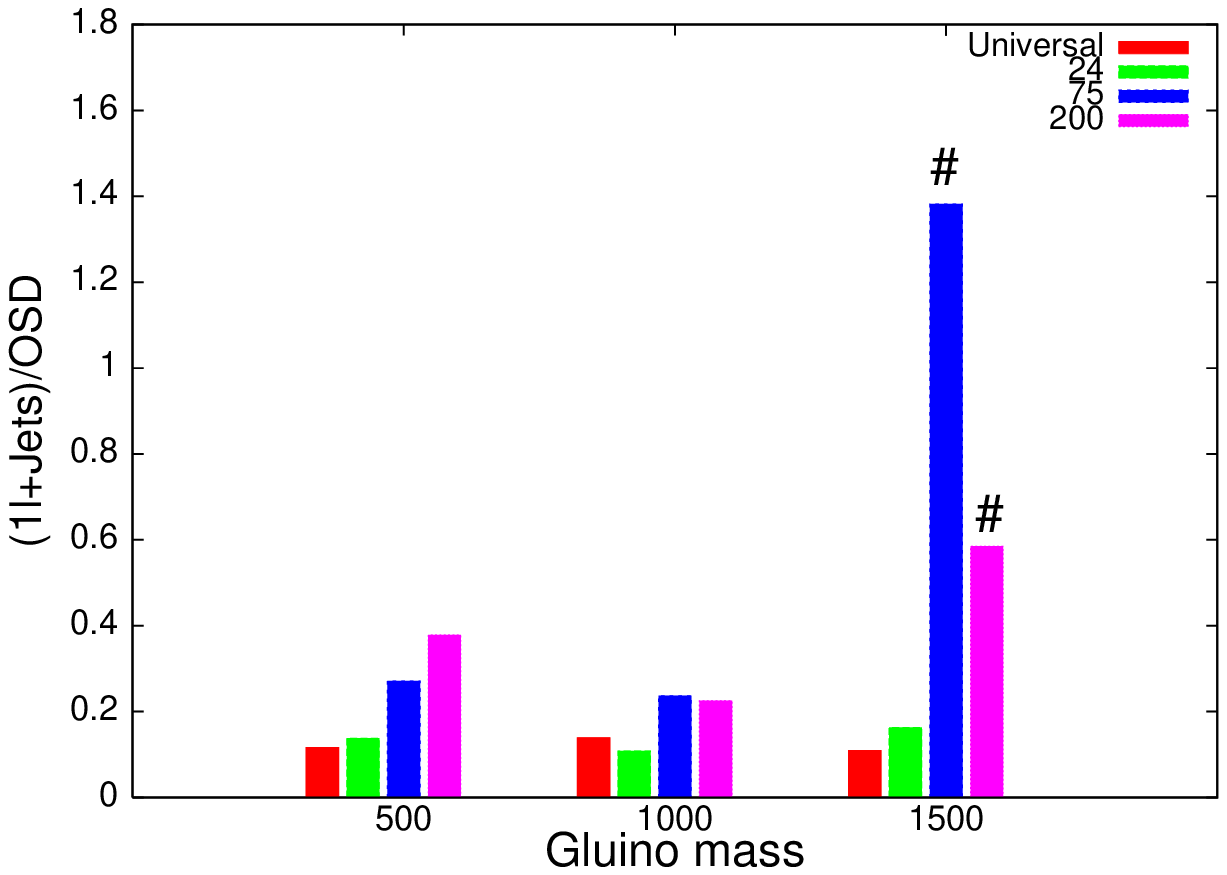,width=6.5cm,height=5.5cm,angle=-0}}
\vskip 10pt
\centerline{\epsfig{file=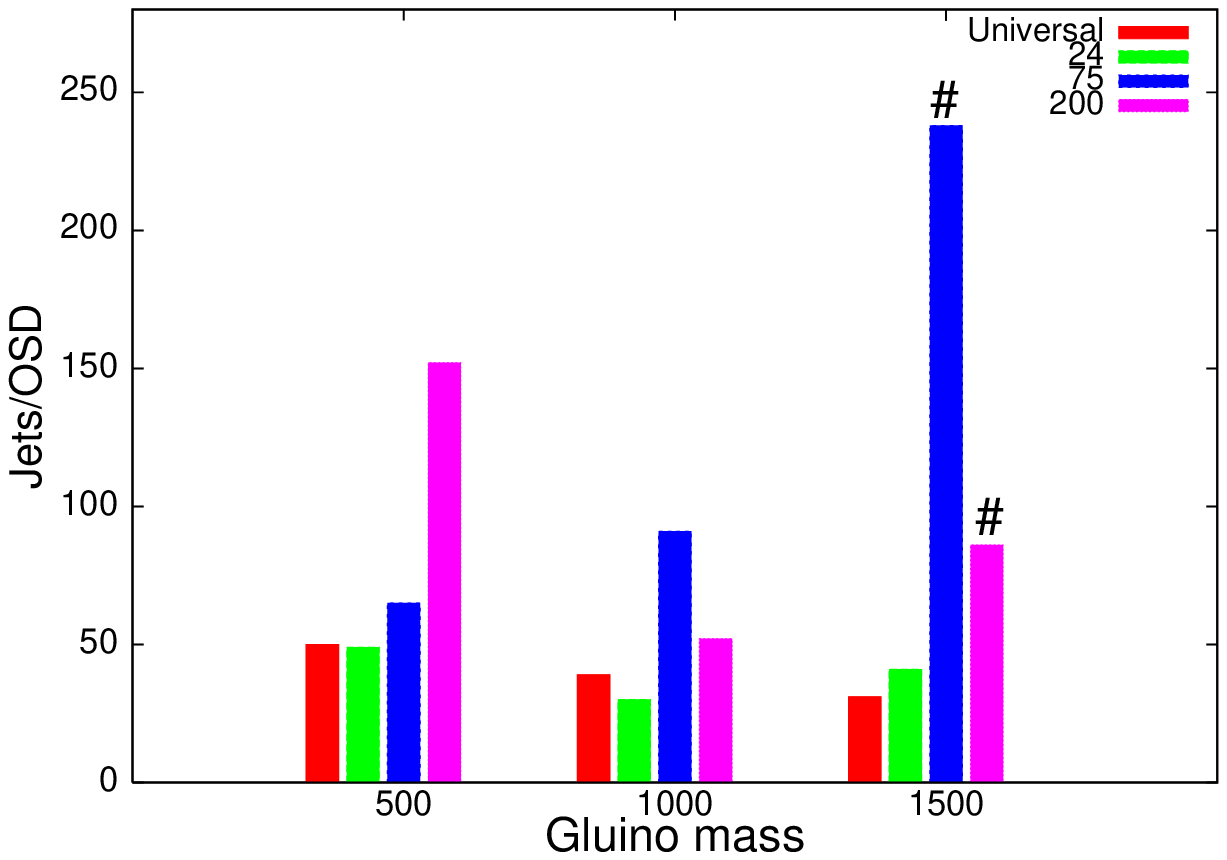,width=6.5 cm,height=5.50cm,angle=-0}
\hskip 20pt \epsfig{file=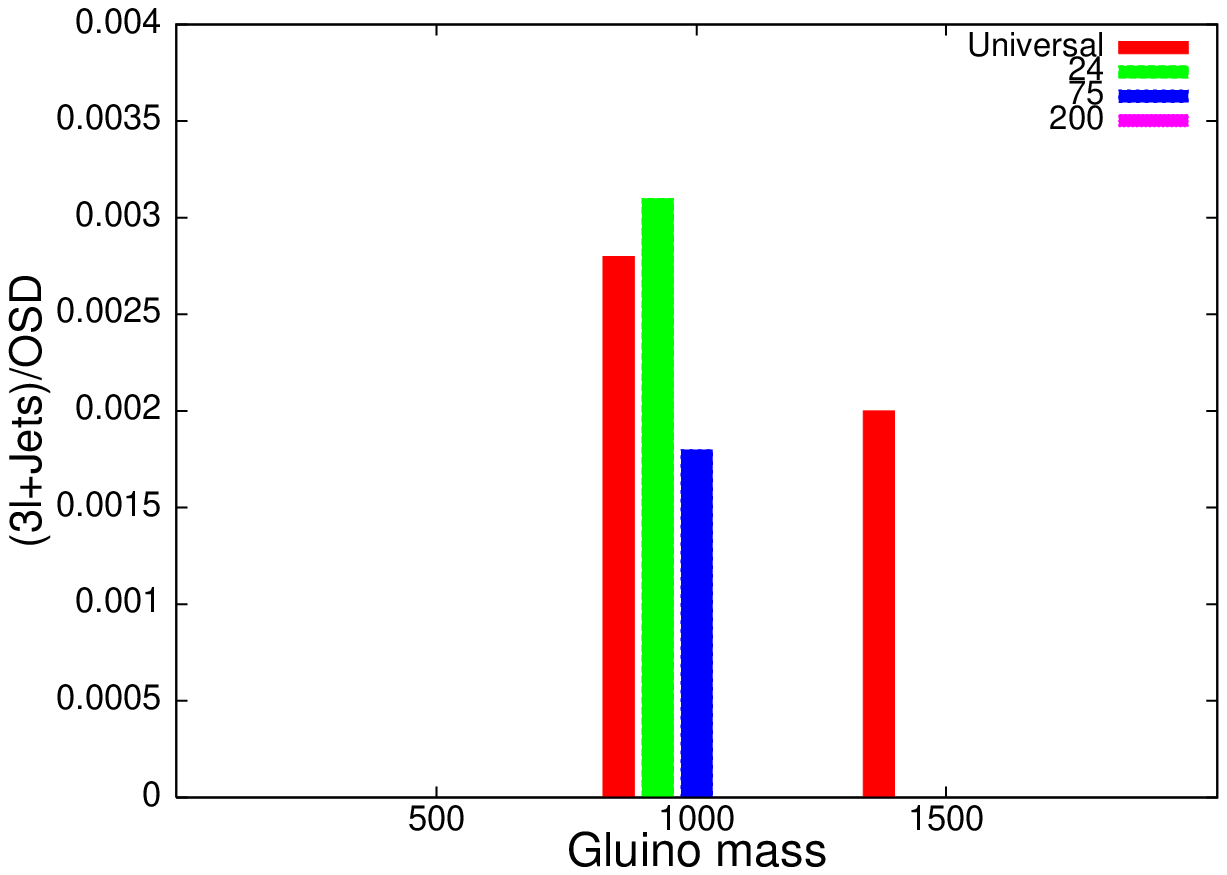,width=6.5cm,height=5.50cm,
angle=-0}}
\caption{ Event ratios for {\bf pMSSM} in SU(5): $m_{\tilde f}=$1000 GeV, 
$\mu=300$ GeV, $\tan{\beta}=40$} 
%distribution}
\end{center}

\end{figure}

\item In general, the $(3\ell + jets)$ channel is a rather useful
discriminator. This is because in the non-universal cases, especially
for {\bf 75} and {\bf 200}, the masses  of $\chi^0_2$ and
$\chi^{\pm}_1$ are rather widely spaced, as opposed to
the case of universality. This can be attributed to the fact that
the ratio $M_2 /M_1$ is different from the universal case,
and, while the gaugino contribution to $\chi^{\pm}_1$
comes exclusively from the Wino,  $\chi^0_2$ has Bino
contributions as well with the altered mass ratios. For {\bf 24}, 
too, the spacing between
$\chi^0_2$ and $\chi^0_1$ is different from the universal case.
Thus the suppression of trileptons for {\bf 75} and {\bf 200} can
be useful, while the maximum number of such events can be
obtained in the universal case. 
All these affect the branching ratios for 
$\chi^0_2  \chi^{\pm}_1 \longrightarrow 3\ell + E_{T}\!\!\!\!/$. 
However, events rates tend to be low in this channel,
as a result of which its ratio with the OSD rates cannot be presented
in a number of cases. However, the rates are in general on the
higher side for $\tan\beta~=~40$ than $5$, because of the
lower mass of the lighter sbottom state in the former case, which
enhances its production and subsequent cascades to $\chi^{\pm}_1$ and
$\chi^0_2$. Besides, the compositions of  $\chi^{\pm}_1$ and
$\chi^0_2$ also is somewhat altered by a different $\tan\beta$.

\begin{figure}[t]
\begin{center}
%\vspace*{-2.2cm}
\centerline{\epsfig{file=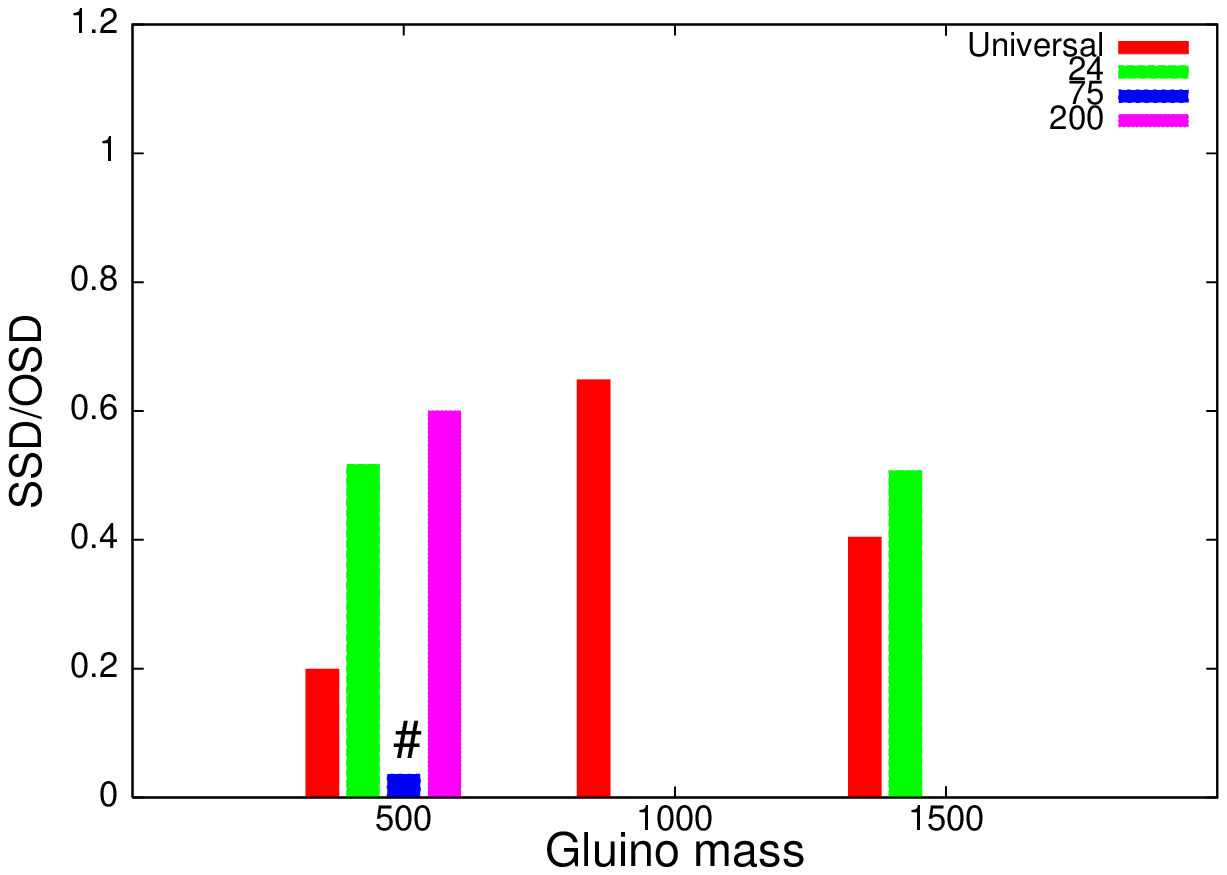,width=6.5 cm,height=5.50cm,angle=-0}
\hskip 20pt \epsfig{file=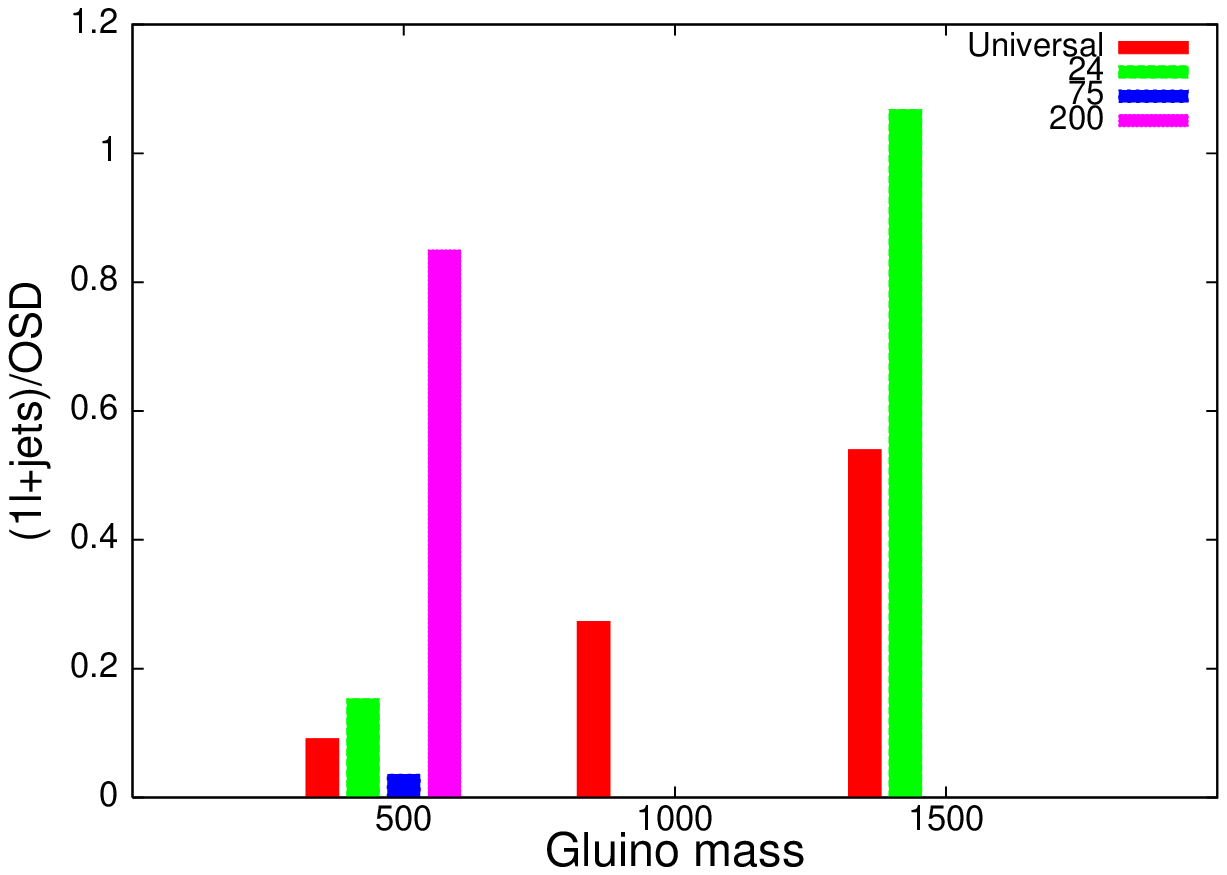,width=6.5cm,height=5.5cm,angle=-0}}
\vskip 10pt
{\epsfig{file=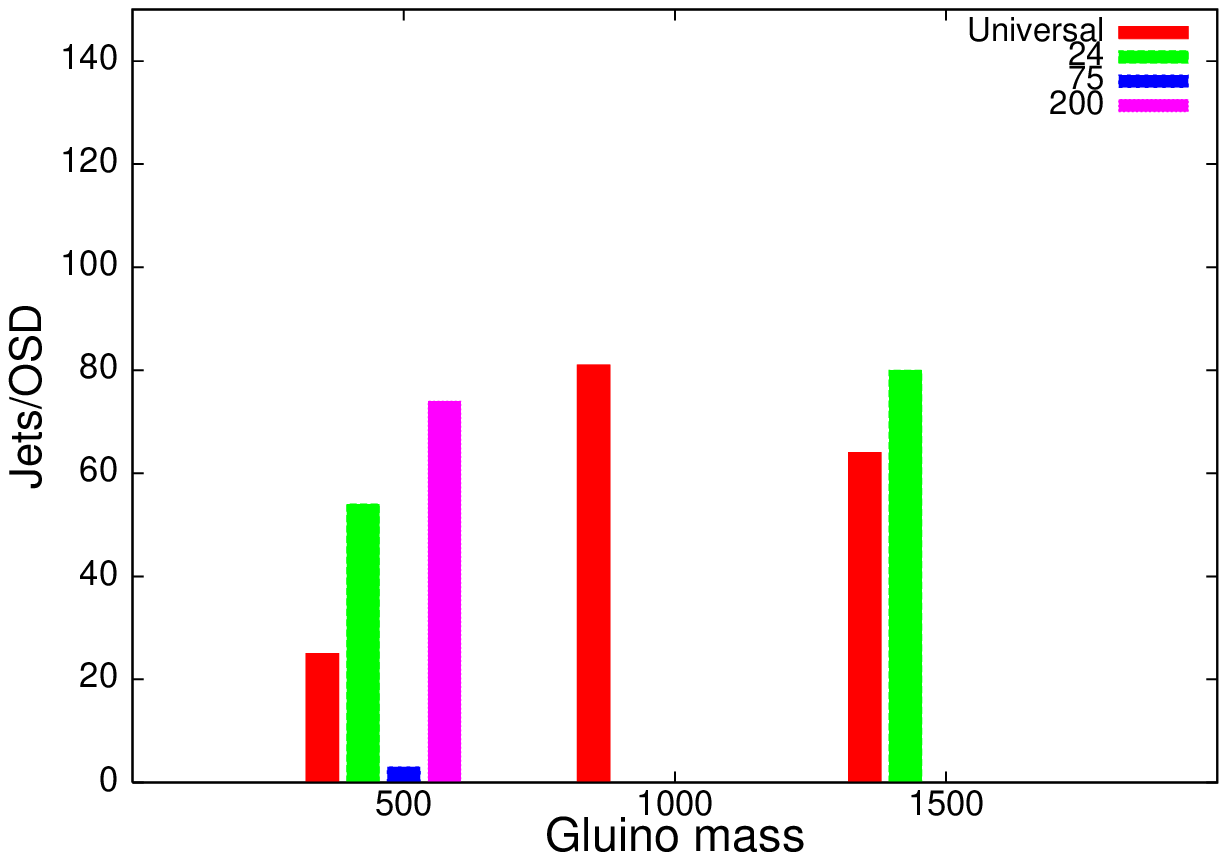,width=6.5 cm,height=5.50cm,angle=-0}}

\caption{ Event ratios for {\bf pMSSM} in SU(5): $m_{\tilde f}=$500 GeV, 
 $\mu=1000$ GeV, $\tan{\beta}=5$} 
%distribution}
\end{center}

\end{figure}

\item SSD/OSD is usually less useful in distinguishing among the 
different cases of non-universality. This is because the modified 
gaugino mass ratios at high scale due to non-singlet GUT-breaking
representations usually tend to affect $m_{\chi^{\pm}_1}$ and
$m_{\chi^0_2}$ similarly, thus having the same impact on both
the SSD and OSD rates.

\item The rates for single lepton events, as in the case of 
trileptons,  are affected significantly once the isolation cut 
between leptons and jets is turned on.

\item The absolute rates for events with jets in the final state
are always way above the backgrounds with the cuts adopted here.
However, the suppression of OSD, SSD and single-lepton channels
for (a) high gluino/squark masses  and (b) relatively higher
chargino/neutralino masses for cases such as {\bf 75} and {\bf 200}
often tend to drown them with backgrounds, as a result of
which the ratios are likely to be useful only when statistics
can be significantly improved. The trilepton events are rather
easy to keep above backgrounds, due to the rather stiff jet $p_T$ cut
and the missing-$E_T$ cut.

\begin{figure}[t]
\begin{center}
%\vspace*{-2.2cm}
\centerline{\epsfig{file=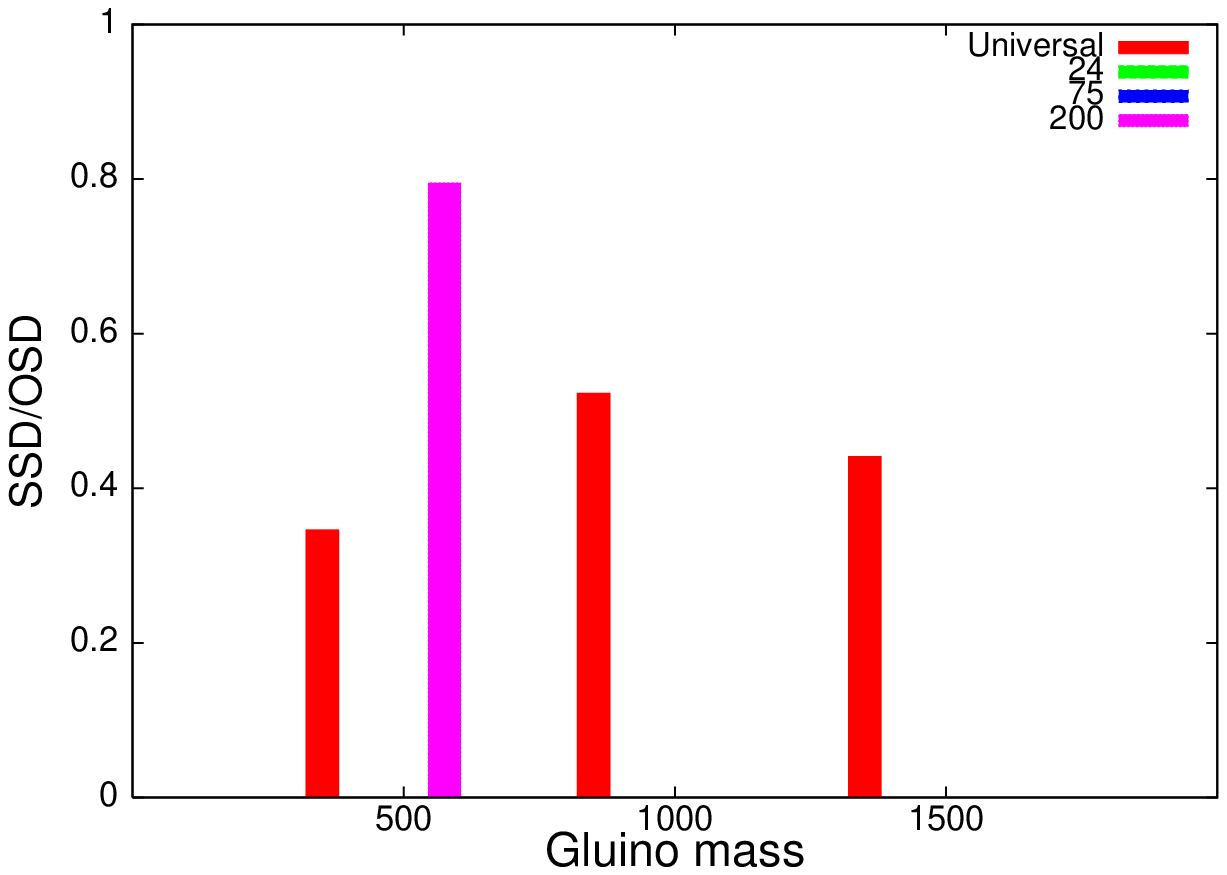,width=6.5 cm,height=5.50cm,angle=-0}
\hskip 20pt \epsfig{file=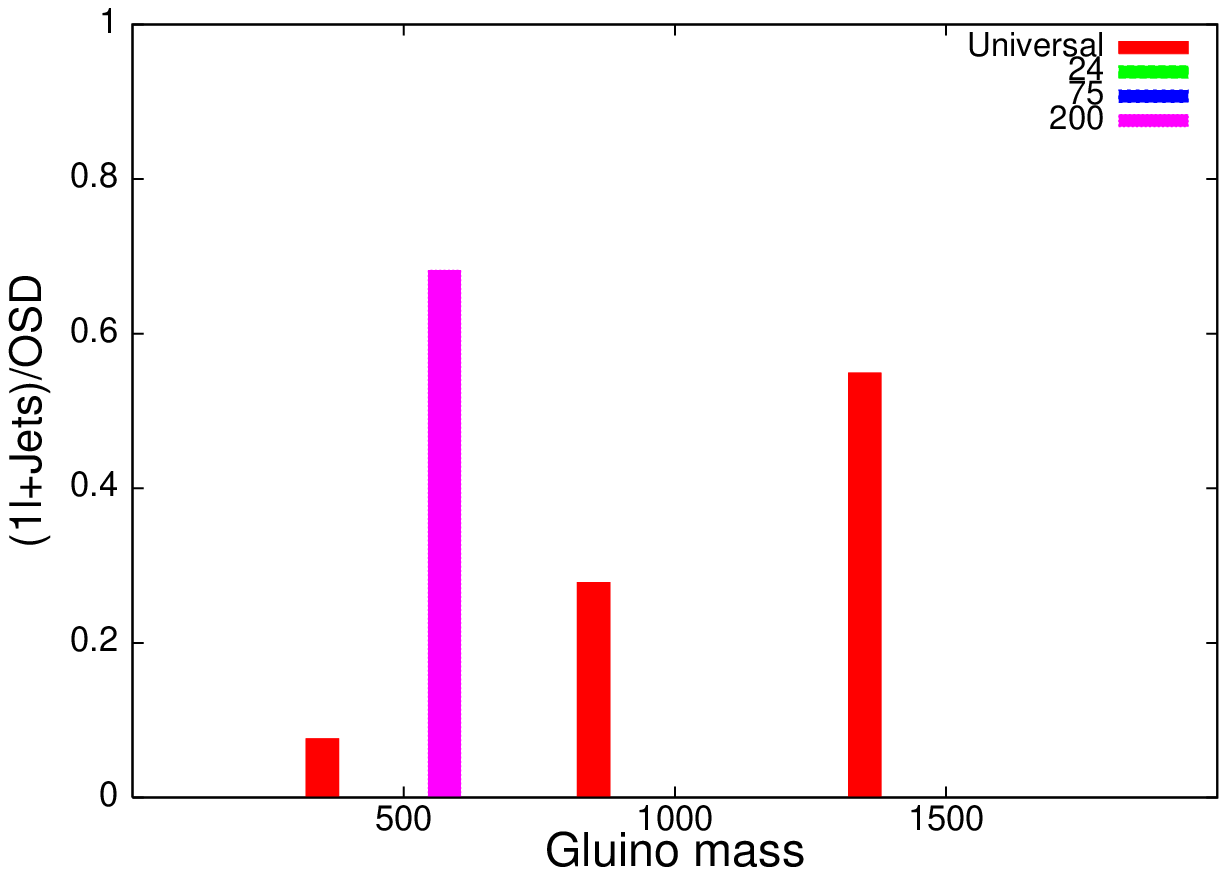,width=6.5cm,height=5.50cm,angle=-0}}
\vskip 10pt
{\epsfig{file=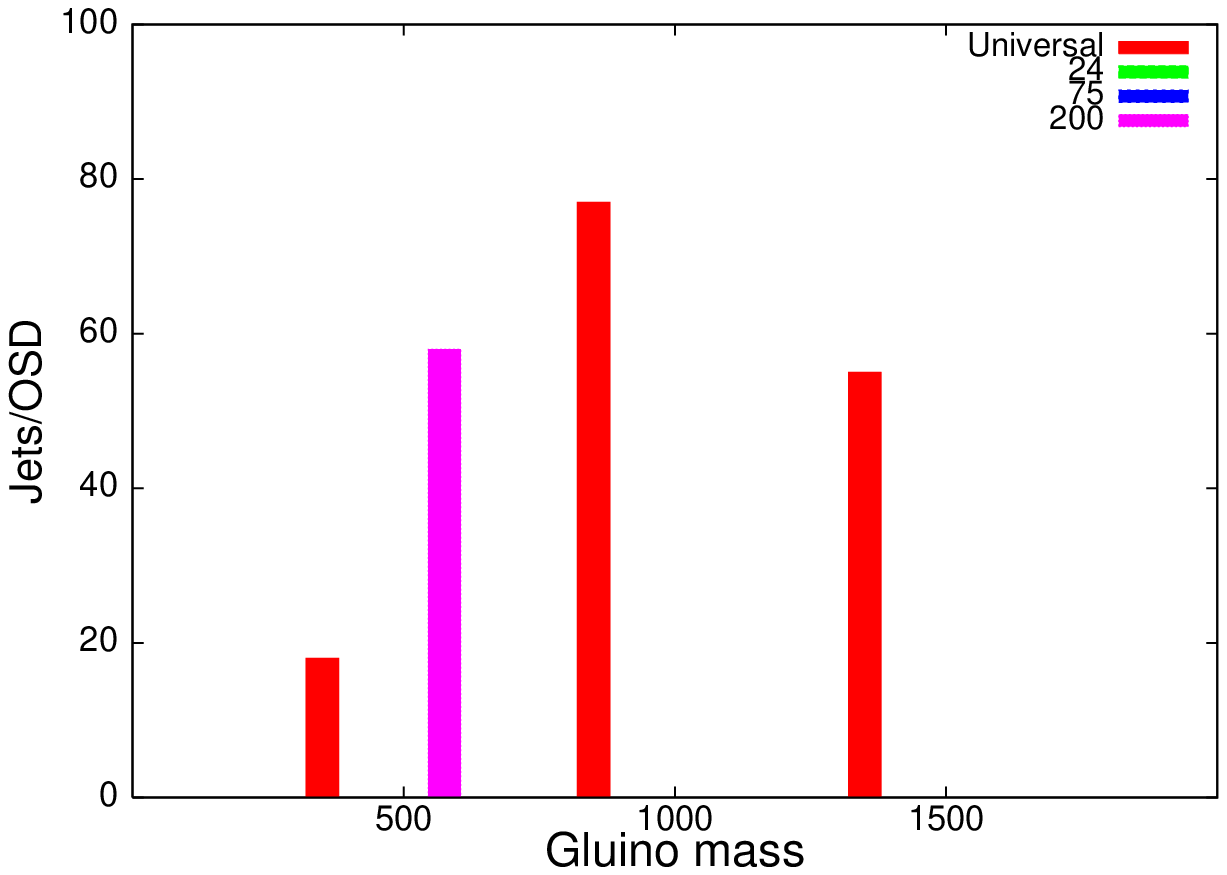,width=6.5 cm,height=5.50cm,angle=-0}}

\caption{ Event ratios for {\bf pMSSM} in SU(5): $m_{\tilde f}=$500 GeV, 
$\mu=$1000 GeV, $\tan{\beta}=40$} 
%distribution}
\end{center}

\end{figure}

\item The SSD and single lepton events (and sometimes the OSD events) 
for $m_{\tilde f}$= 1000 Gev,
and gluino mass in the range of 1000 GeV or higher, are relatively
background-prone for {\bf 75} and {\bf 200}. The reason for this is
higher values of the chargino and neutralino masses and the suppression
of leptonic final states by heavy sleptons.

\item For $\mu$= 1000 GeV, $m_{\tilde f}$= 500 GeV and
$m_{\tilde {g}}\gg$ 500 GeV,  most of the 
non-universal scenarios give inconsistent spectrum, because 
both the gaugino and Higgsino components of the lightest neutralino
tend to make it heavier than some sfermion(s). For 
$m_{\tilde {g}}$= 500 GeV, too, this happens for $\tan\beta$= 40,
as it lowers the lighter stau mass below that of $\chi^0_1$.

\begin{figure}[t]
\begin{center}
%\vspace*{-2.2cm}
\centerline{\epsfig{file=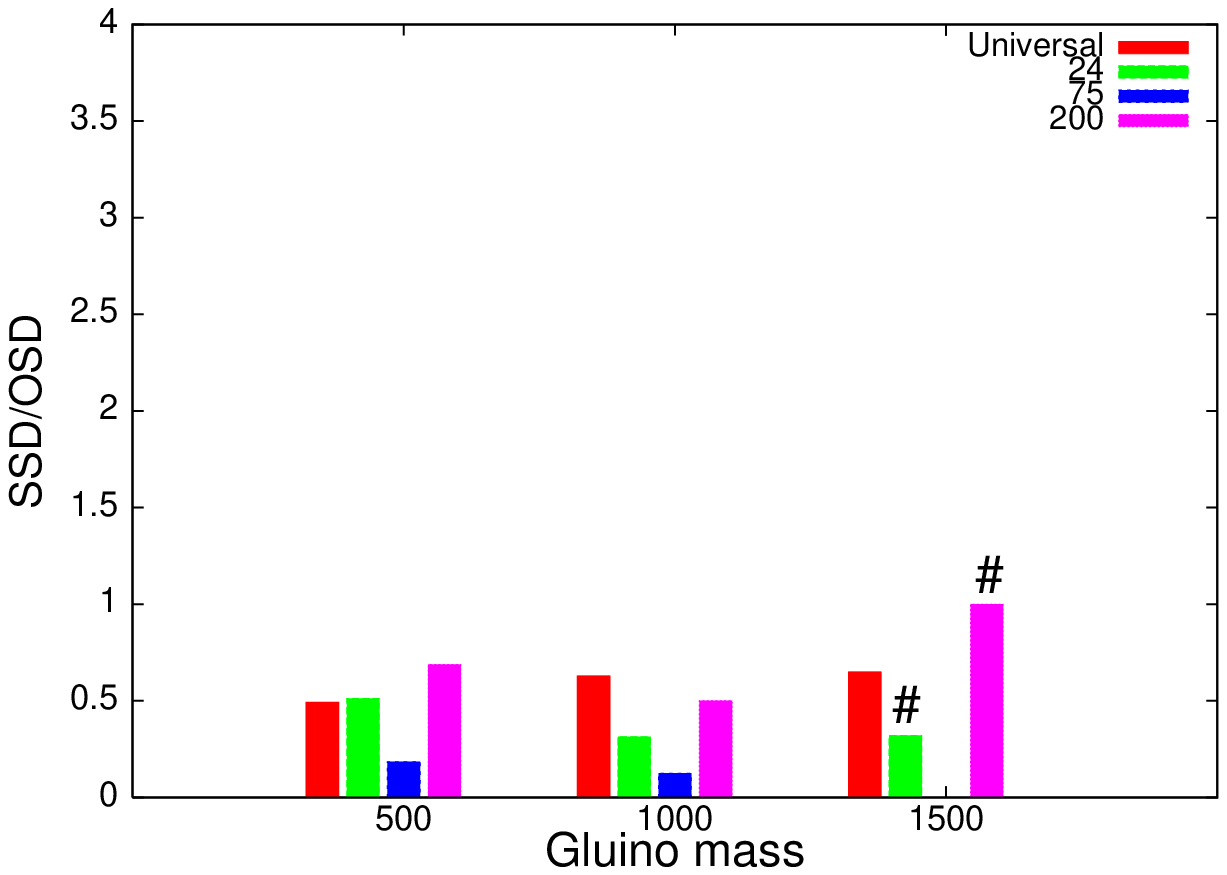,width=6.5 cm,height=5.50cm,angle=-0}
\hskip 20pt \epsfig{file=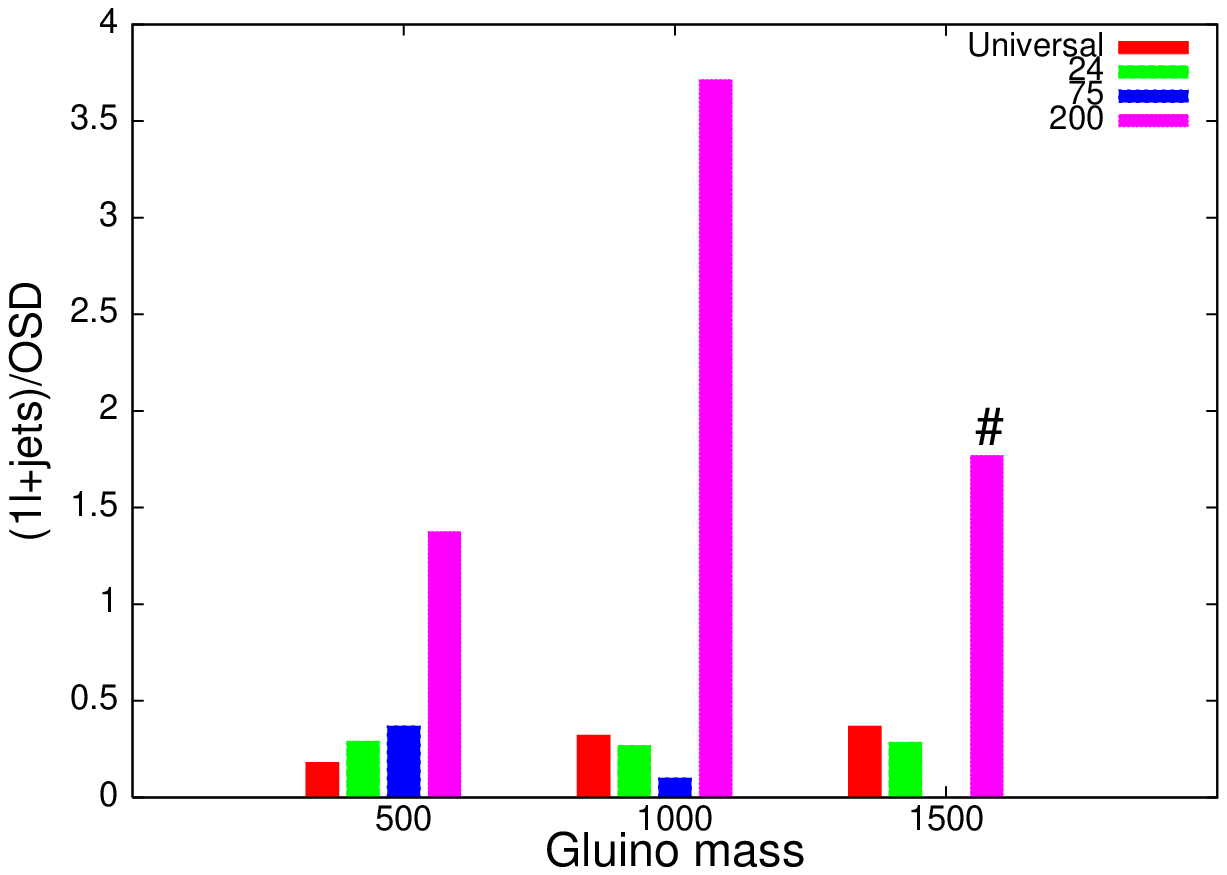,width=6.5cm,height=5.5cm,angle=-0}}
\vskip 10pt
\centerline{\epsfig{file=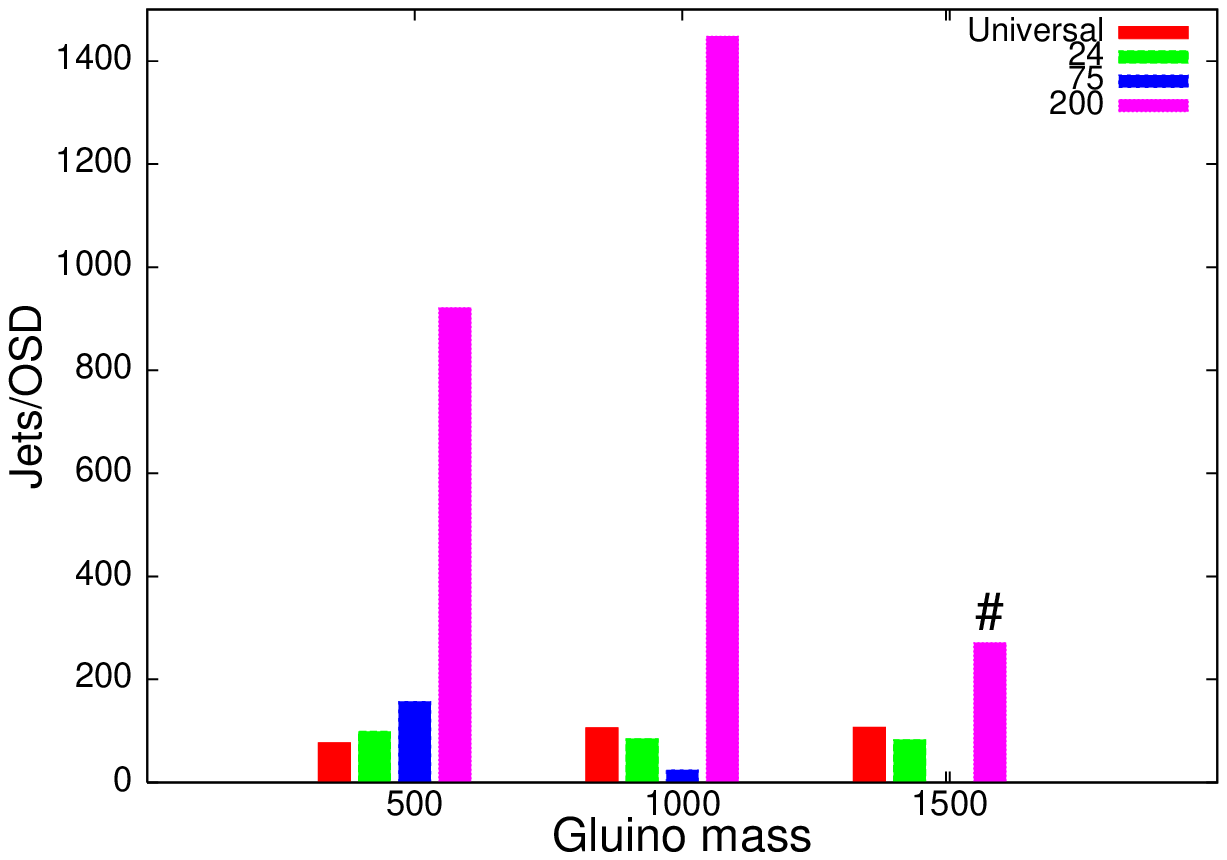,width=6.5 cm,height=5.50cm,angle=-0}
\hskip 20pt \epsfig{file=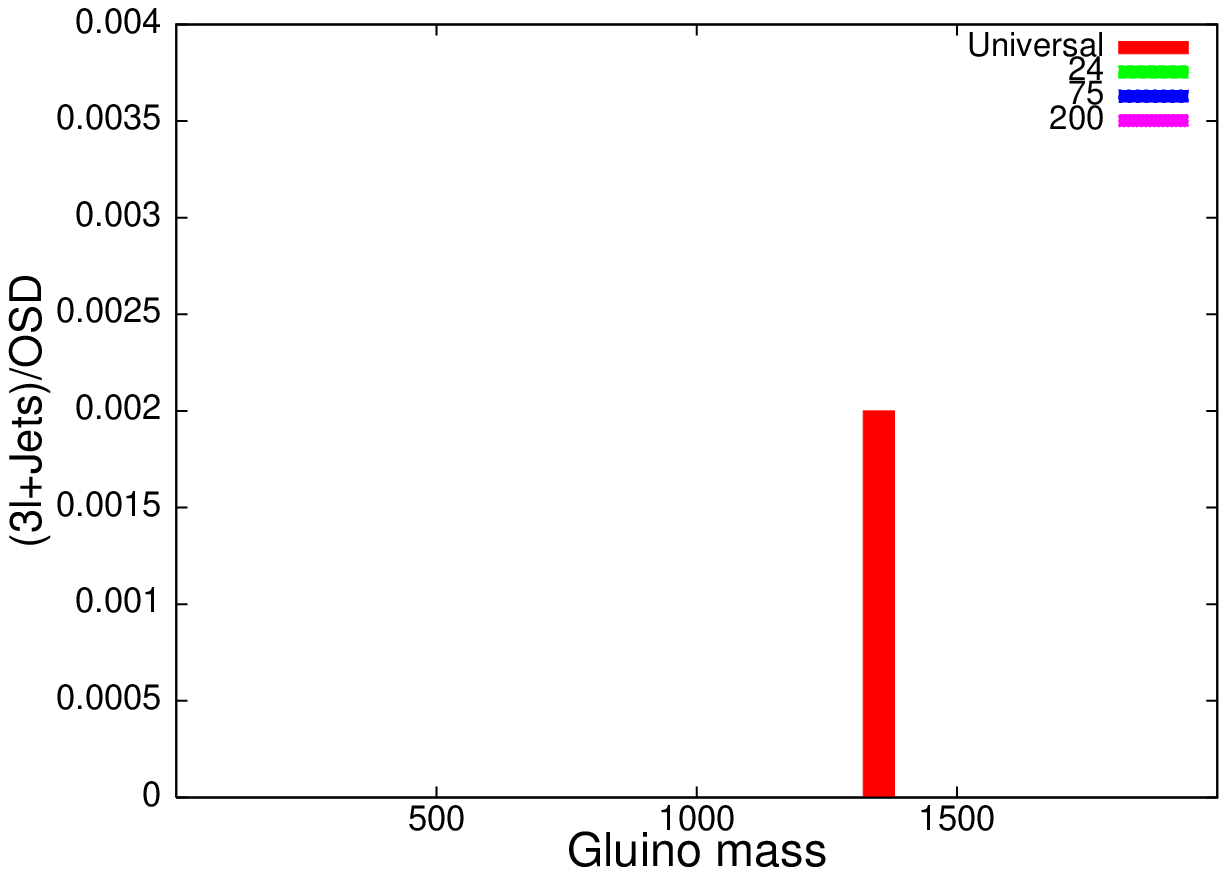,width=6.5cm,height=5.50cm,angle=-0}}
\caption{ Event ratios for {\bf pMSSM} in SU(5): $m_{\tilde f}=$1000 GeV, 
 $\mu=$1000 GeV, $\tan{\beta}=5$} 
%distribution}
\end{center}

\end{figure}

\item For $\mu$ increased from 300 GeV to 1000 GeV in the universal case,
particularly with gauginos on the lower side, the Higgsino component
in the lighter charginos/neutralinos decreases and enhances the
probability of leptons arising from cascades. Thus, say, the ratio
$jets$/OSD is smaller for higher $\mu$. This feature, however, is
not always there (for example for non-universality, ostensibly
due to the more complicated gaugino mass ratios as well as the
different hierarchy between the gluino and chargino/neutralino 
masses. 

\item It should be noted (in the contexts of both SU(5) and SO(10) )
that no observation is predicted in some channels for
certain representations and in certain regions of the parameter space.
Such `null observations', however, can themselves be of use in
distinguishing among scenarios. 
\end{enumerate}

\begin{figure}[t]
\begin{center}
%\vspace*{-2.2cm}
\centerline{\epsfig{file=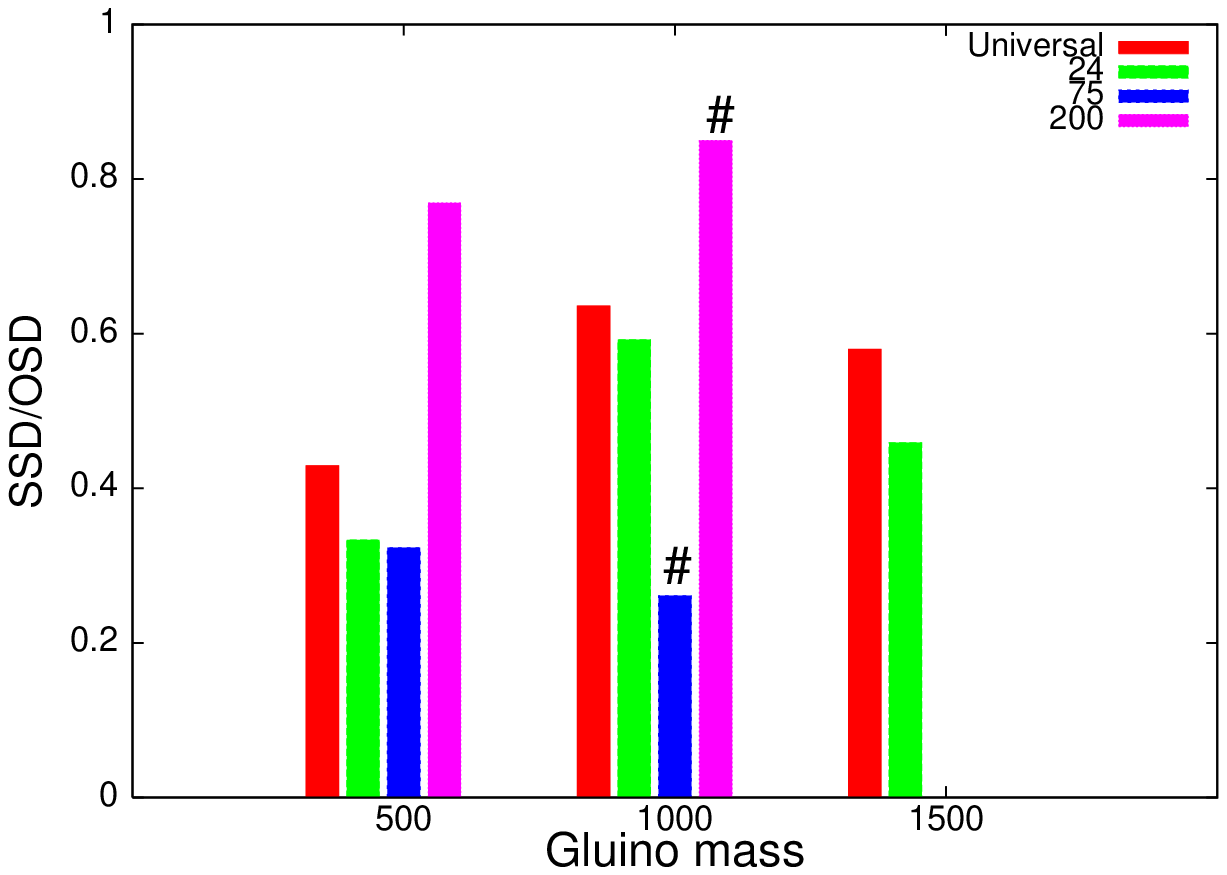,width=6.5 cm,height=5.50cm,angle=-0}
\hskip 20pt \epsfig{file=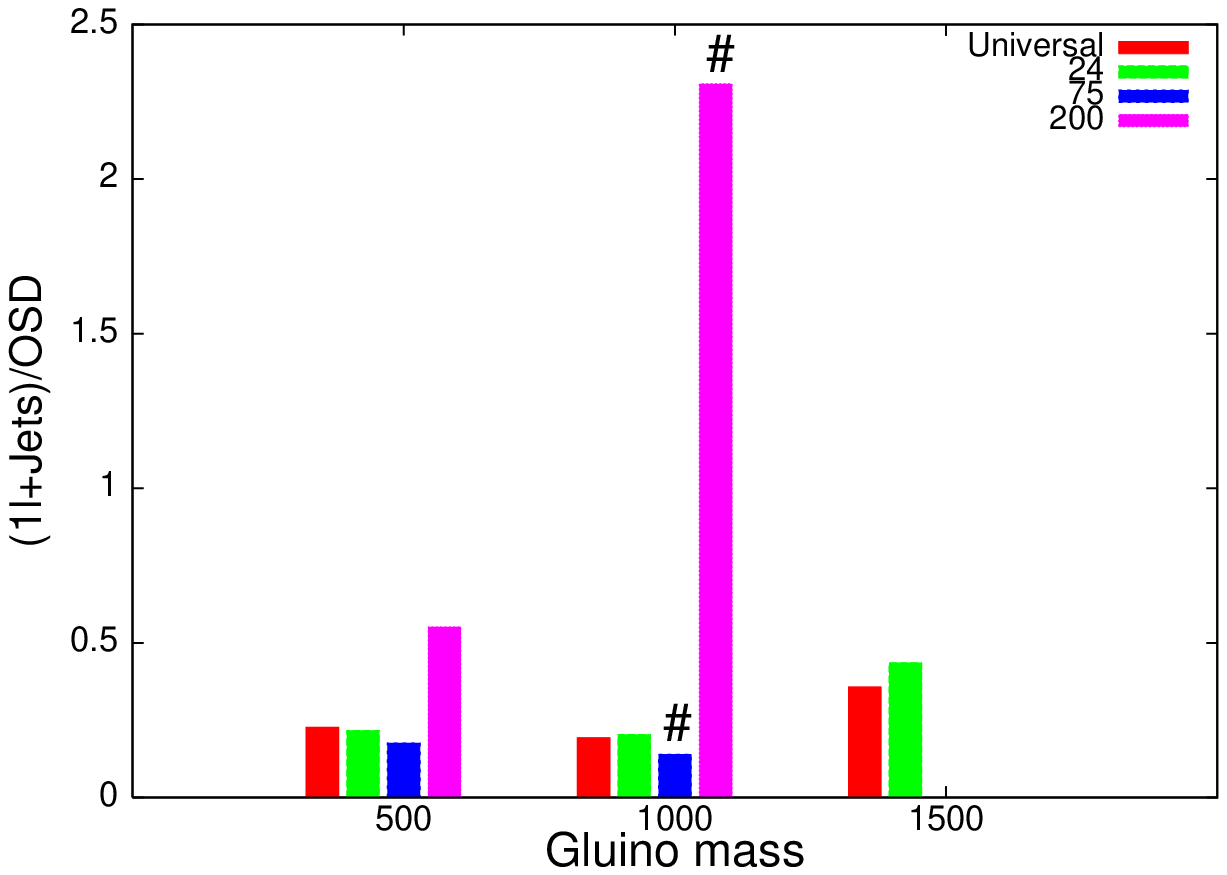,width=6.5cm,height=5.50cm,angle=-0}}
\vskip 10pt
\centerline{\epsfig{file=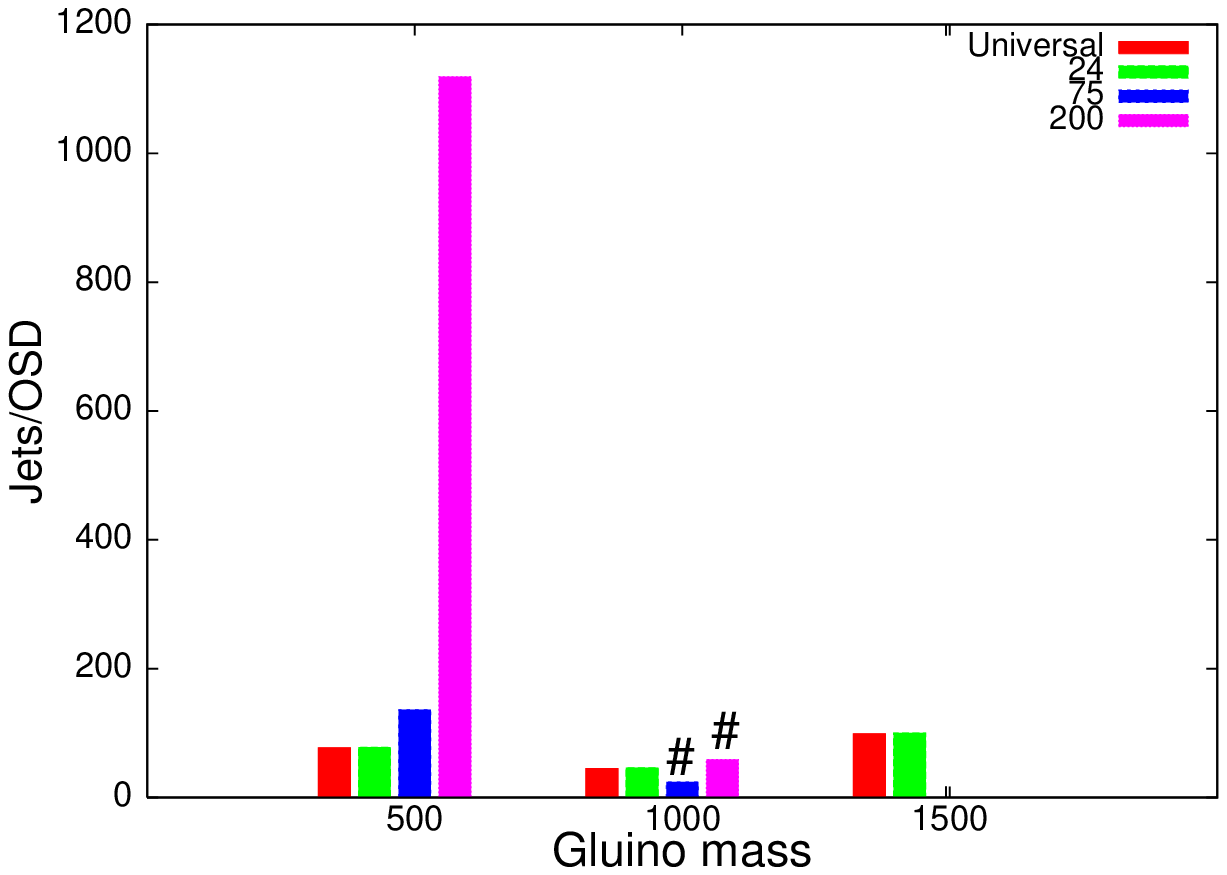,width=6.5 cm,height=5.50cm,angle=-0}
\hskip 20pt \epsfig{file=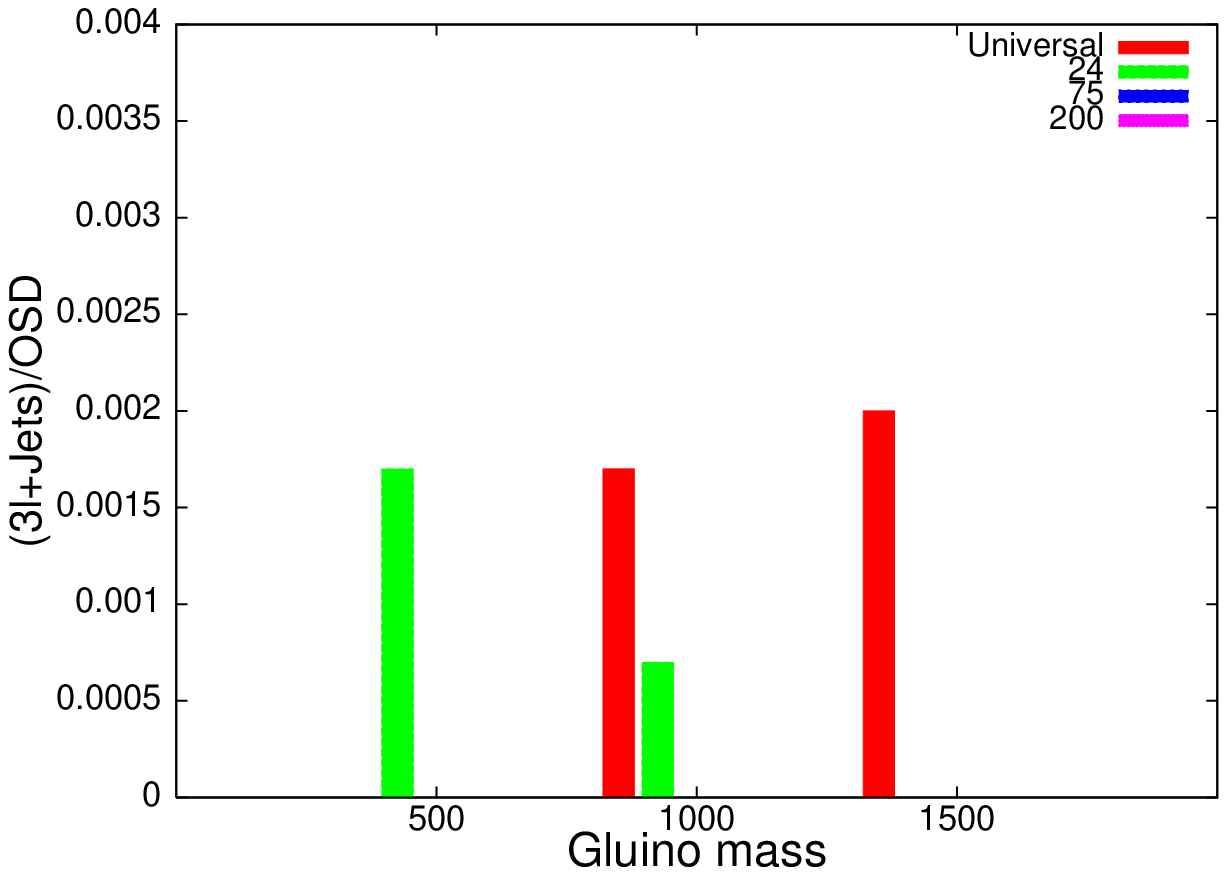,
width=6.5cm,height=5.50cm,angle=-0}}
\caption{ Event ratios for {\bf pMSSM} in SU(5): $m_{\tilde f}=$1000 GeV,
 $\mu=$1000 GeV, $\tan{\beta}=40$} 
%distribution}
\end{center}

\end{figure} 

In the region of the parameter space illustrated in figure 1, the 
$3\ell+jets$ channel search gives null result for all the representations. 
For  $m_{\tilde g}$=500 GeV, one can distinguish the case of {\bf 75}  
from others from the ratio $(1\ell+jets)$/OSD, and {\bf 200} from $jets$/OSD. 
It is very difficult to distinguish the universal and {\bf 24}   
from any of the plots. However, as has been mentioned already, one can 
do so from the absolute number 
in the OSD channel search where {\bf 24} gives a 
significantly larger number. 
 For $m_{\tilde g}$= 1000 GeV, the ratios for both {\bf 75} and {\bf 200} 
are distinctly larger than those for {\bf 24} and the universal 
case, when one considers SSD/OSD, $(1\ell+jets)$/OSD and $jets$/OSD. However, 
distinguishing between {\bf 75} and {\bf 200} is difficult 
not only in this ratio space but also from the absolute rates. 
Distinction between the remaining two representation 
is possible through SSD/OSD and also through the absolute rates in 
the OSD channel, where the universal case gives 
sufficiently larger number than {\bf 24}. This is because the charginos and 
higher neutralinos become sufficiently heavy in the latter case.
For $m_{\tilde g}$= 1500 GeV, the leptonic signals corresponding to
{\bf 75} are beset with backgrounds, thus putting the ratio  SSD/OSD
at the mercy of statistics. {\bf 200} can be 
separated through SSD/OSD or $(1\ell+jets)$/OSD, while
{\bf 75} is distinguishable from {\bf 1} and {\bf 24} quite clearly  
with the help of $(1\ell+jets)$/OSD. 
However, the distinction between {\bf 24} and 
the universal case is still difficult. Figure 2 differs from the figure 1 
only in $\tan \beta$, whose effect on (3l+Jets)/OSD has already
been discussed. The SSD/OSD values in this case  shows a different 
behaviour from $\tan\beta ~=~$5 for $m_{\tilde g}$= 1000 GeV, the
ratio showing a rather flat character with respect to gluino mass 
variation. Moreover, the ratio $(1\ell+jets)$/OSD also shows a significant
enhancement for {\bf 75}. 

\begin{figure}[t]
\begin{center}
%\vspace*{-2.2cm}
\centerline{\epsfig{file=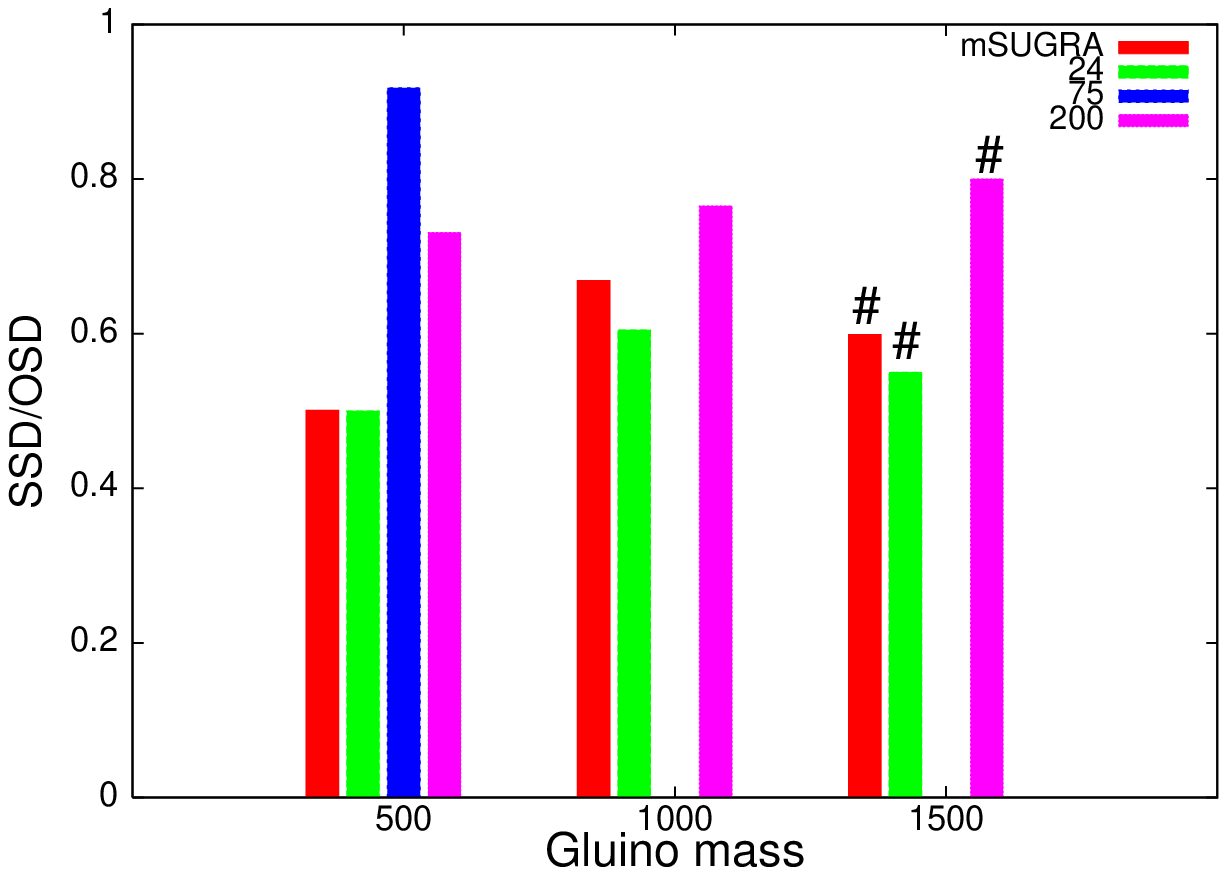,width=6.5 cm,height=5.50cm,angle=-0}
\hskip 20pt \epsfig{file=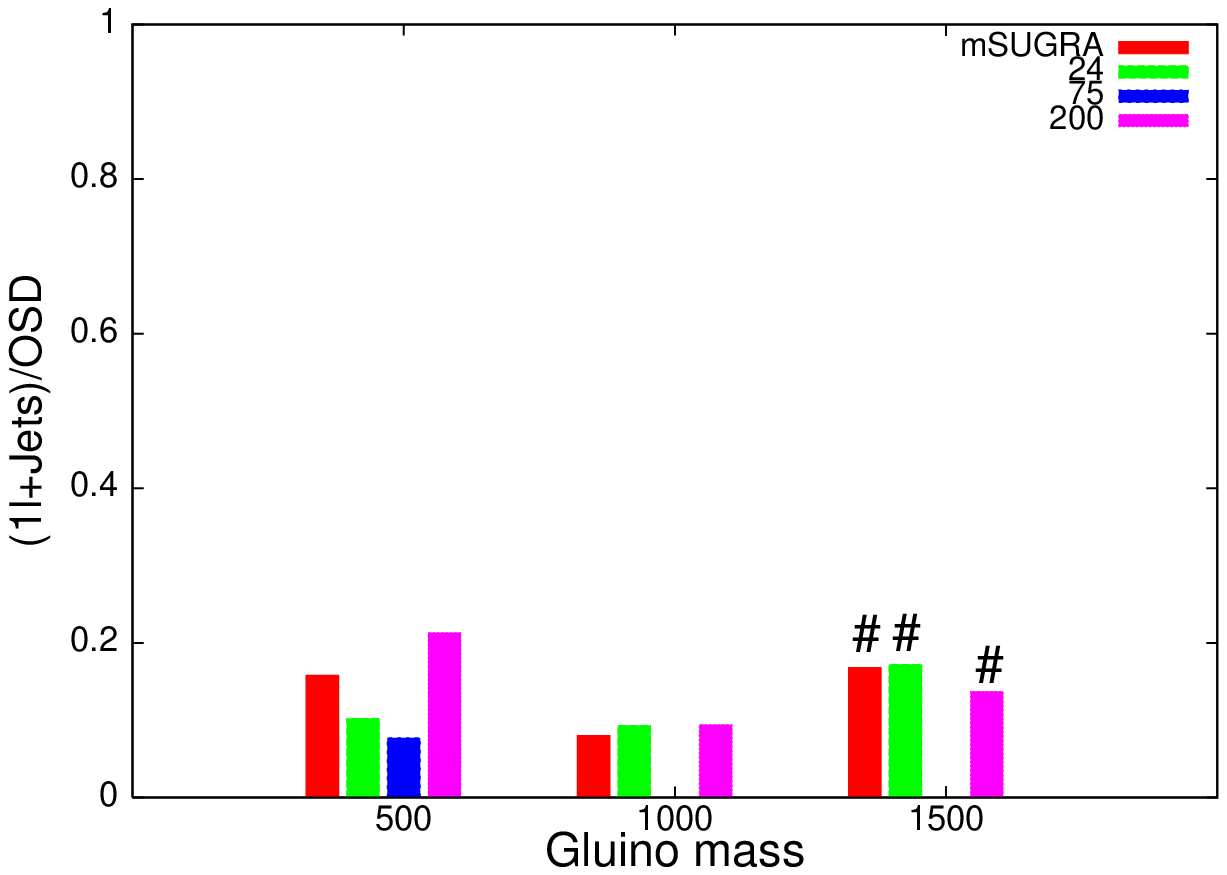,width=6.5cm,height=5.50cm,angle=-0}}
\vskip 10pt
\centerline{\epsfig{file=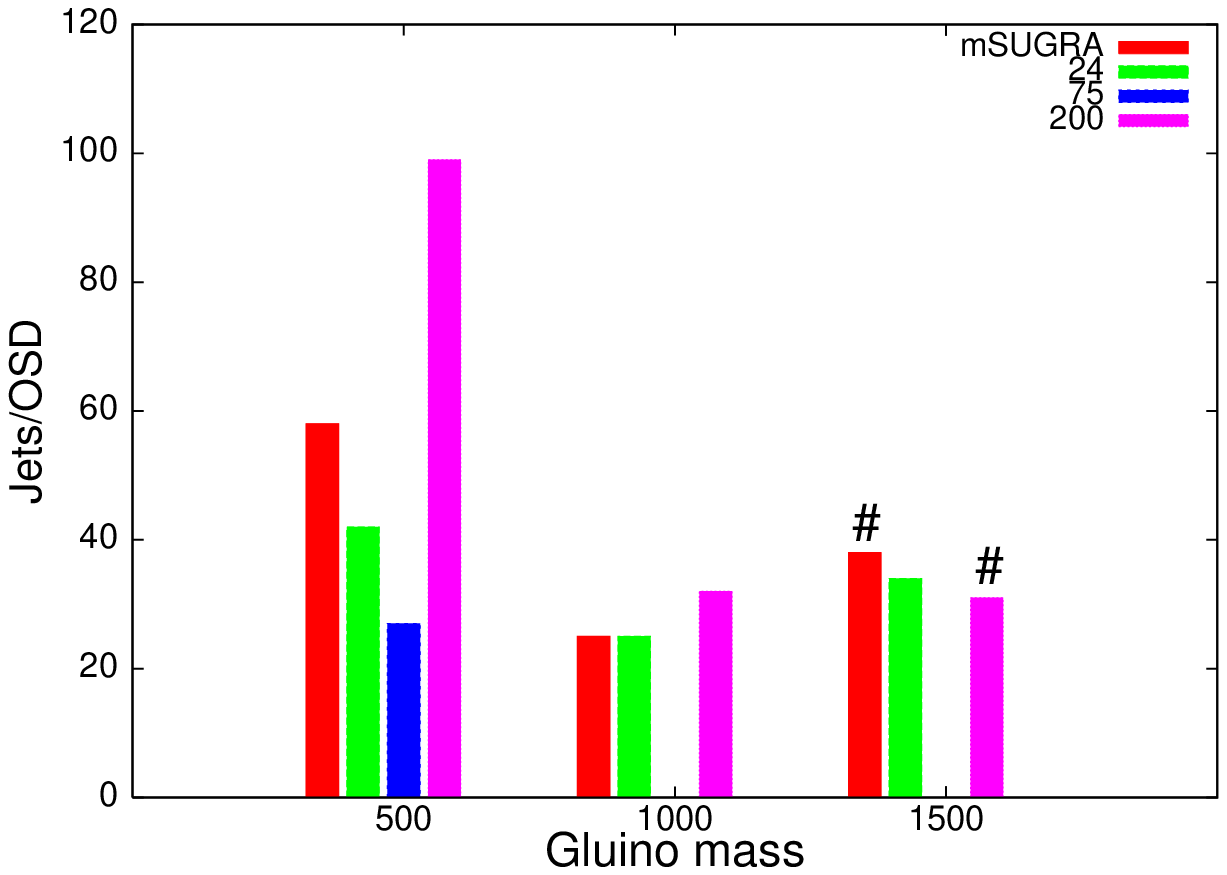,width=6.5 cm,height=5.50cm,angle=-0}
\hskip 20pt \epsfig{file=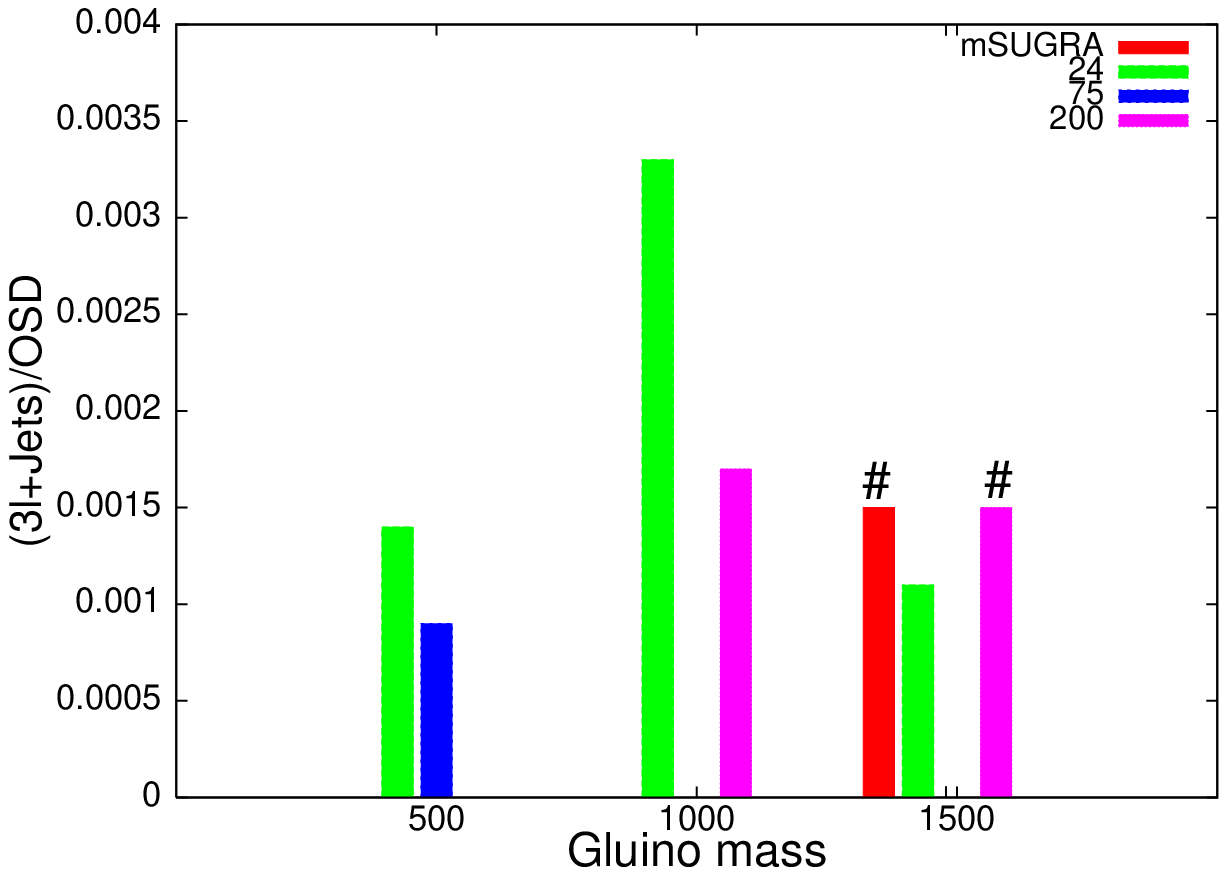,width=6.5cm,
height=5.50cm,angle=-0}}
\caption{ Event ratios for SU(5) {\bf SUGRA} with non-universal gaugino masses: 
$m_{0}=506$ GeV, $\tan{\beta}=5$, $sgn(\mu)=+$, $A_{0}=0$} 
%distribution}
\end{center}

\end{figure} 

Figures 3 and 4 differ from figures 1 and 2 in terms of  $m_{\tilde f}$ only. 
For $m_{\tilde g}$= 500 GeV, 
the ratios $(1\ell+jets)$/OSD and $jets$/OSD for {\bf 75} and {\bf 200} are 
well separated from others for $\tan\beta$= 5, while the distinction
between these two representations from the ratios is difficult.
For  $\tan\beta$= 40, however,  SSD/OSD and  $jets$/OSD make
such distinction possible. Similar conclusions can be drawn for
higher gluino masses as well, except that the $(3\ell+jets)$ channel
emerges as a successful discriminator for $m_{\tilde g}$= 1000 GeV.

\begin{figure}[t]
\begin{center}
%\vspace*{-2.2cm}
\centerline{\epsfig{file=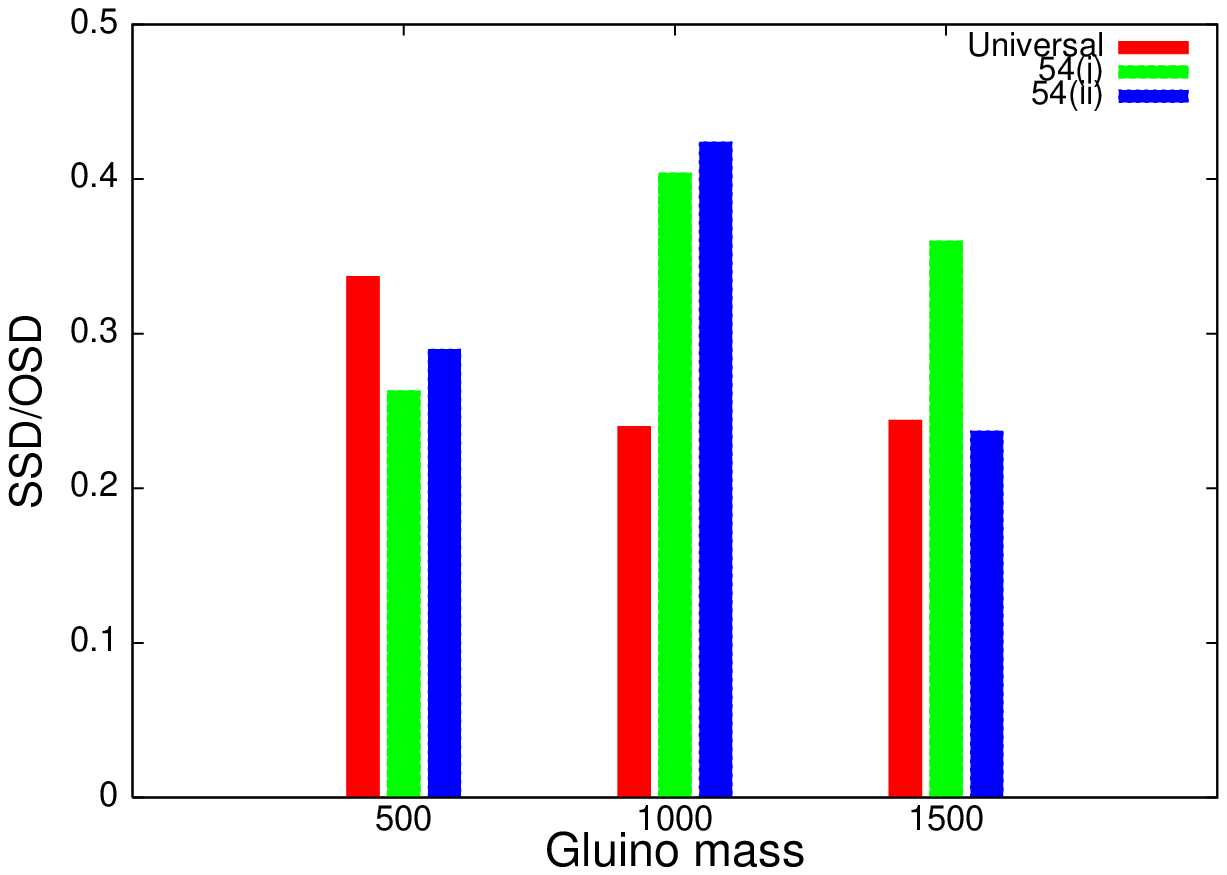,width=6.5 cm,height=5.50cm,angle=-0}
\hskip 20pt \epsfig{file=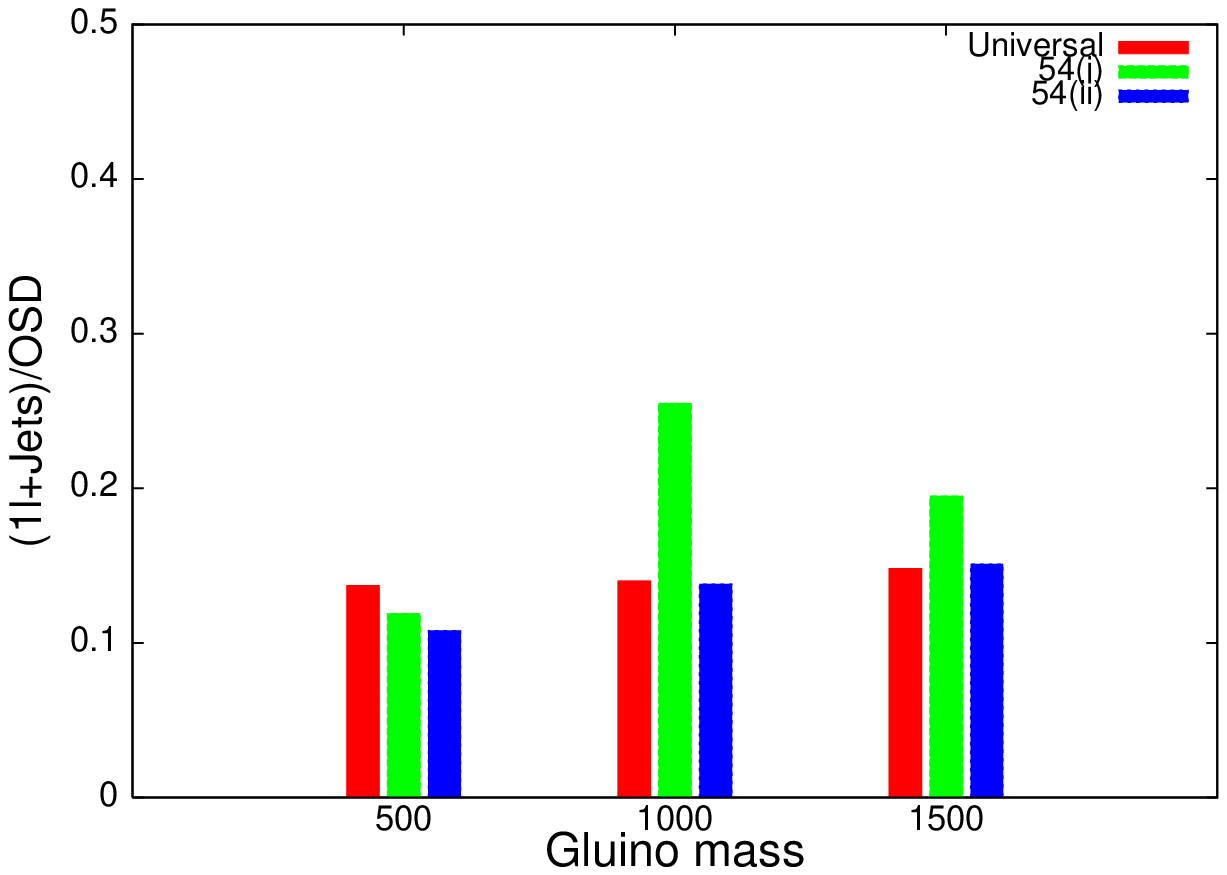,width=6.5cm,height=5.50cm,angle=-0}}
\vskip 10pt
{\epsfig{file=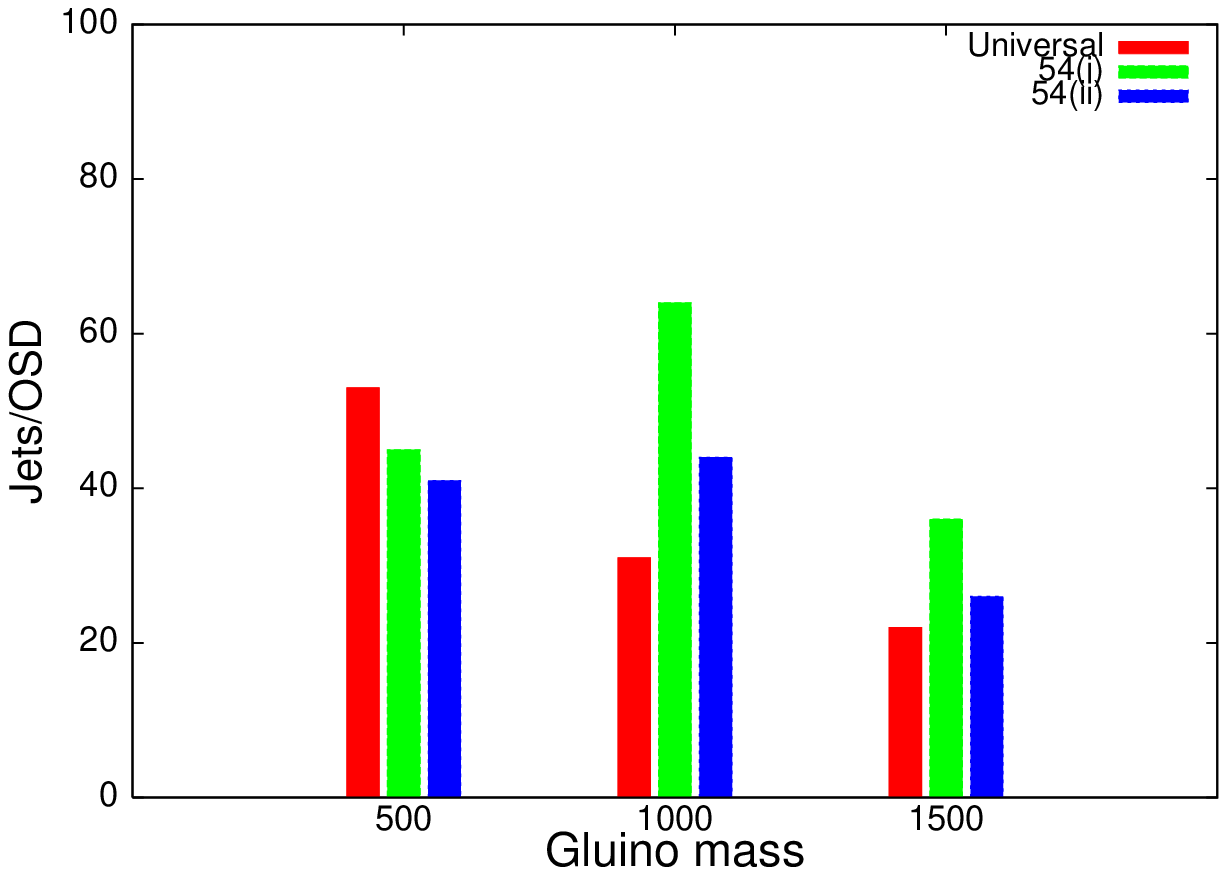,width=6.5 cm,height=5.50cm,angle=-0}}

\caption{ Event ratios for {\bf pMSSM} in SO(10): $m_{\tilde f}=1000$ GeV, 
 $\mu=300$ GeV, $\tan{\beta}=5$} 
%distribution}
\end{center}

\end{figure}

The predictions corresponding to a high value of $\mu$ are shown 
in  figures 5 and 6. This scenario often does not allow a 
consistent spectrum except for a low gluino mass, because, with
sfermion masses on the the low side, the lightest neutralino
is mostly not the LSP. The situation is found to be 
worse for $\tan\beta$= 40. However, all the aforementioned ratios
provide rather easy ways of discriminations among the
different representations for those cases which survive.

Figures 7 and 8 show predictions with both the sfermion masses and
$\mu$ at 1000 GeV. For both the values of $\tan\beta$, {\bf 200}
is clearly differentiable, for cases  where consistent spectra that can
rise above the background are possible. While the ratio SSD/OSD can act as
a fair discriminator for  $\tan\beta$= 40, the $single-lepton$
channel or $jets$/OSD do better for $\tan\beta$= 5,      
The signals for {\bf 24} and the universal case still
require knowledge of the absolute event rates. For $\tan\beta$= 40,
these two representations can be distinguished through $(3\ell+jets)$/OSD,
which does not give sufficient event rates for the universal case for
$m_{\tilde g}$= 500 GeV, while the same thing happens to {\bf 24}
for $m_{\tilde g}$= 1500 GeV. Both of these cases yield measurable
$(3\ell+jets)$/OSD  rates for $m_{\tilde g}$= 1000 GeV, but are sufficiently
apart numerically.

\begin{figure}[t]
\begin{center}
%\vspace*{-2.2cm}
\centerline{\epsfig{file=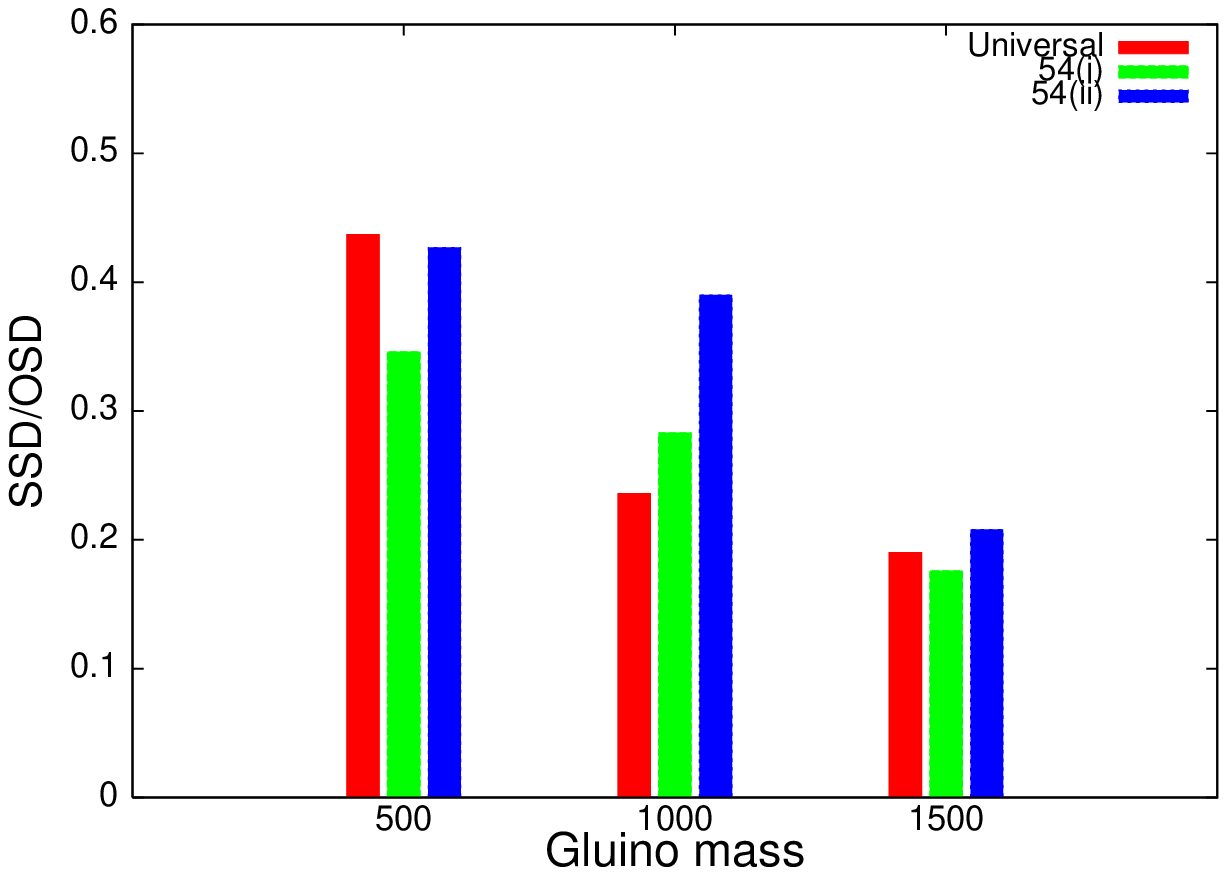,width=6.5 cm,height=5.50cm,angle=-0}
\hskip 20pt \epsfig{file=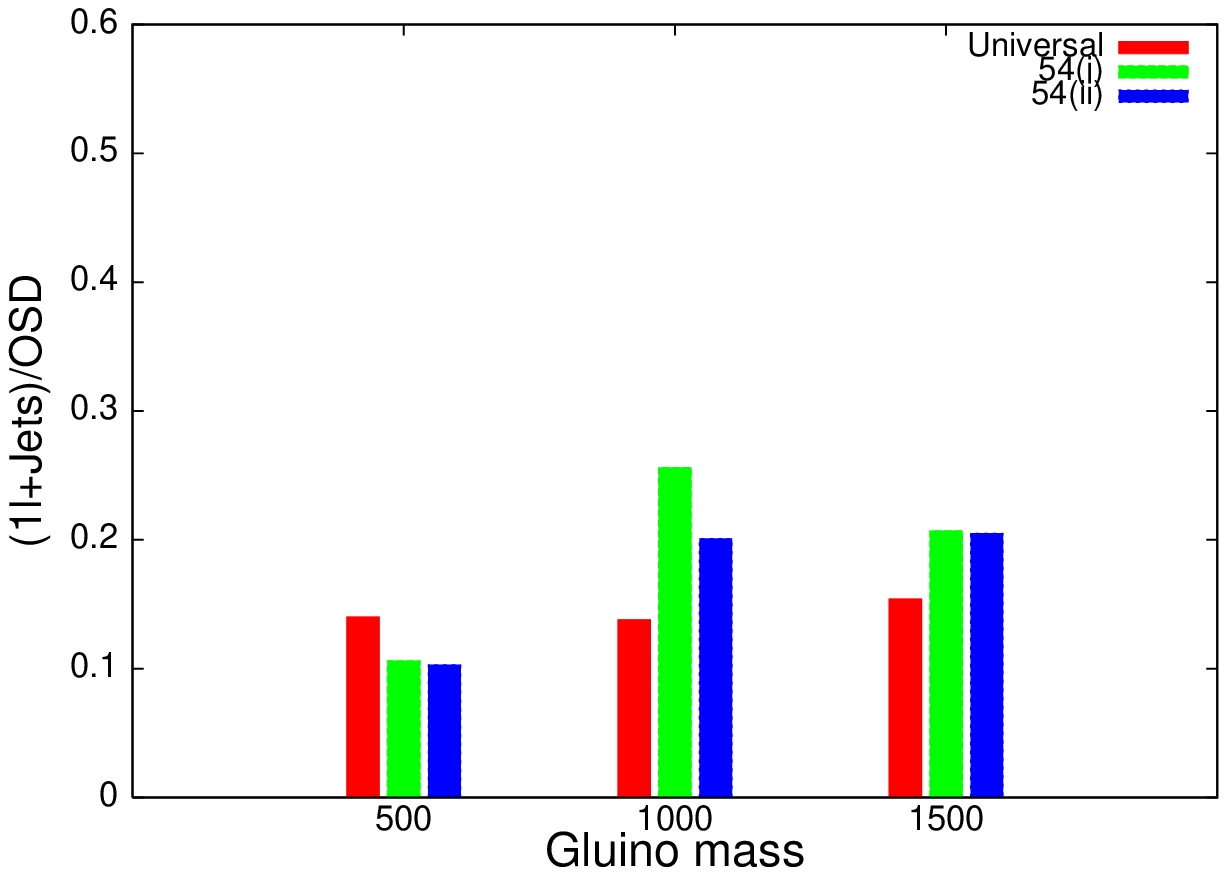,width=6.5cm,height=5.50cm,angle=-0}}
\vskip 10pt
\centerline{\epsfig{file=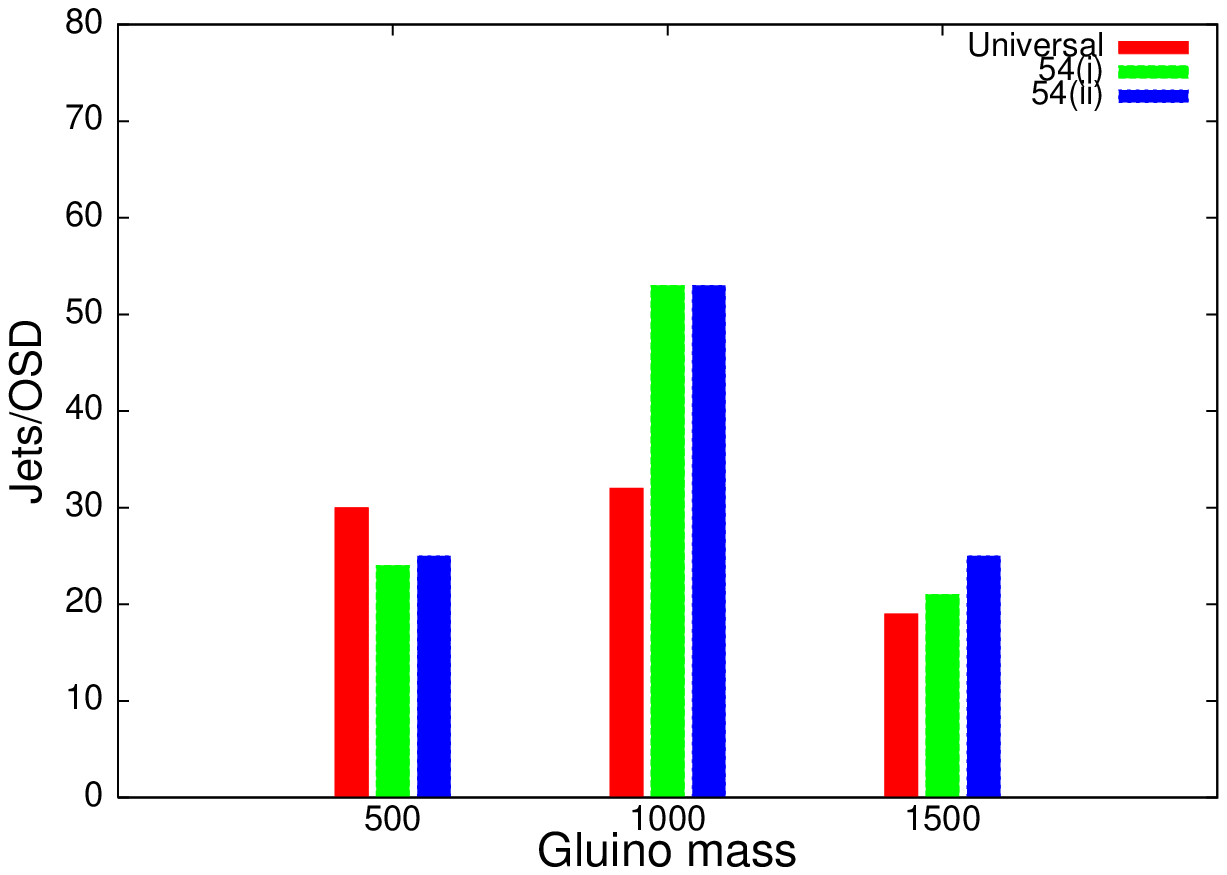,width=6.5 cm,height=5.50cm,angle=-0}
\hskip 20pt \epsfig{file=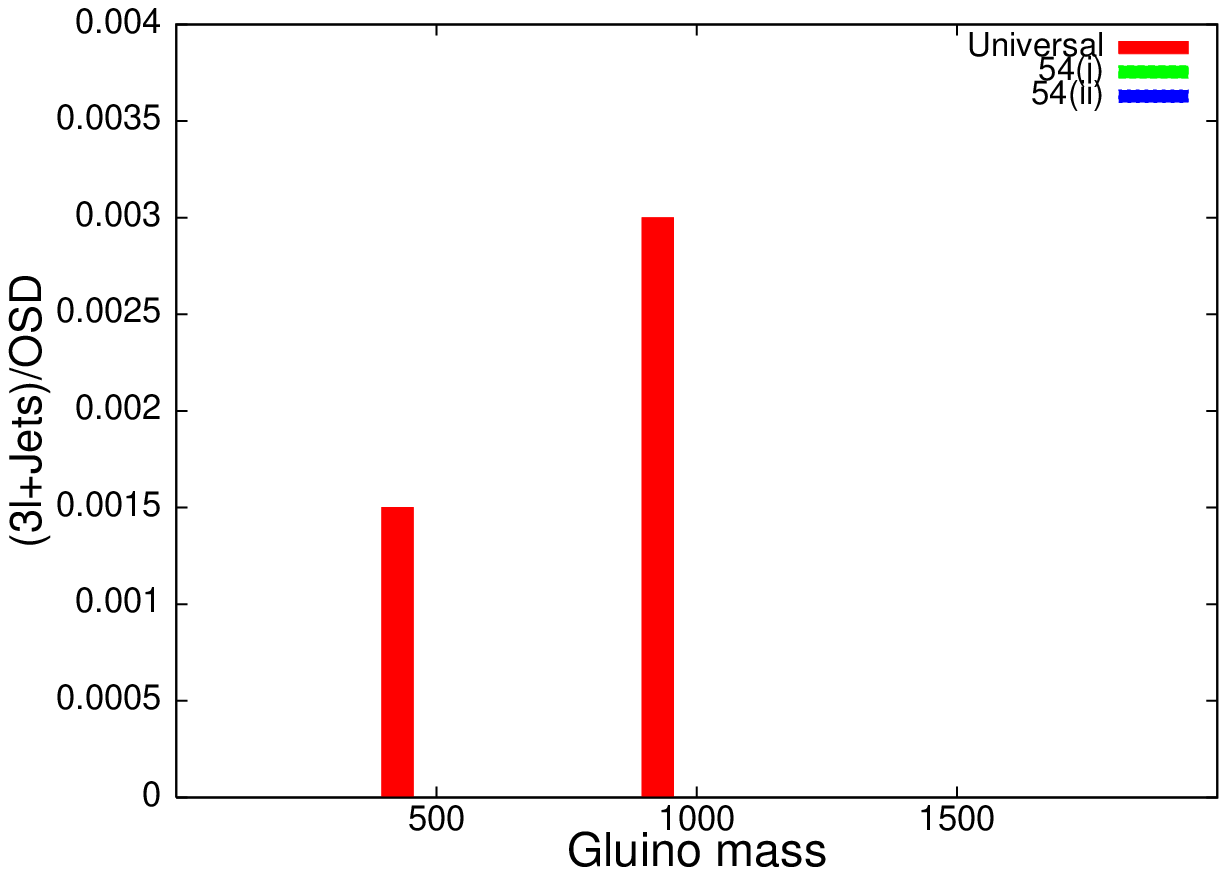,
width=6.5cm,height=5.50cm,angle=-0}}
\caption{ Event ratios {\bf pMSSM} in SO(10): $m_{\tilde f}=500$ GeV, 
 $\mu=300$ GeV, $\tan{\beta}=40$} 
%distribution}
\end{center}
\end{figure}

Figure 9 contains some illustrative numbers for SUGRA with non-universal
gaugino masses at high scale. It may be noted that, corresponding to
$m_{\tilde{g}}$= 1000 GeV, the values of the lighter charginos/neutralinos
become too small to be allowed by LEP results, whereas for 
$m_{\tilde{g}}$= 1500 GeV, no spectrum is generated for 
{\bf 75} since it cannot implement radiative electroweak symmetry
breaking (the gaugino contributions being responsible for
rendering all scalar mass-squared values positive). 
For $m_{\tilde g}=500$ GeV, {\bf 75} is allowed, and 
can easily be distinguished from either the SSD/OSD
or the $jets$/OSD ratio. Identification of 200 is also
possible through $jets$/OSD. {\bf 24}  and {\bf 75}  may 
be separated from {\bf 200} and the universal case with the help
of the ratio $(3\ell+jets)$/OSD. On the whole, for gluino mass on the lower
side, all the four GUT breaking schemes can be distinguished
from each other through the ratios  SSD/OSD, in conjunction 
with non-observation (or otherwise) of $(3\ell+jets)$/OSD. This is in
a sense a gratifying conclusion, since the one can make useful
inference even while avoiding the overall uncertainties of
events containing jets only.
$(1\ell+jets)$/OSD is quite suppressed in all the cases and are numerically
quite uniform, so that it is not of much help. 
For  $m_{\tilde{g}}$= 1000 GeV,
{\bf 24} and {\bf 200} can be separated quite visibly from 
$(3\ell+jets)$/OSD, while non-observation of $(3\ell+jets)$ events (with the
other final states observed) will point towards {\bf 75} since
{\bf 75} is inadmissible for the reason mentioned above and 
observations in all other channels indicate 24. The results
presented for  $m_{\tilde g}$= 1500 GeV are not
numerically very different from each other; however, for all 
representations excepting {\bf 24}, the OSD events do not rise
beyond $2\sigma$ above the backgrounds for an integrated
luminosity of 300 fb$^{-1}$. For  {\bf 24}, all of the $jets$,
OSD and $trilepton$ channels rise above backgrounds, and thus
the ratios $jets$/OSD and $(3\ell+jets)$/OSD should be able
to make it stand out.

\begin{figure}[t]
\begin{center}
%\vspace*{-2.2cm}
\centerline{\epsfig{file=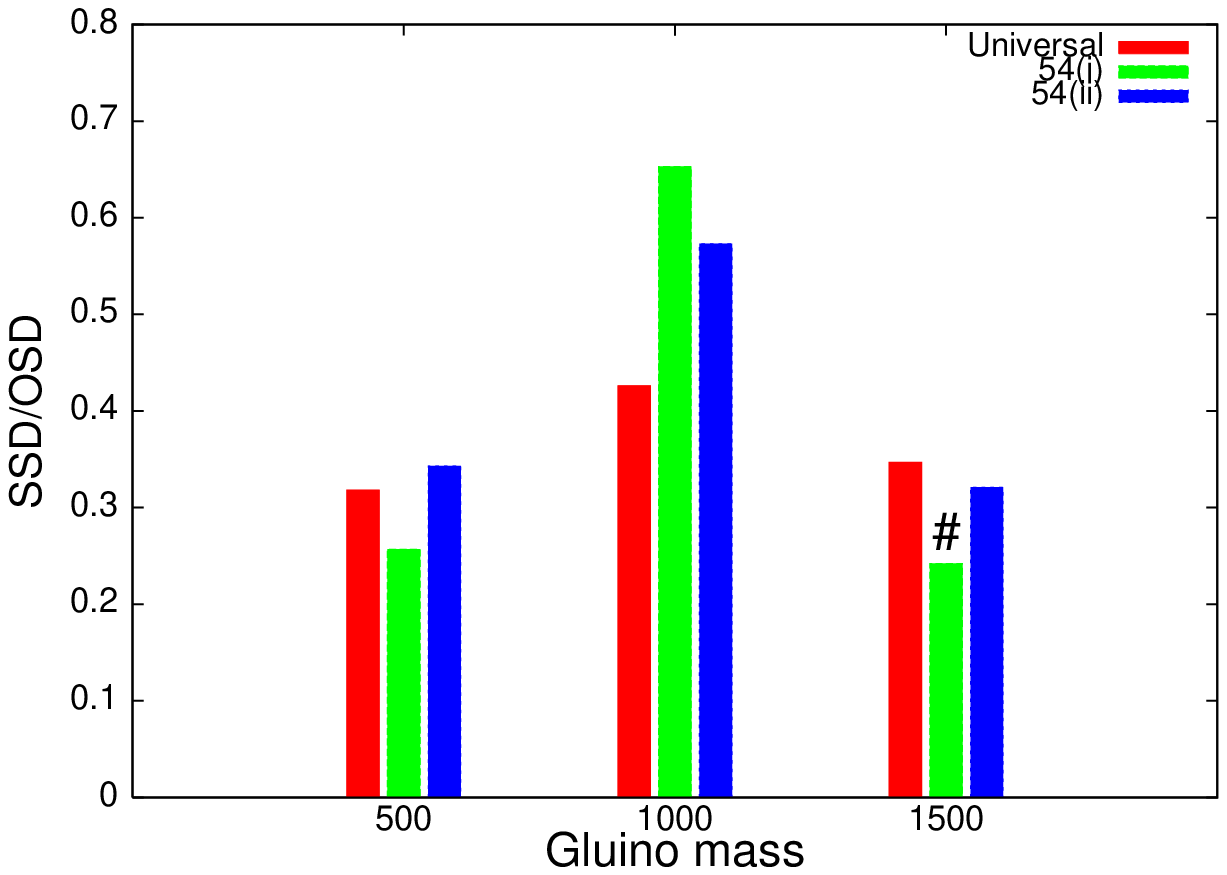,width=6.5 cm,height=5.50cm,angle=-0}
\hskip 20pt \epsfig{file=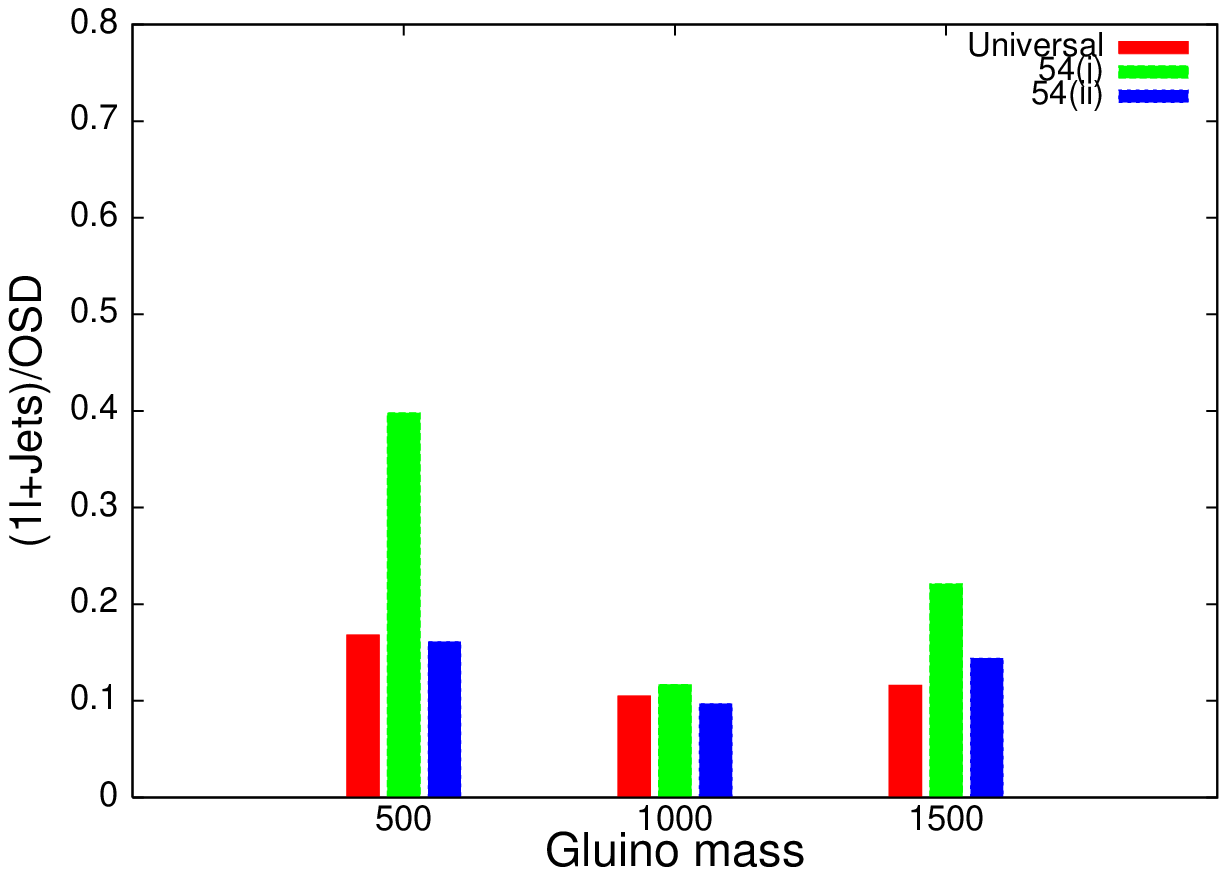,width=6.5cm,height=5.50cm,angle=-0}}
\vskip 10pt
\centerline{\epsfig{file=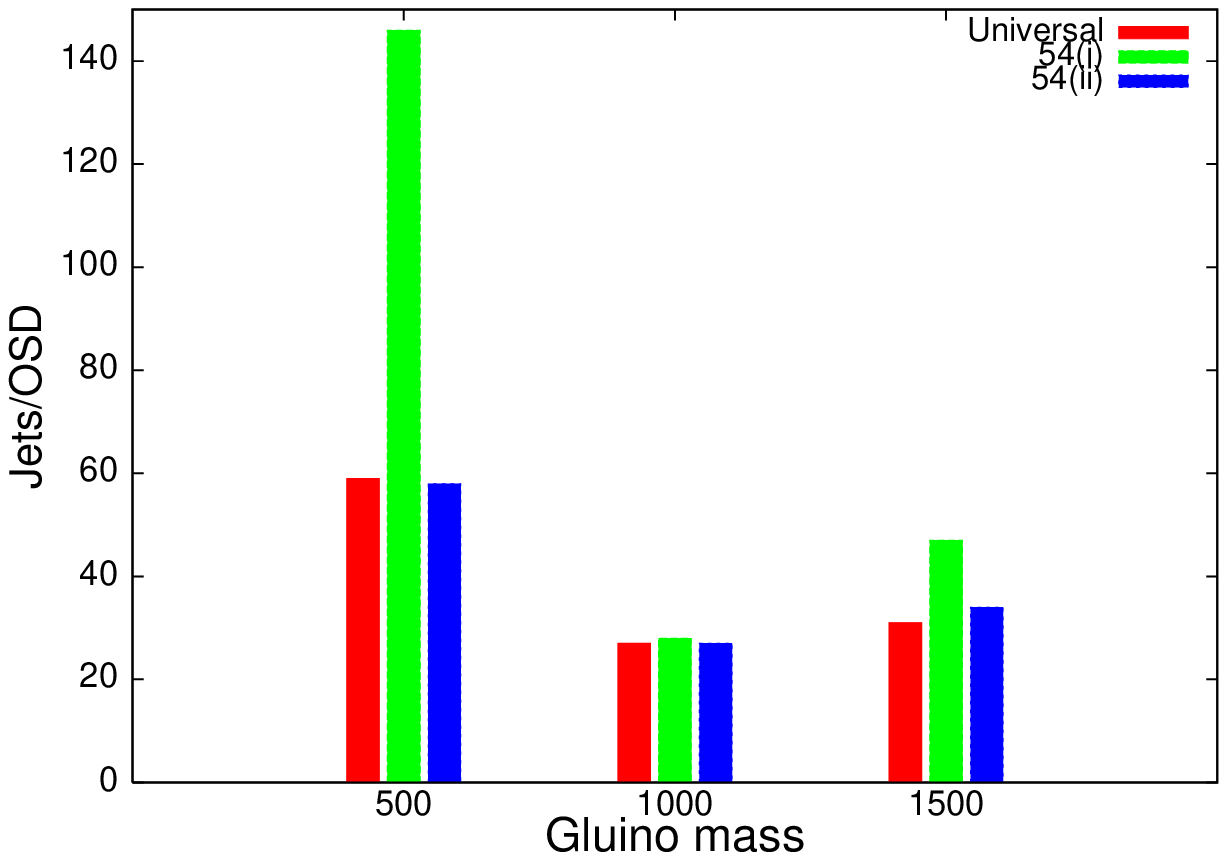,width=6.5 cm,height=5.50cm,angle=-0}
\hskip 20pt \epsfig{file=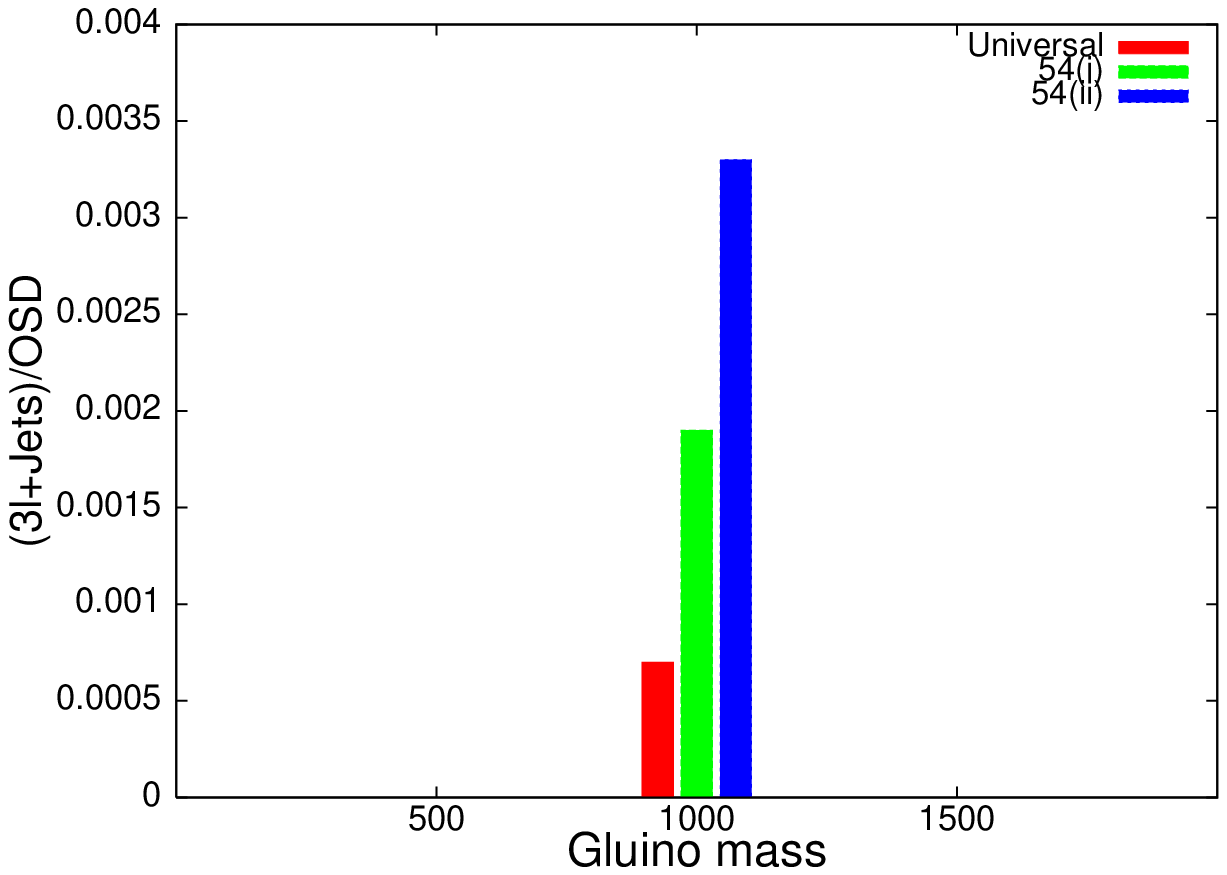,width=6.5cm,height=5.50cm,
angle=-0}}
\caption{ Event ratios for {\bf pMSSM} in SO(10): $m_{\tilde f}=1000$ GeV, 
 $\mu=300$ GeV, $\tan{\beta}=5$} 
%distribution}
\end{center}

\end{figure}

\subsection{ Non-universal SO(10) in pMSSM :}

In this subsection we analyse some cases of gaugino non-universality 
arising in SO(10) scenarios. As has been mentioned earlier,
the gaugino mass ratios at high scale in this case depend not only
on the chain of SO(10) breaking but also on the presence of an intermediate
breaking scale. Considering all of these will thus
lead to a plethora of possibilities. Here we take an illustrative case
of SO(10) breaking through the lowest non-singlet representation,
namely {\bf 54}, and consider two breaking chains:
(i) via $SU(2) \times SO(7)$ (denoted by {\bf 54(i)}) and 
(ii) via $SU(4) \times SU(2)_{L} \times SU(2)_{R}$
(denoted by {\bf 54(ii)}). We also assume that there is 
no intermediate scale involved in the GUT breaking process \cite{Chamoun}
(See in section 2).

Our style of analysis remains the same as in the case of SU(5). In figures 
10 - 17 we present the predictions in the {\bf pMSSM} framework, with
the same sequence in choosing the low-energy parameters as in figures
1 - 8. Figure 18 contains our predictions for SUGRA in SO(10), 
the parameters being chosen in the same fashion as in figure 9.  

\begin{figure}[t]
\begin{center}
%\vspace*{-2.2cm}
\centerline{\epsfig{file=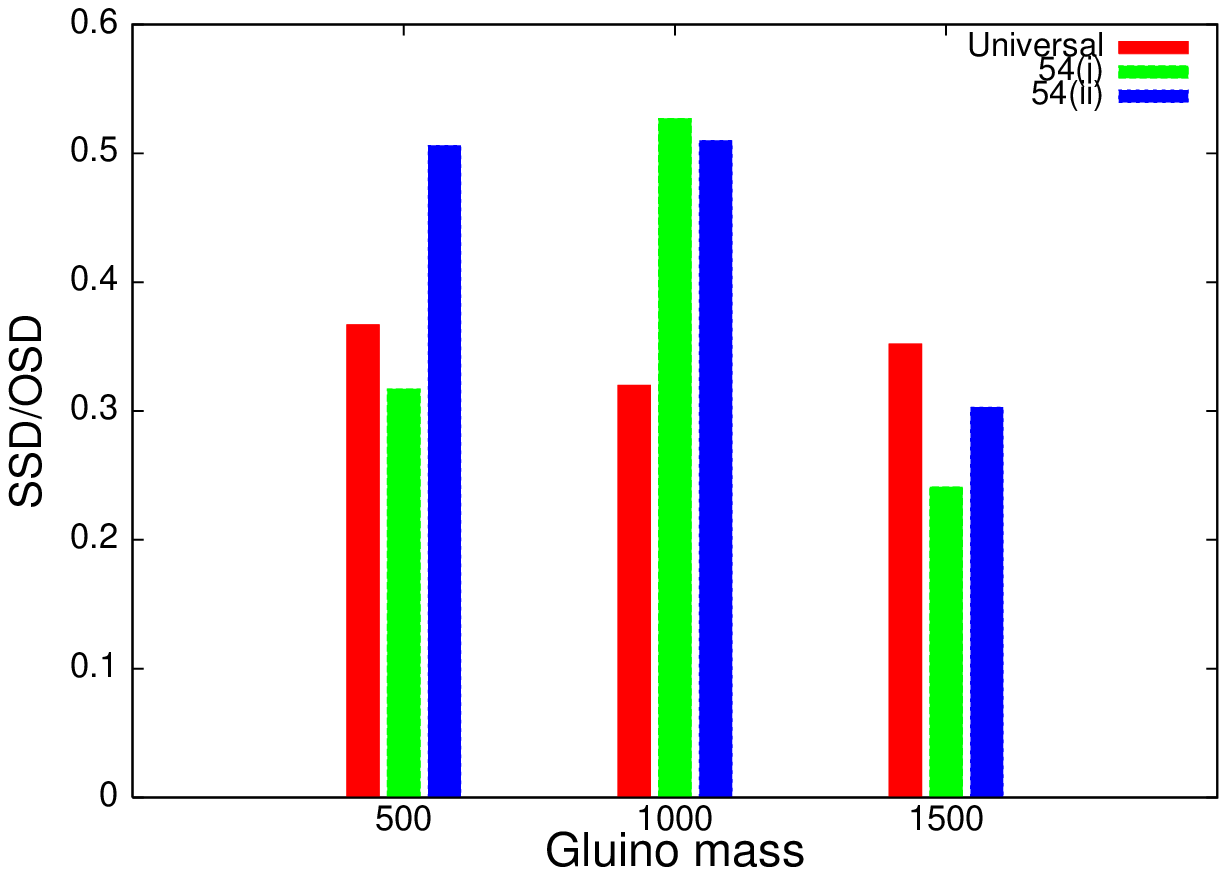,width=6.5 cm,height=5.50cm,angle=-0}
\hskip 20pt \epsfig{file=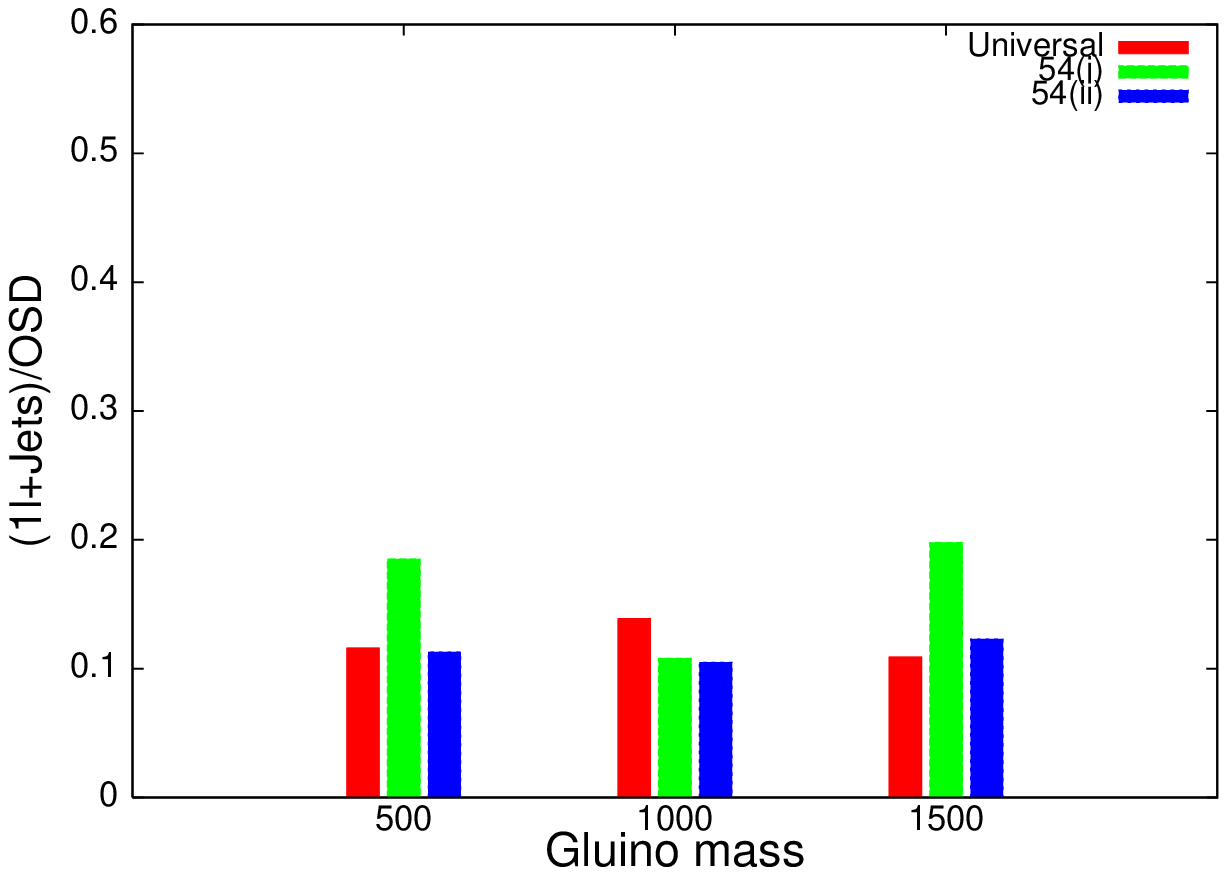,width=6.5cm,height=5.50cm,angle=-0}}
\vskip 10pt
\centerline{\epsfig{file=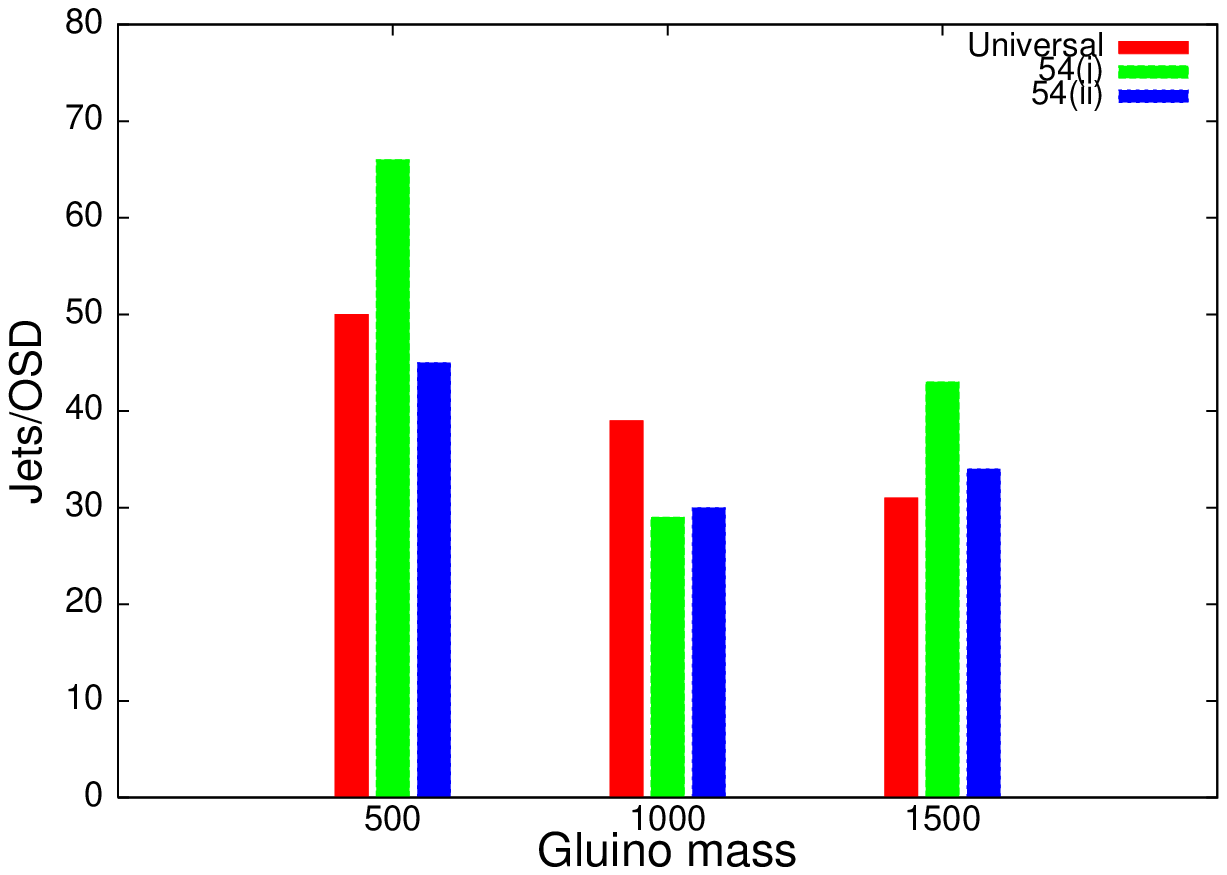,width=6.5 cm,height=5.50cm,angle=-0}
\hskip 20pt \epsfig{file=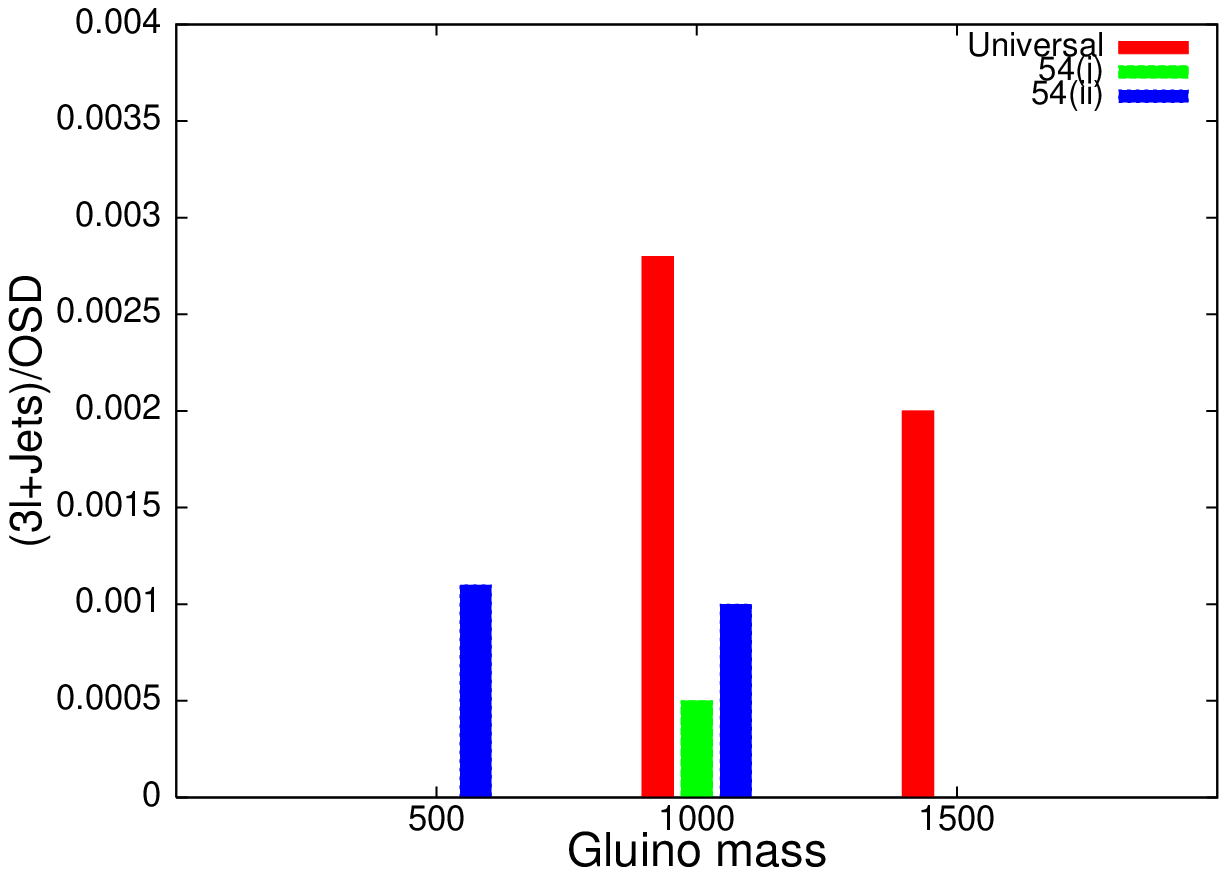,width=6.5cm,height=5.50cm,
angle=-0}}
\caption{ Event ratios for {\bf pMSSM} in SO(10): $m_{\tilde f}=$1000 GeV,
 $\mu=$300 GeV, $\tan{\beta}=40$} 
%distribution}
\end{center}

\end{figure} 

The observations on the different cases take very similar
lines as those in the case of SU(5). However,
the following general features are noticed from figures
10 - 17:

\begin{enumerate}
\item The case of {\bf 54(i)} is largely distinguishable from the other
cases through one channel or the other. This possibility is more 
pronounced for sfermion masses at 1000 GeV. 

\begin{figure}[t]
\begin{center}
%\vspace*{-2.2cm}
\centerline{\epsfig{file=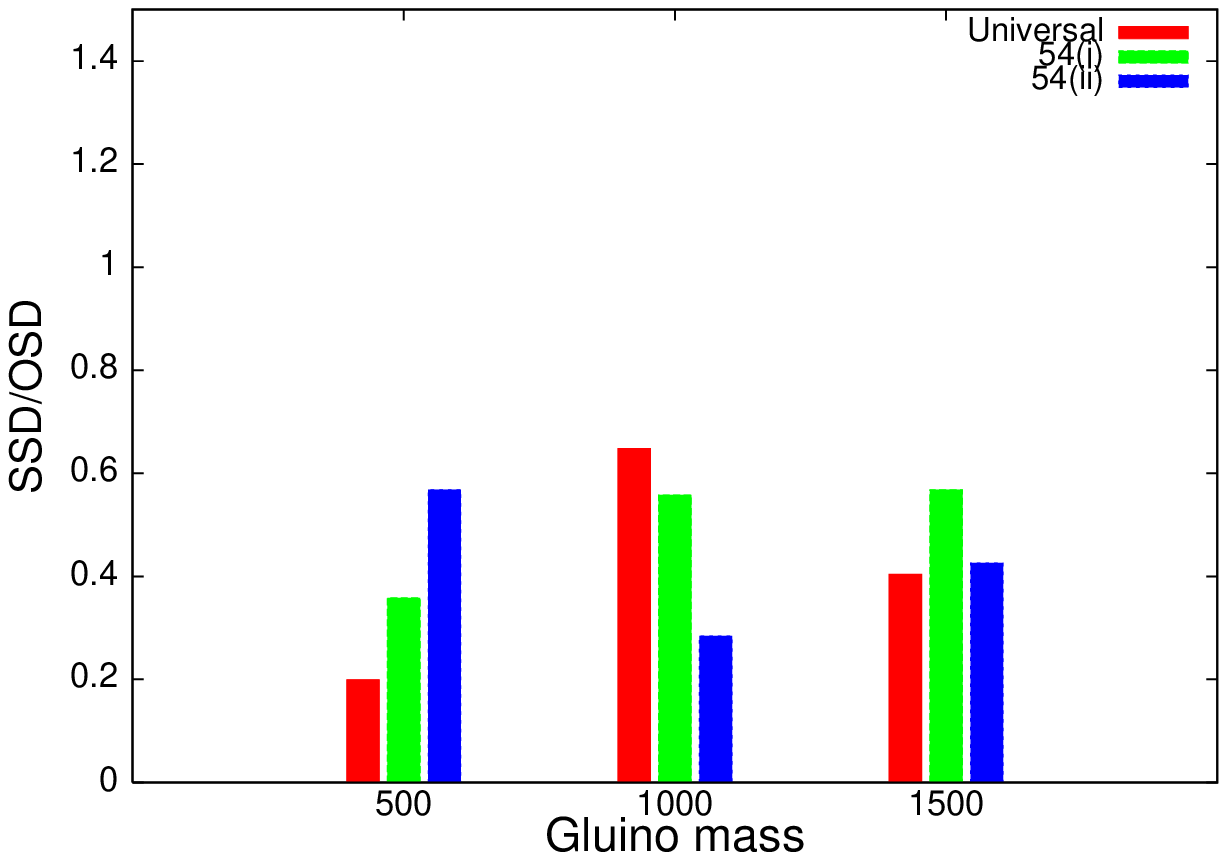,width=6.5 cm,height=5.50cm,angle=-0}
\hskip 20pt \epsfig{file=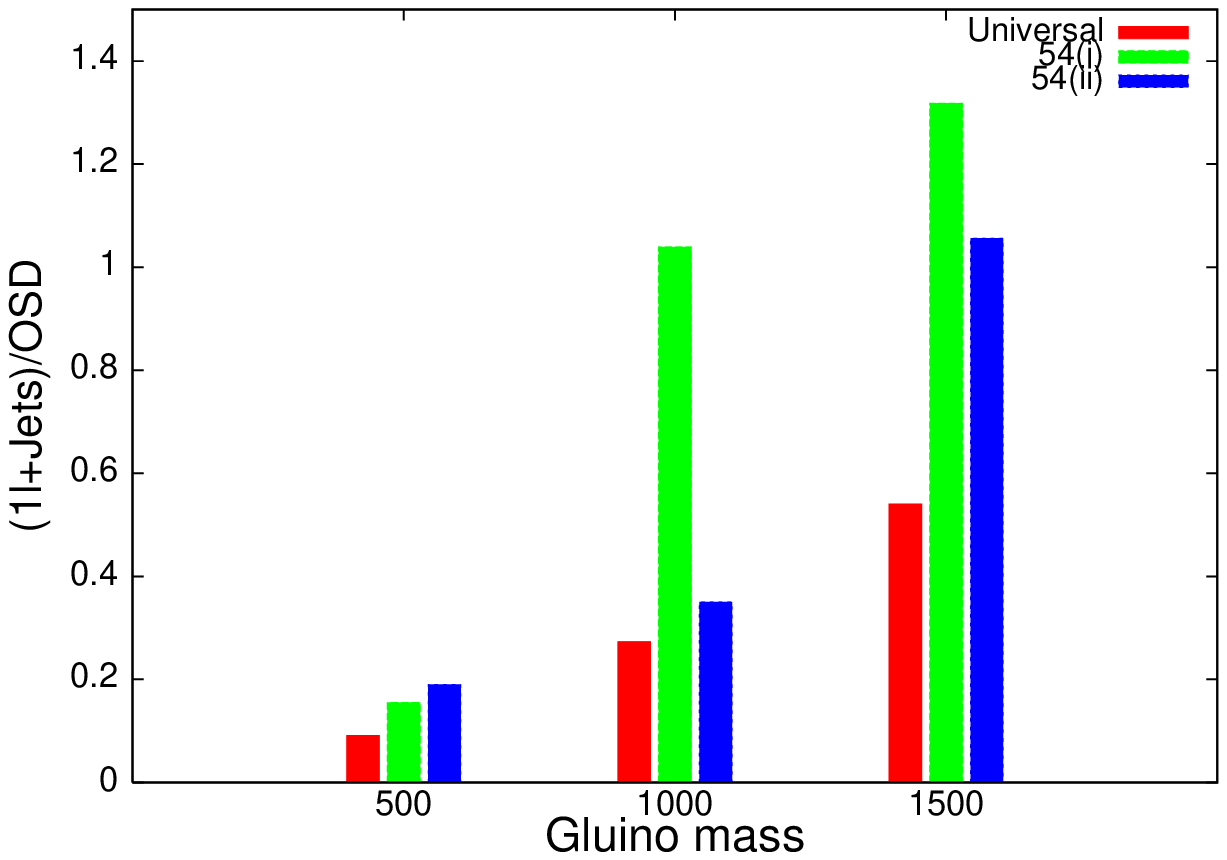,width=6.5cm,height=5.50cm,angle=-0}}
\vskip 10pt
{\epsfig{file=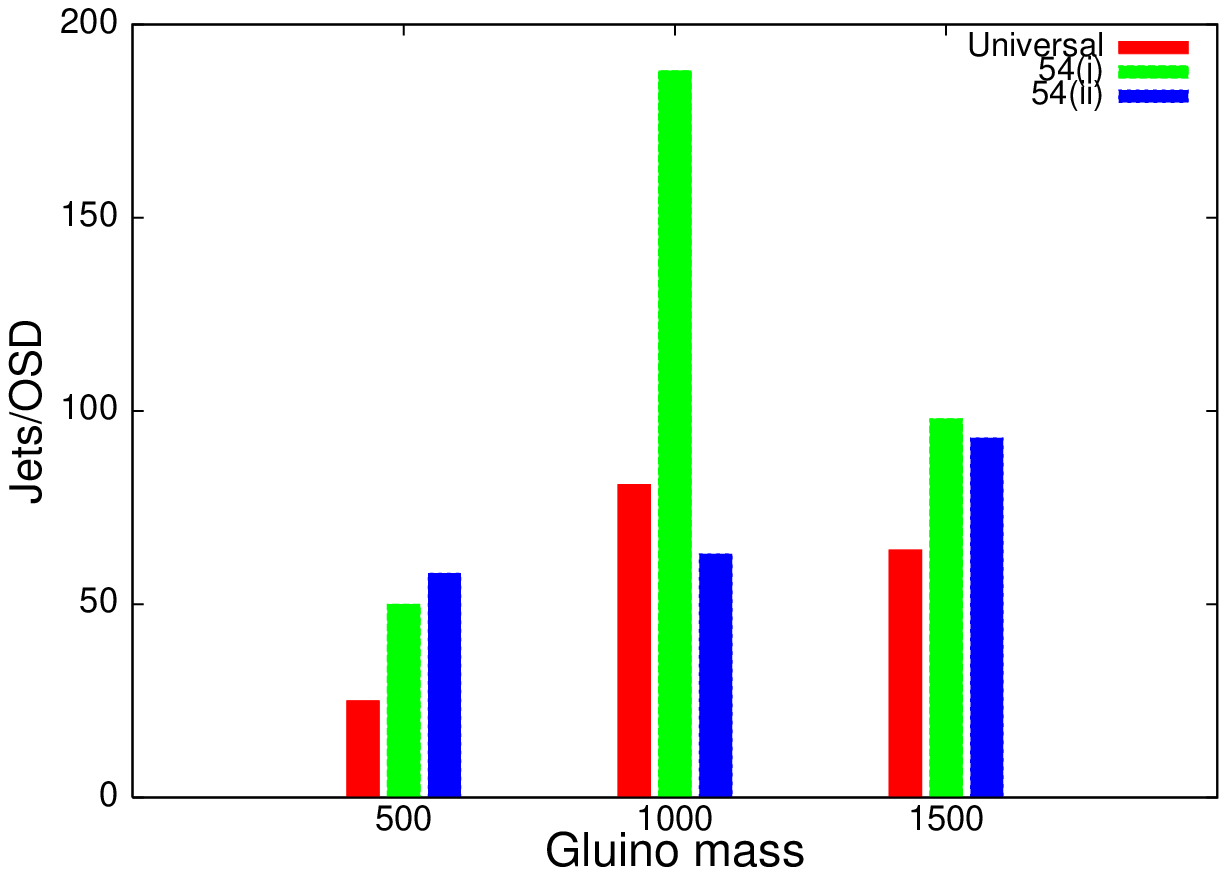,width=6.5 cm,height=5.50cm,angle=-0}}

\caption{ Event ratios for {\bf pMSSM} in SO(10): $m_{\tilde f}=$500 GeV, 
$\mu=$1000 GeV, $\tan{\beta}=5$} 
%distribution}
\end{center}

\end{figure} 

\item The universal case, on the other hand, shows very similar behaviour
as in {\bf 54(ii)}. This is because the chargino/neutralino spectra
do not show much variation between these two cases, as a result
of which the absolute cross-sections, too, do not provide much of a handle. 
The most effective discrimination is possible through the 
ratio $(3\ell+jets)$/OSD channel. 

\item The dependence on $\tan \beta$ is less than in the case
of SU(5). This is because, as can be seen from Appendix A, the charginos
and higher neurtralinos  are heavier here than in the corresponding
cases with SU(5). As a result, the sbottom decaying into them 
(which initiates 
cascades leading to leptons in the final state) are relatively suppressed,
thus denying one the enhancement that could be seen through enhanced
sbottom production rates for $\tan\beta$= 40.

\item The problem with backgrounds for single-lepton and SSD signals
with heavy gluinos and sfermions,
already pointed out in the case of SU(5) persists for {\bf 54(i)}.

\item The suppression of single lepton events is still observed,
especially for low $\mu$.

\item The results corresponding to the universal case for each point in 
figures 10-17 are identical with those for SU(5) {\bf pMSSM} with the same 
parameter values. Similarly, the results for universal {\bf SUGRA} in figure 18
 are identical with those in figure 9.

\end{enumerate}

\begin{figure}[t]
\begin{center}
%\vspace*{-2.2cm}
\centerline{\epsfig{file=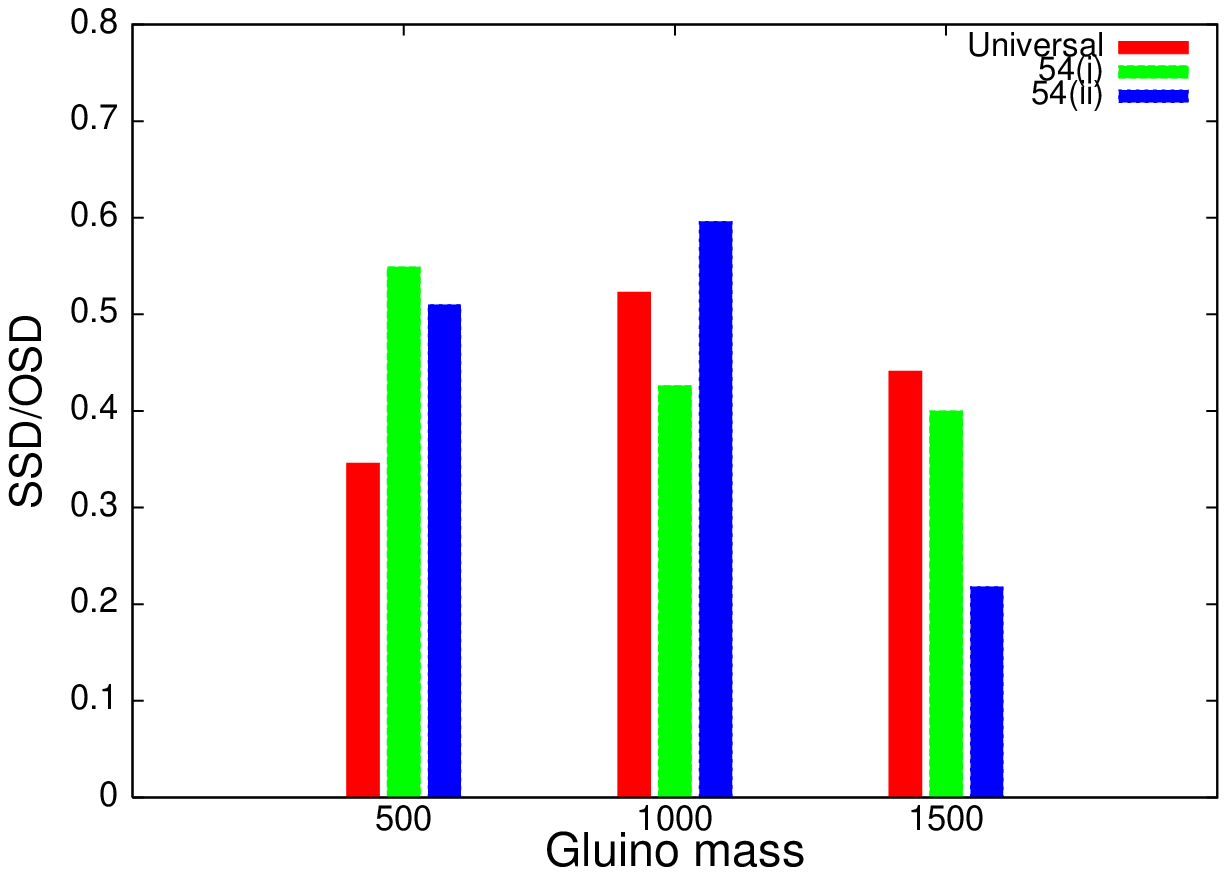,width=6.5 cm,height=5.50cm,angle=-0}
\hskip 20pt \epsfig{file=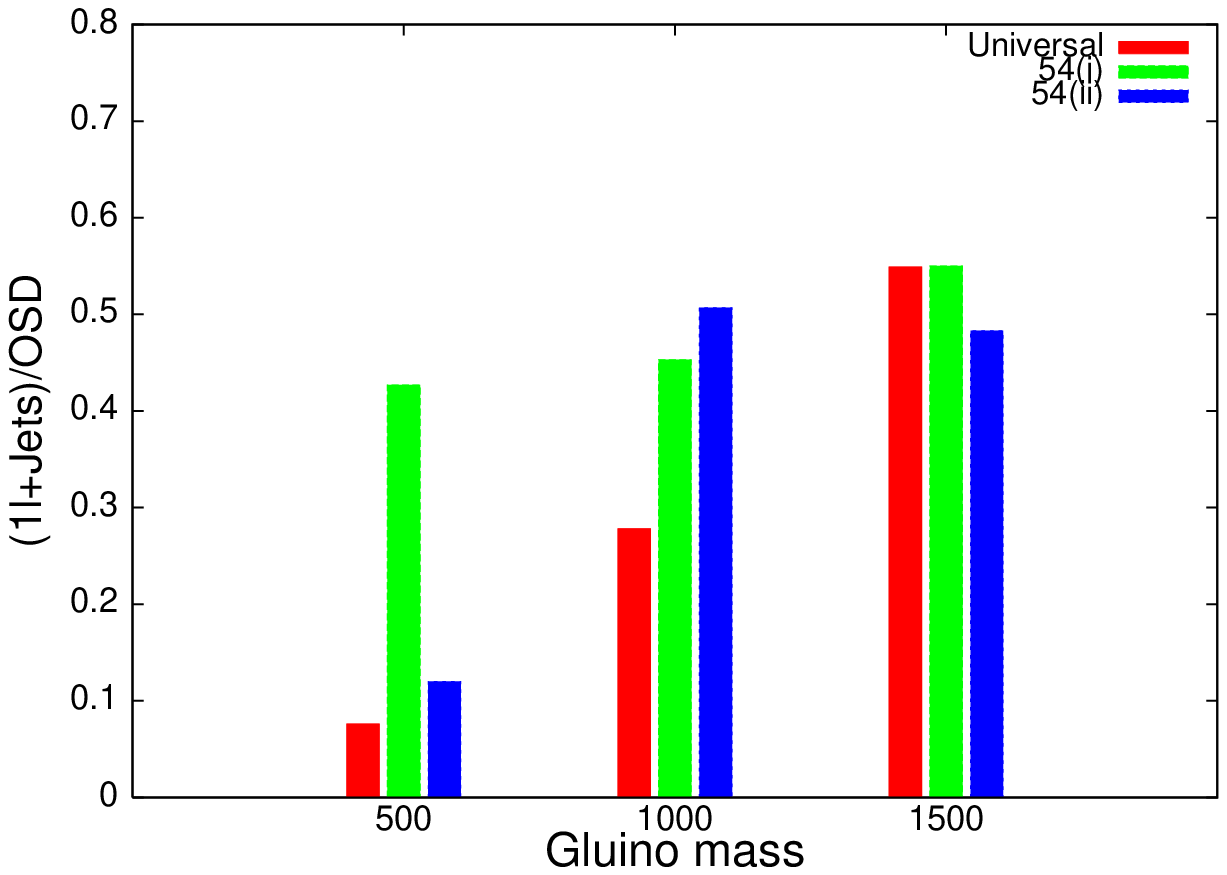,width=6.5cm,height=5.50cm,angle=-0}}
\vskip 10pt
\centerline{\epsfig{file=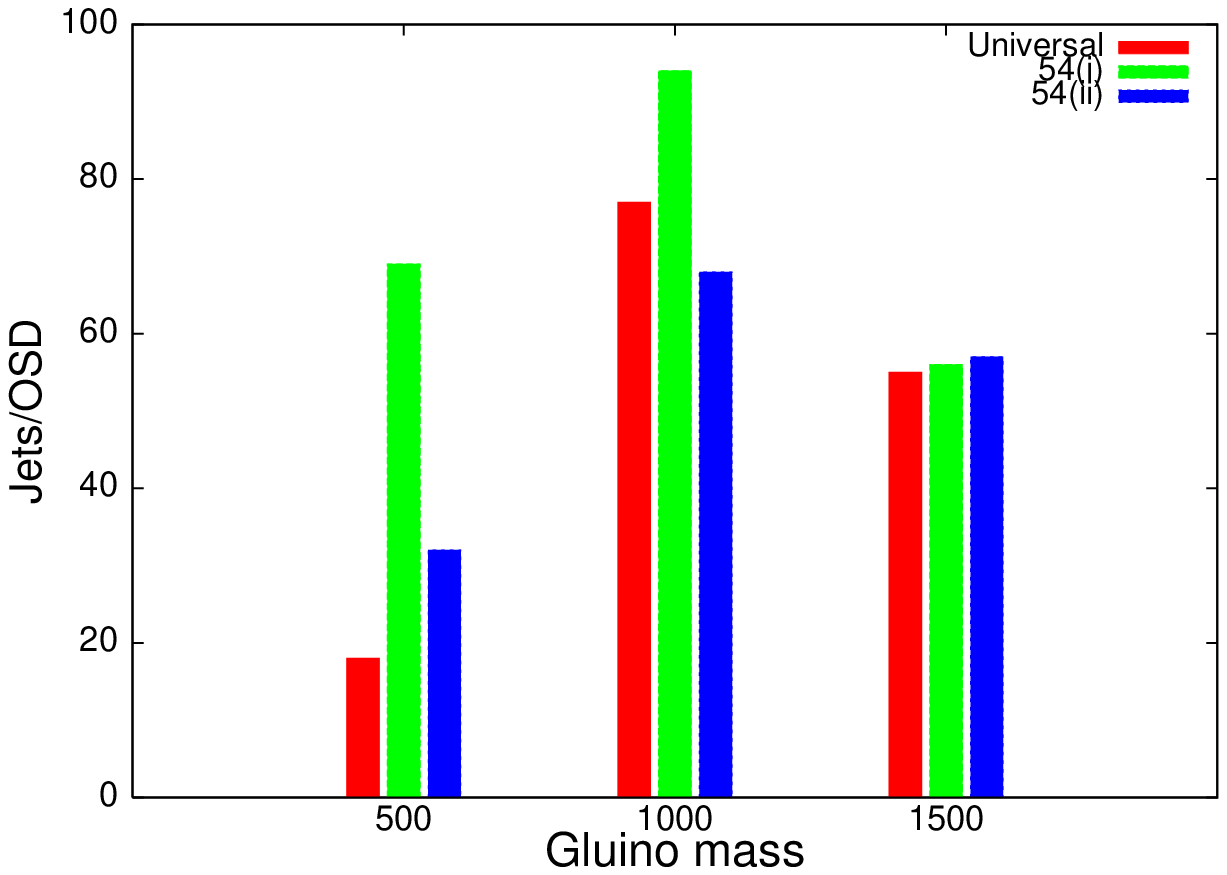,width=6.5 cm,height=5.50cm,angle=-0}
\hskip 20pt \epsfig{file=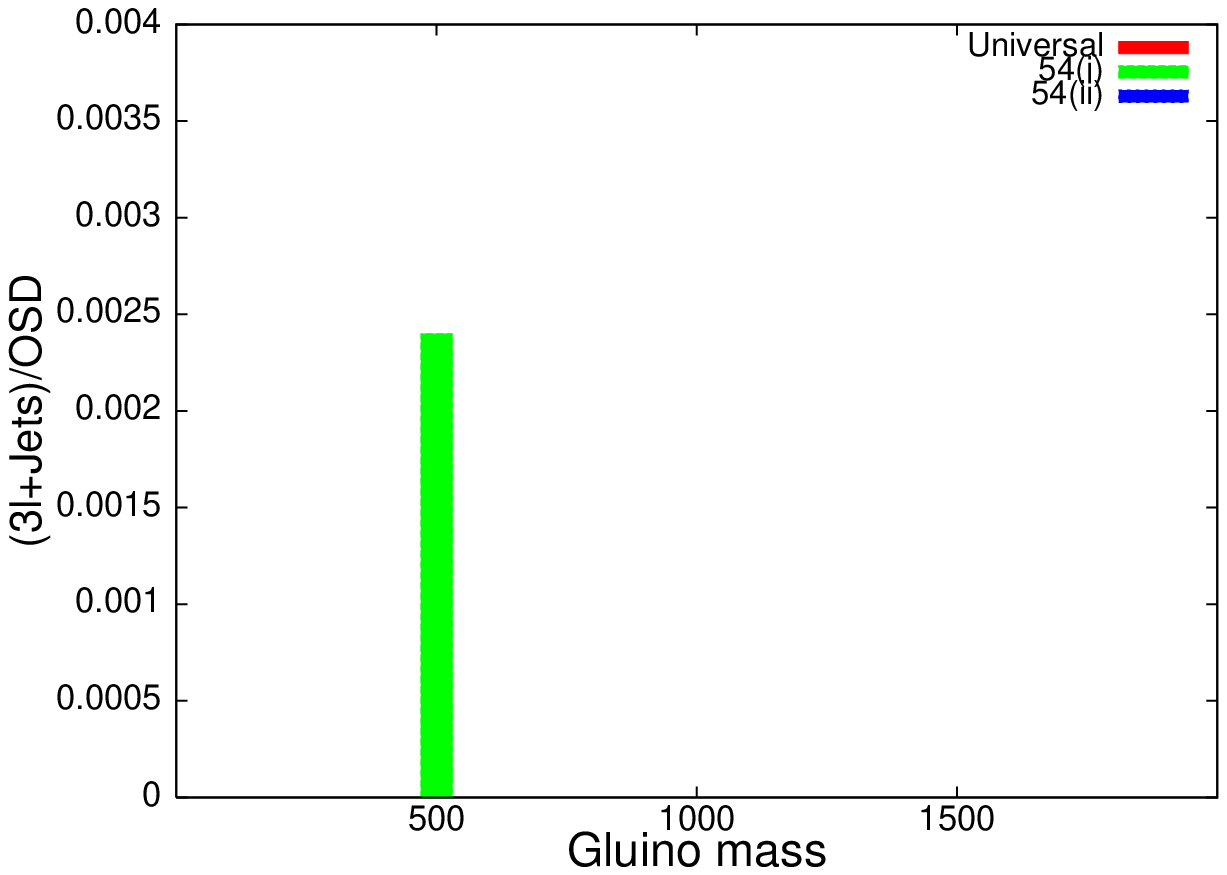,width=6.5cm,height=5.50cm,
angle=-0}}
\caption{ Event ratios for {\bf pMSSM} in SO(10): $m_{\tilde f}=$500 GeV,
  $\mu=$1000 GeV, $\tan{\beta}=40$} 
%distribution}
\end{center}

\end{figure}

\begin{figure}[t]
\begin{center}
%\vspace*{-2.2cm}
\centerline{\epsfig{file=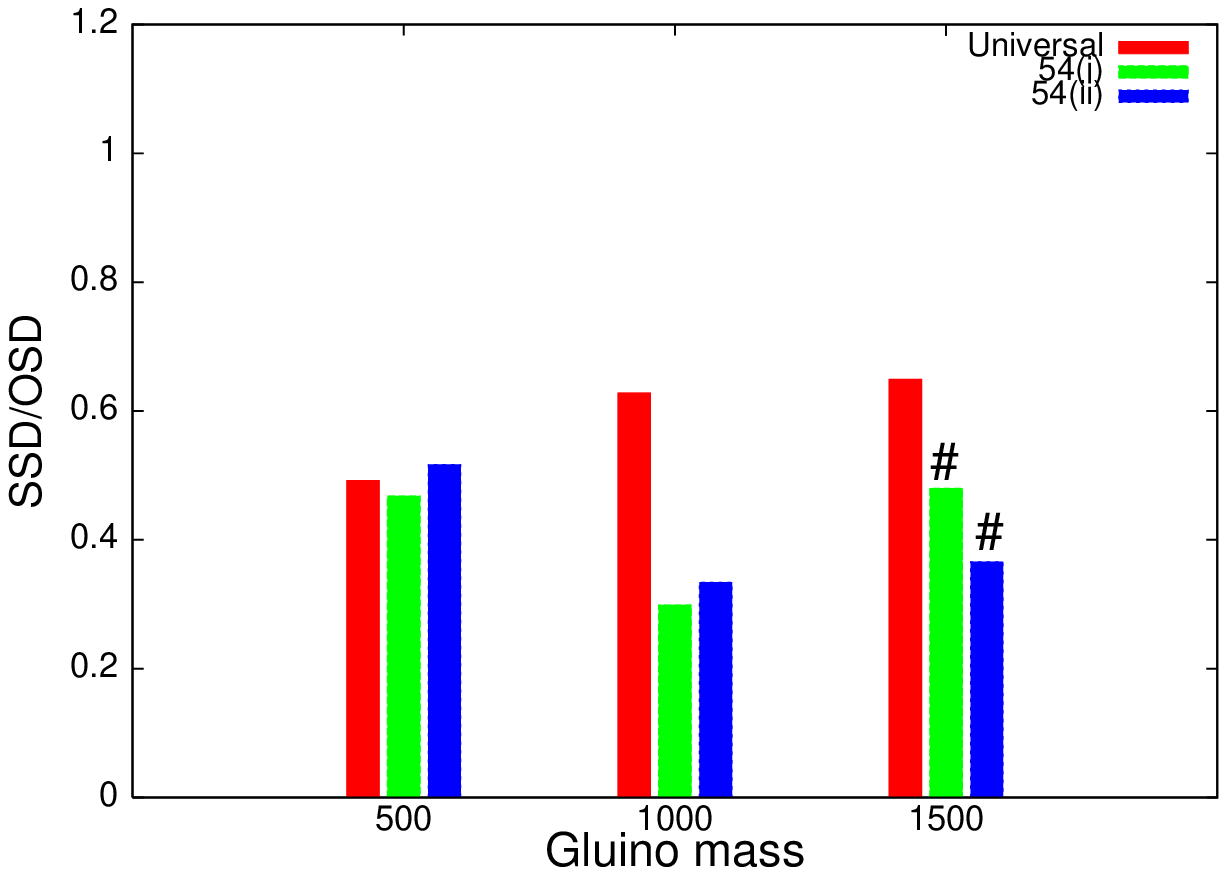,width=6.5 cm,height=5.50cm,angle=-0}
\hskip 20pt \epsfig{file=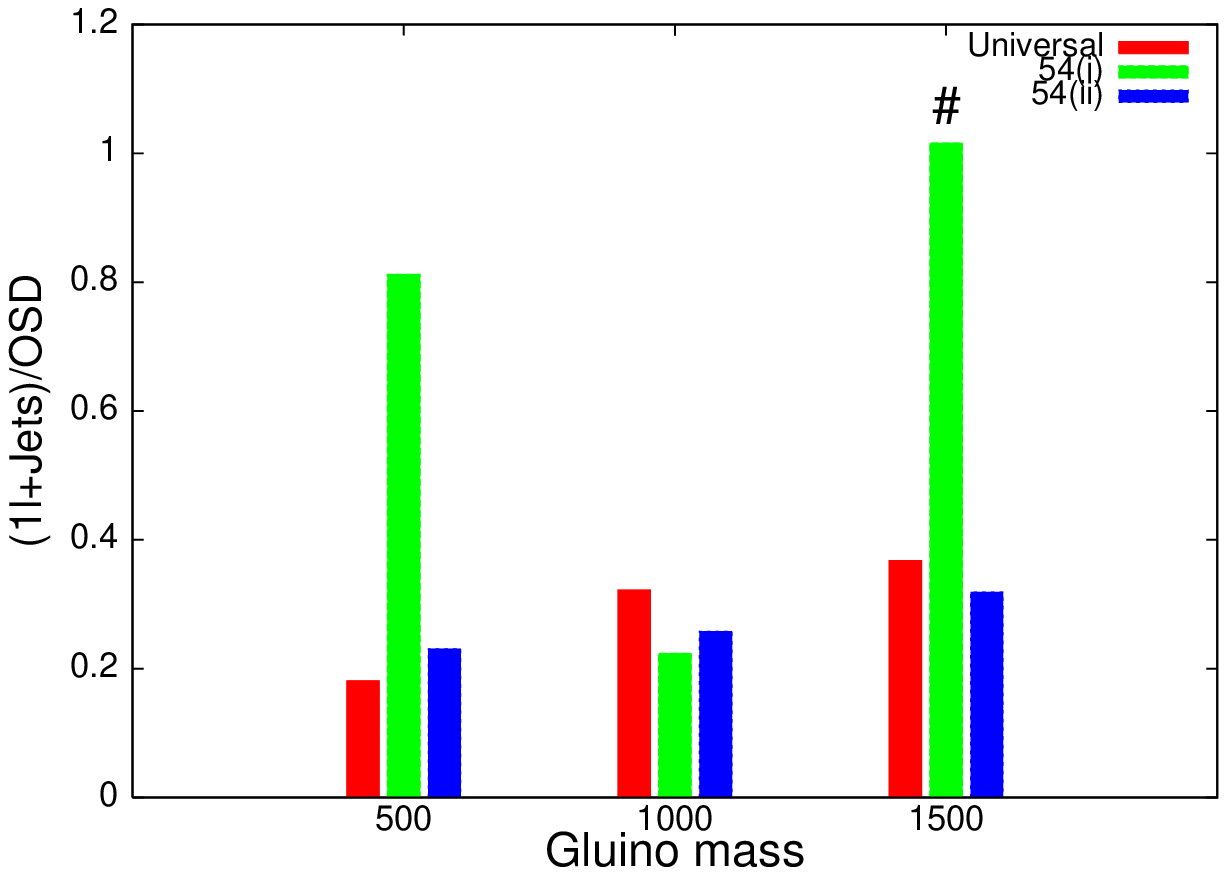,width=6.5cm,height=5.50cm,angle=-0}}
\vskip 10pt
\centerline{\epsfig{file=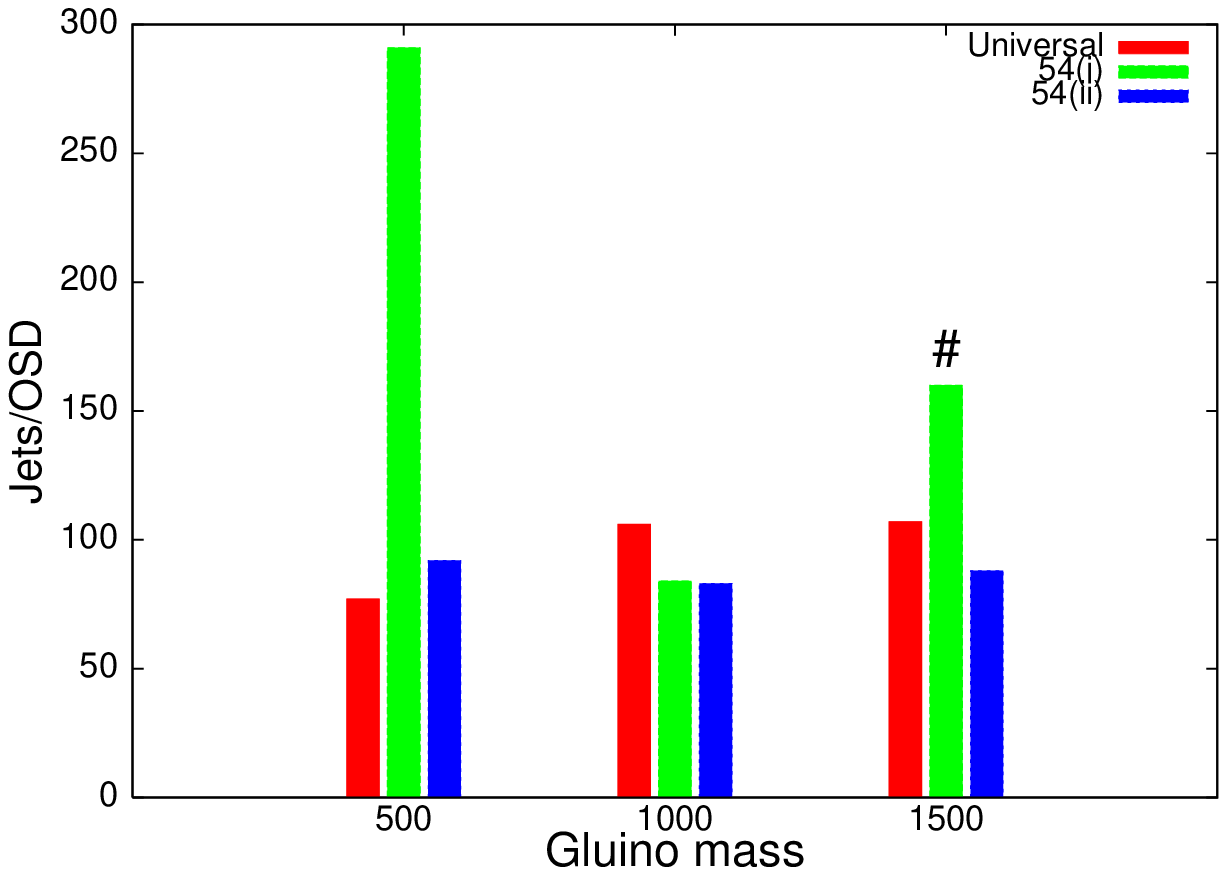,width=6.5 cm,height=5.50cm,angle=-0}
\hskip 20pt \epsfig{file=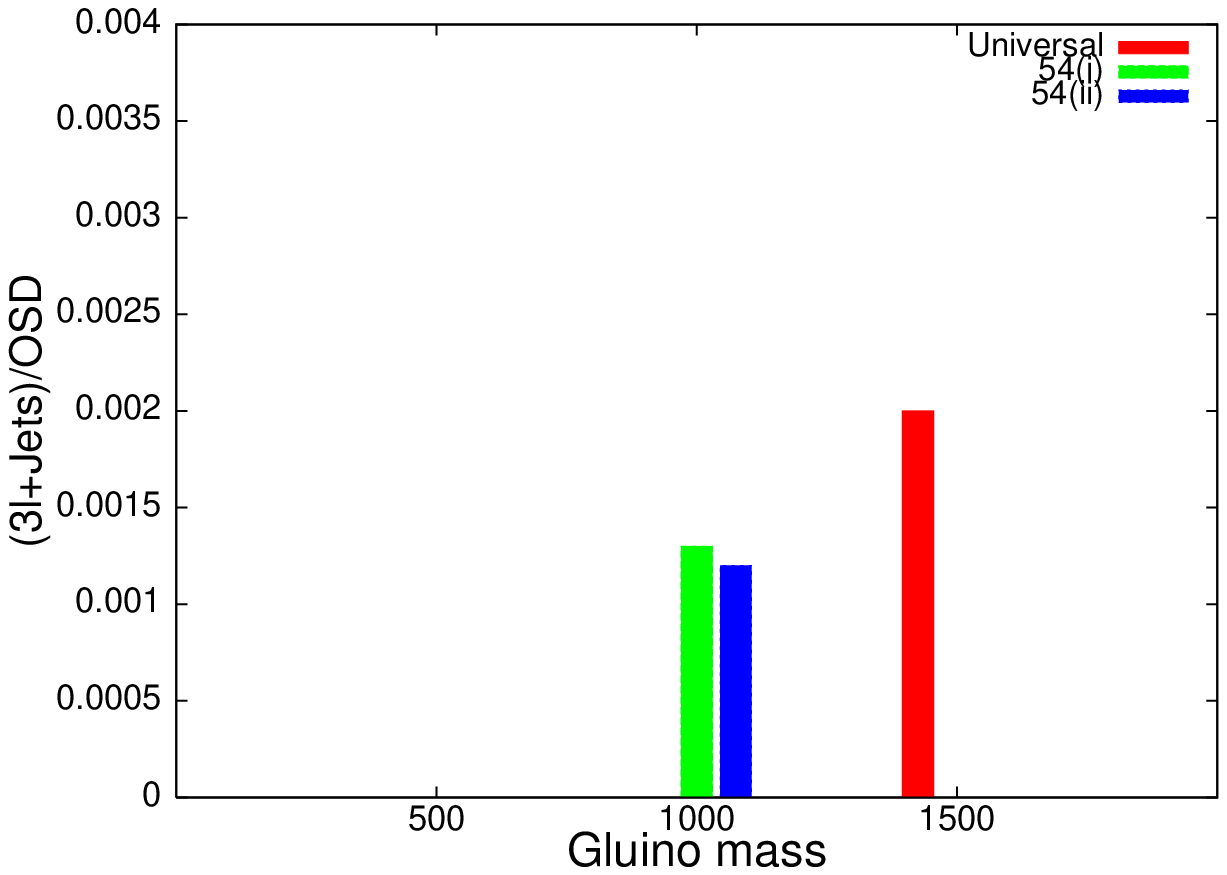,width=6.5cm,height=5.50cm,
angle=-0}}
\caption{ Event ratios for {\bf pMSSM} in SO(10): $m_{\tilde f}=$1000 GeV,
 $\mu=$1000 GeV, $\tan{\beta}=5$} 
%distribution}
\end{center}

\end{figure}

\begin{figure}[t]
\begin{center}
%\vspace*{-2.2cm}
\centerline{\epsfig{file=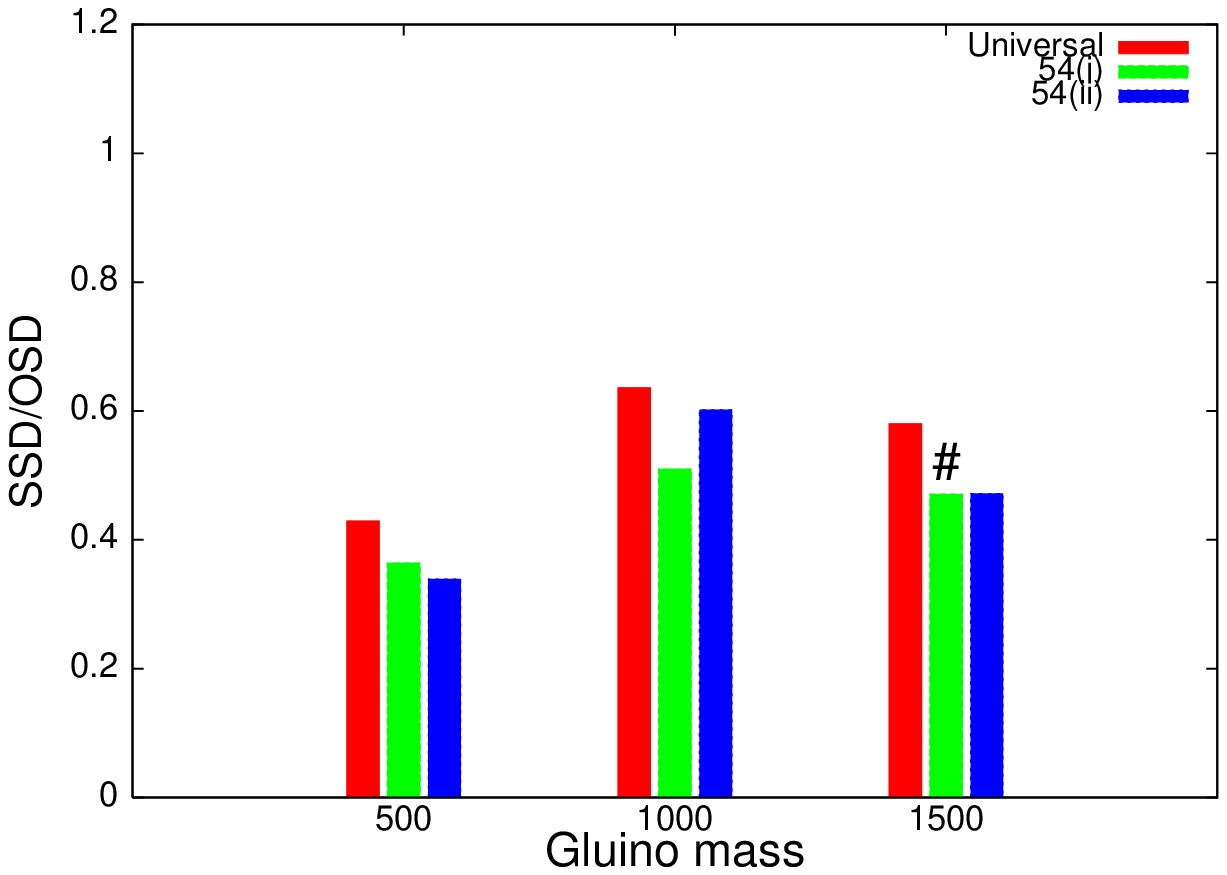,width=6.5 cm,height=5.50cm,angle=-0}
\hskip 20pt \epsfig{file=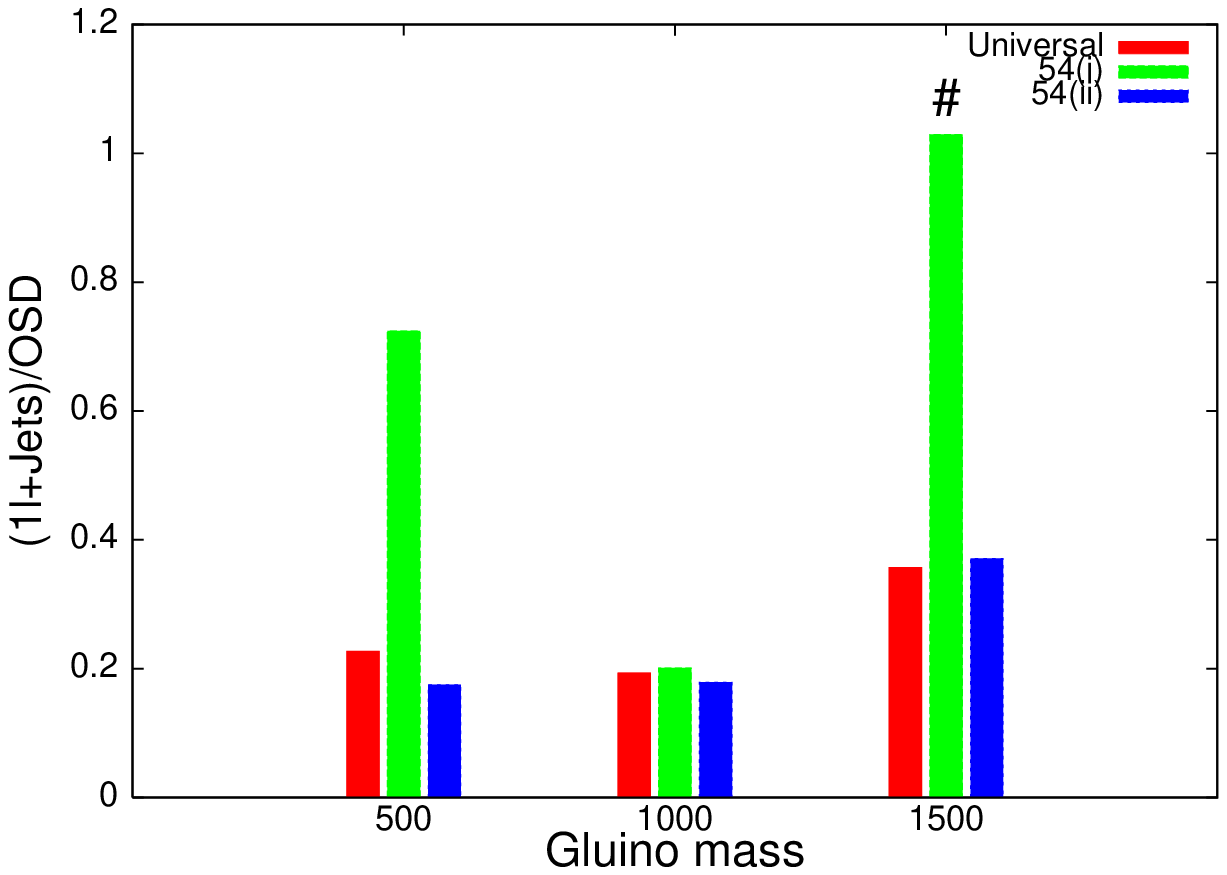,width=6.5cm,height=5.50cm,angle=-0}}
\vskip 10pt
\centerline{\epsfig{file=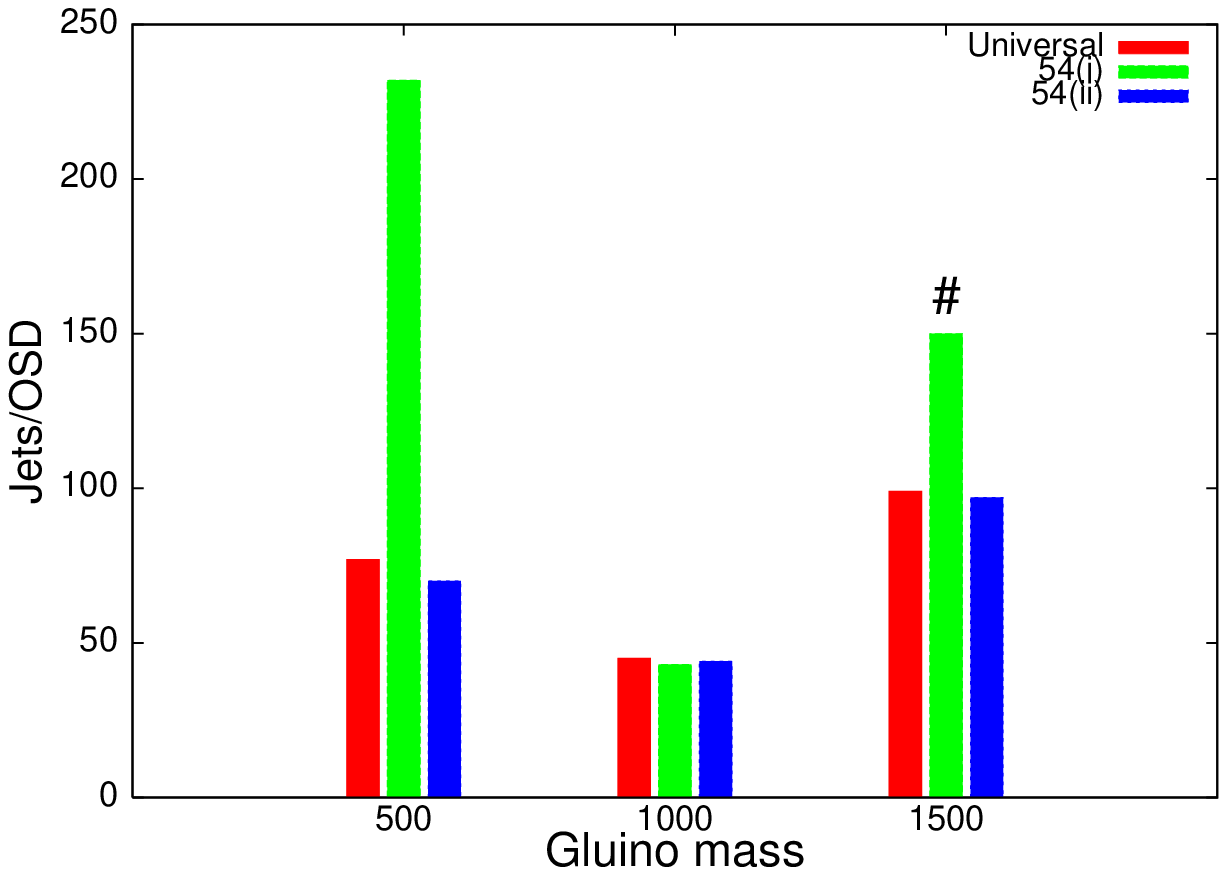,width=6.5 cm,height=5.50cm,angle=-0}
\hskip 20pt \epsfig{file=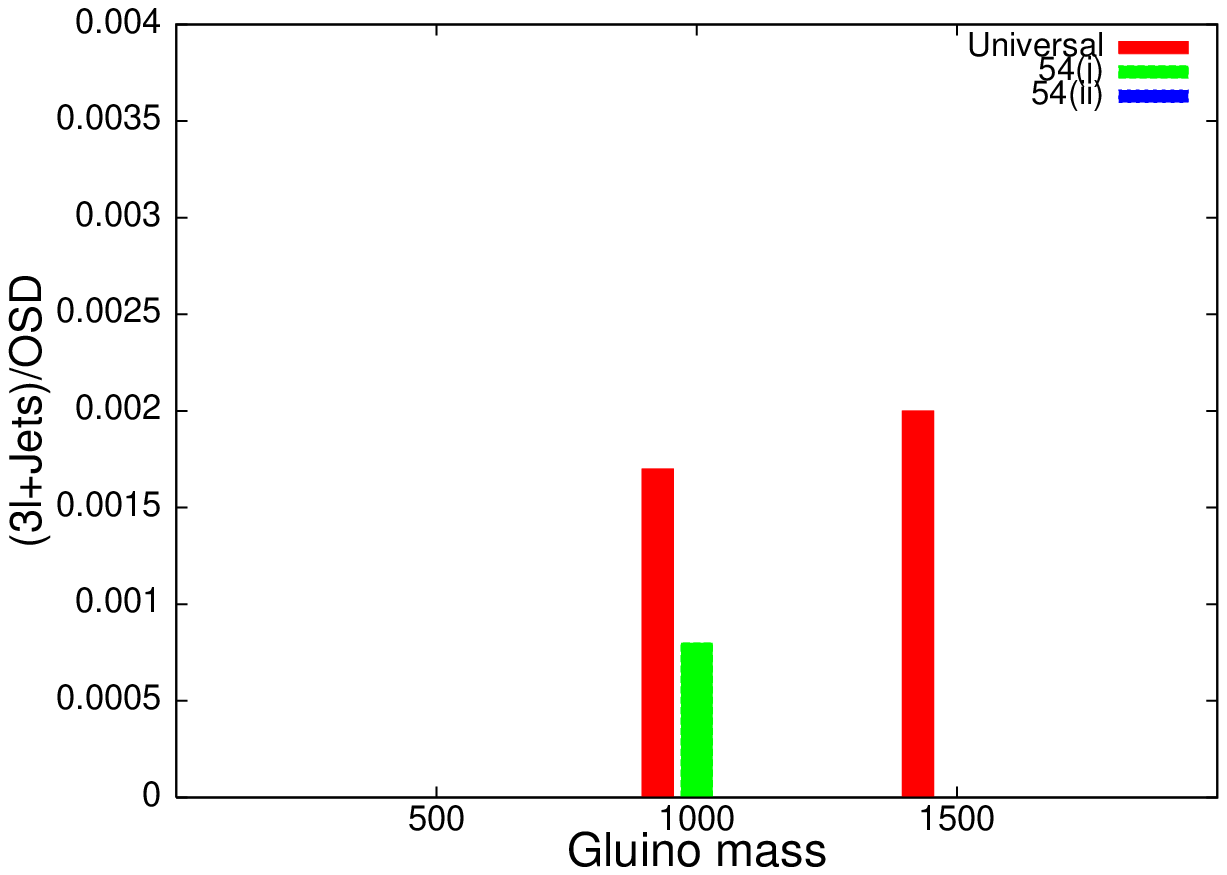,width=6.5cm,height=5.50cm,
angle=-0}}
\caption{ Event ratios for {\bf pMSSM} in SO(10): $m_{\tilde f}=$1000 GeV, 
 $\mu=$1000 GeV, $\tan{\beta}=40$} 
%distribution}
\end{center}

\end{figure}

Figure 18 contains the predictions for SO(10) in a  non-universal SUGRA 
setting. For
$m_{\tilde g}$=500 GeV,  the universal case may be separated from others
through $(1\ell+jets)$/OSD or $jets$/OSD as well as through null observation 
of $(3\ell+jets)$/OSD, while {\bf 54(i)} may be distinguished 
from SSD/OSD or through $jets$/OSD. 
Thus, as opposed to the {\bf pMMSM} case, all the three schemes
of GUT breaking studied here are separable from each other. For
$m_{\tilde g}$= 1000 GeV, any observation of $(3\ell+jets)$/OSD points uniquely
to {\bf 54(ii)}, while the separation of the universal case and 
{\bf 54(i)} is difficult. For $m_{\tilde g}$= 1500 GeV, 
the  observables are in general drowned by backgrounds,
excepting the case of {\bf 54(ii)} with $jets$/OSD and $(3\ell+jets)$/OSD.

\begin{figure}[t]
\begin{center}
%\vspace*{-2.2cm}
\centerline{\epsfig{file=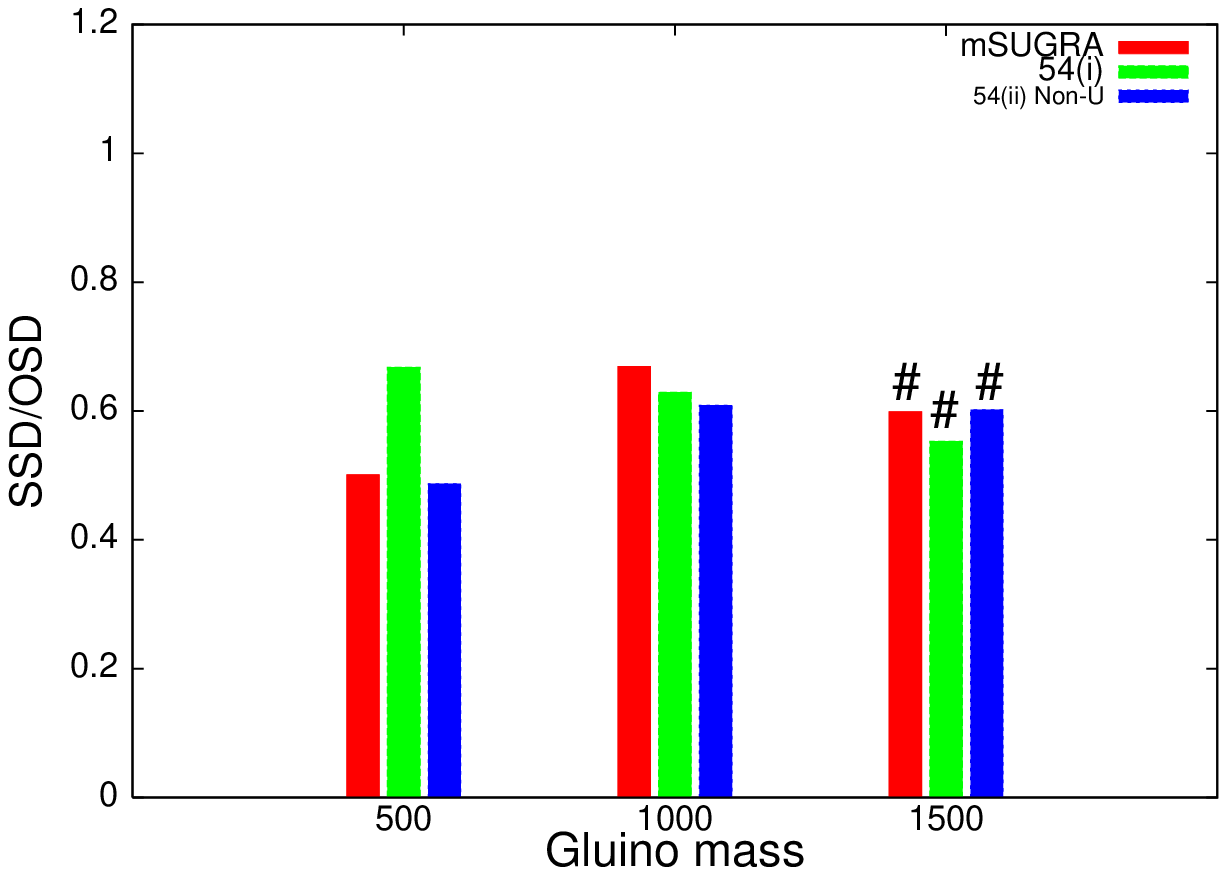,width=6.5 cm,height=5.50cm,angle=-0}
\hskip 20pt \epsfig{file=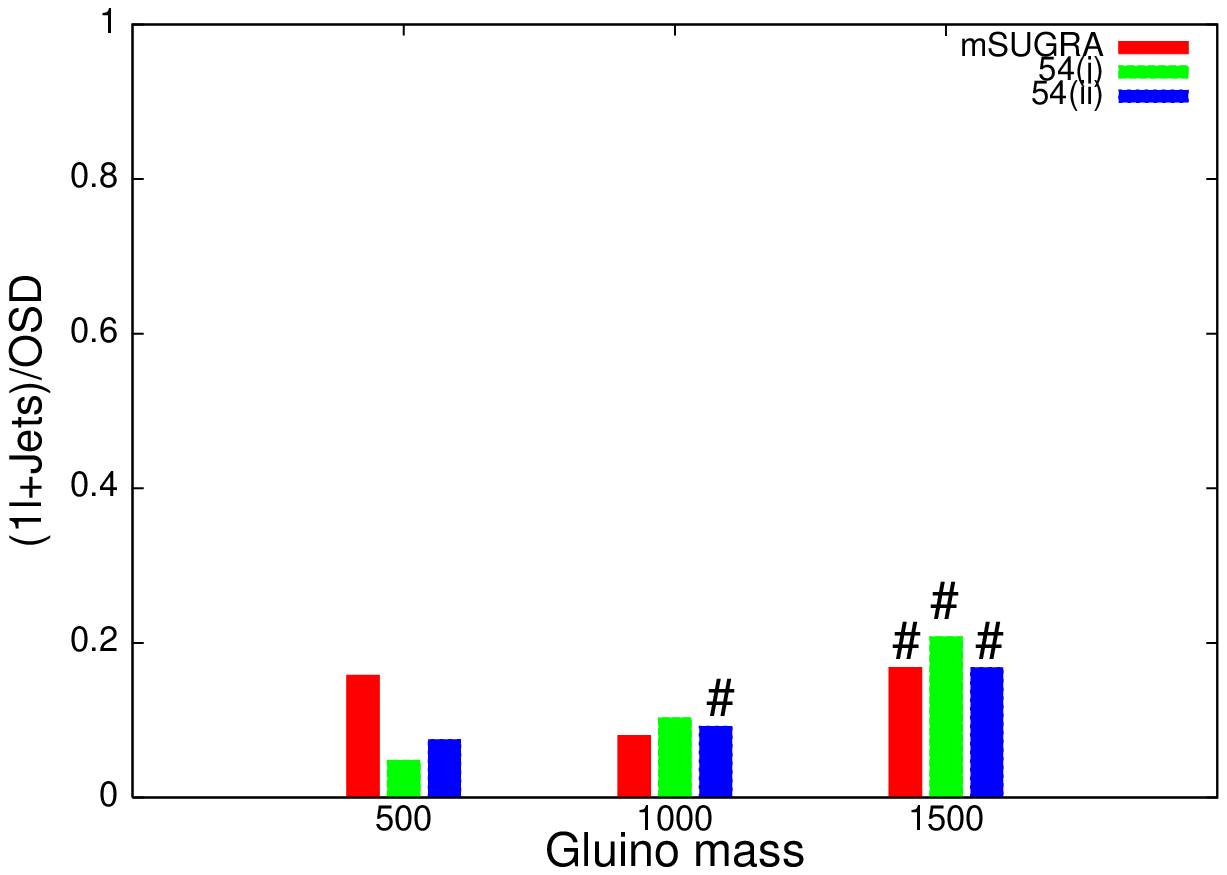,width=6.5cm,
height=5.50cm,angle=-0}}
\vskip 10pt
\centerline{\epsfig{file=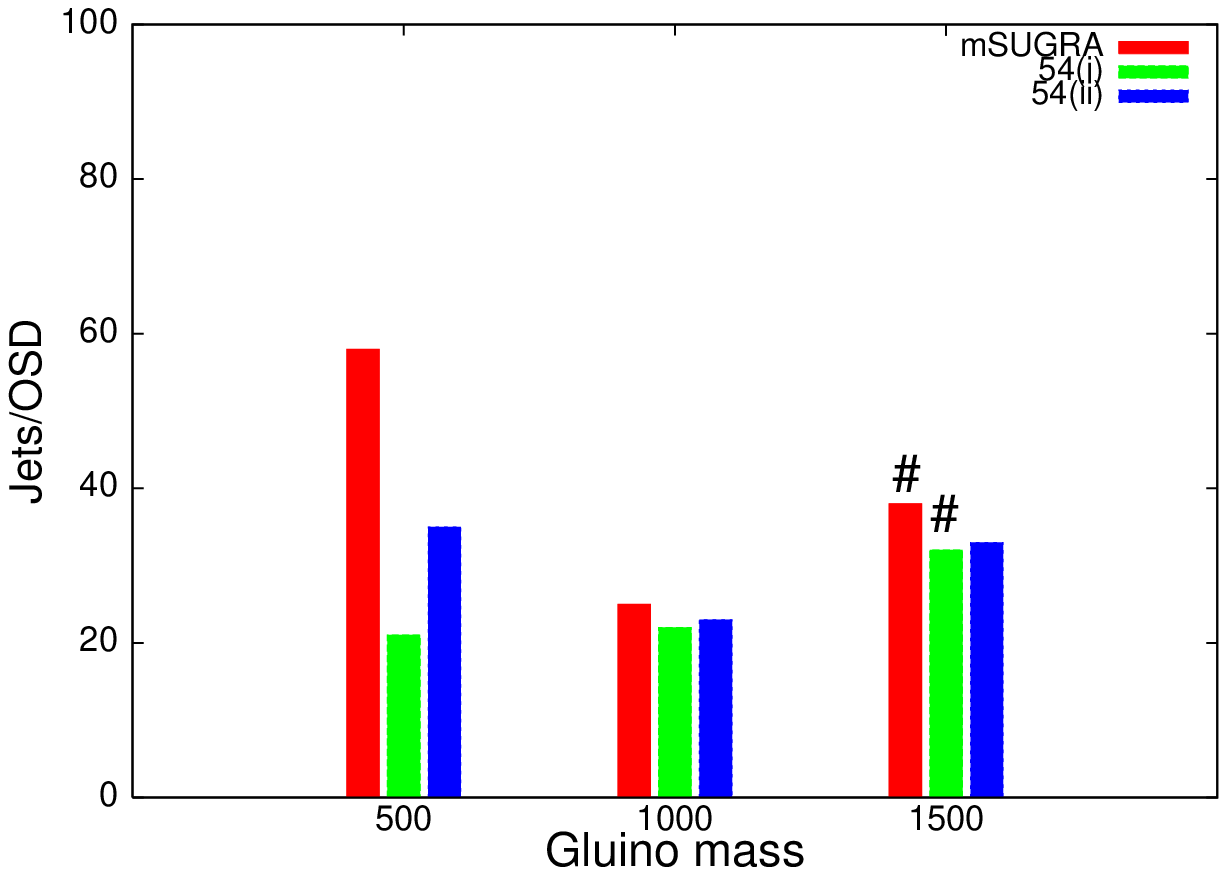,width=6.5 cm,height=5.50cm,angle=-0}
\hskip 20pt \epsfig{file=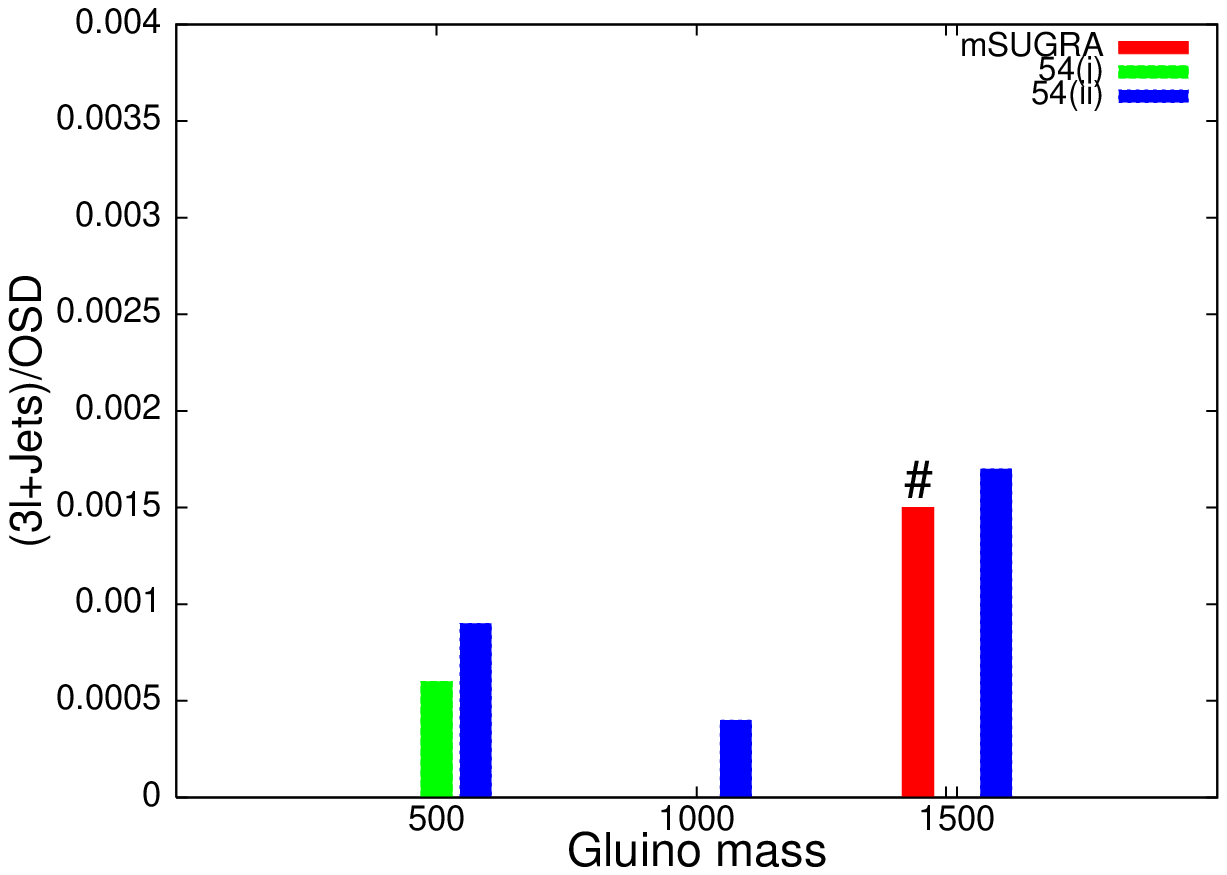,width=6.5cm,height=5.50cm,
angle=-0}}
\caption{ Event ratios for SO(10) {\bf SUGRA} with non-universal gaugino masses:
 $m_{0}=506$ GeV, $\tan{\beta}=5$, $sgn(\mu)=+$, $A_{0}=0$} 
%distribution}
\end{center}

\end{figure}

\section {Summary and conclusions}

We have carried out a multichannel analysis of SUSY signals, including 
$jets$ + ${E}_T\!\!\!\!\!\!/~$, SSD, OSD, 
$trileptons +jets + {E}_T\!\!\!\!\!\!/~$ and 
$single~ lepton + jets + {E}_T\!\!\!\!\!\!/~$, 
for a number of non-universal
representations breaking the SU(5) and SO(10) GUT groups, and
compared them with those corresponding to universal gaugino masses. 
While all representations of SU(5) have been considered, we have
confined ourselves to two breaking chains of SO(10) through {\bf 54}.
Both a phenomenological SUSY spectrum for the remaining particles
and one arising from a SUGRA scenario have been studied in
this context.

We have found it most useful to discriminate among the various
cases with the help of ratios of event rates for the various
signals mentioned above, although the absolute event rates have 
also been presented, and can be used for specific cases. In any case
 the absolute event rates provide additional information which can be 
gainfully used in one's analysis. In general, 
it is found that the GUT-breaking representations
are rather clearly differentiable over a substantial region of
the parameter space in the case of
{\bf 75} and  {\bf 200} of SU(5) and {\bf 54 (i)}  of SO(10). 
For these kinds of gaugino non-universality, distinction between
an SU(5) and an SO(10) SUSY is also rather easy.
For the {\bf 24} of SU(5), {\bf 54(ii)} of SO(10) and the universal case,
such distinction is relatively difficult in many cases from the
event ratios, and one may have to use the absolute event rates 
for them. However, even in these cases the ratio $(3\ell+jets)$/OSD 
can be useful in discrimination, especially in separating
the universal case.
In general, distinction is relatively easy for high values
of  $\mu$, since a low $\mu$ enhances the Higgsino component of
low-lying charginos and neutralinos, thus tending to partially
obliterate the clear stamps of various gaugino mass patterns
as manifested in the physical states. It is also interesting to note 
that for the non-minimal SUGRA scenario, 
at $m_{\tilde g}$= 1500 GeV, only {\bf 24} in SU(5) and both 
the breaking chains of {\bf 54} in SO(10) give excess signal 
over background in almost all channels, while others including mSUGRA 
are always overwhelmed by background in OSD channel.

In the effort to learn about gaugino non-universality, one
is also required to have an idea of the gluino and sfermion
masses, and it is expected that various kinematic distributions 
(ranging from $p_T$ to effective mass) will throw light on them
in such a study. The role of such distributions
(especially of missing $p_T$ and lepton $p_T$)
is also important when judgment has to be made on the basis
of the mass separation between the two lightest neutralinos,
which is a possible discriminator between {\bf 24} of SU(5)
and the universal case. While the value of $\tan\beta$,
another quantity affecting the observables, can be obtained from
studies of the SUSY Higgs sector and Yukawa couplings,
extraction of the value of the $\mu$ is a more challenging task.
One is likely to face this challenge in ascertaining the nature of
gaugino non-universality, if any, unless the magnitude $\mu$ is 
determined by radiative electroweak symmetry breaking, as is
expected in a SUGRA scenario.

It should also be noted that, in an illustrative study like this,
we have used leading order cross-sections only. Higher order effects
need to be taken into account in order to complete the study,
although the use of ratios suggested by us can cancel the $K$-factors.
However, our preliminary investigation serves to show that, once
data from the LHC are available, a detailed look at them can indeed
indicate whether some SUSY signals are consistent with specific 
scenarios embedded in a GUT setting. Our study is thus commensurate
with the `inverse problem' approach to LHC data.

On the whole, the exploration of gaugino non-universality is an
extremely important task in understanding the underlying nature
of a SUSY scenario. Therefore, further elaborate studies in this
direction need to be undertaken in a signal-based manner.\\

\noindent
{\bf Acknowledgment:}
This work was supported by funds made available by the Department of 
Atomic Energy, Government of India, under a Five-Year Plan Project.
Computational work for this study were partially carried out at the
cluster computing facility at the Harish-Chandra Research Institute
(http://cluster.mri.ernet.in). 
We would like to thank Sudhir Kumar Gupta for his help
in finalising our code. SB thanks Priyotosh Bandyopadhyay, Shamik Banerjee and
Utpal Chattopadhyay 
for many useful discussions. The authors also like to acknowledge
the hospitality of the Theoretical Physics Department, 
Indian Association for the Cultivation of Science,
Kolkata, where a part the project was carried out.

\newpage
\noindent
{\large {\bf APPENDIX A}}\\

\noindent
Here we list the neutralino and chargino masses in the region of the 
parameter space covered by us for all the representations. Tables A1-A8 
represent mass spectra in {\bf pMSSM} framework in SU(5) and SO(10), 
while table A9 is for the {\bf SUGRA} framework. In tables A1-A8, 
we depict the spectra for three gluino masses namely
$m_{\tilde g}$= 500 GeV, 1000 GeV and 1500 GeV and fixed 
$\mu$, $m_{\tilde f}$ and $\tan \beta$. The entries marked NA do not give 
consistent spectra having a neutralino LSP or are disallowed by LEP limits.\\

\noindent

\begin{center}

\begin{tabular}{|c|c|c|c|c|c|c|c|}

\multicolumn{8}{c}{Table A1 : Neutralino and Chargino spectra (GeV) for SU(5) and SO(10) {\bf pMSSM}}\\
\multicolumn{8}{c}{$m_{\tilde f}=~500$ GeV, $\mu$= 300 GeV, $\tan \beta$= 5} \\
\multicolumn{8}{c}{(Figures 1 and 10)}\\

\hline
 $m_{\tilde g}$ & Model & $m_{\tilde {\chi^{0}}_{1}}$ & 
$m_{\tilde {\chi^{0}}_{2}}$ &$m_{\tilde {\chi^{0}}_{3}}$& 
$m_{\tilde {\chi^{0}}_{4}}$ & $m_{\tilde {\chi^{\pm}}_{1}}$ 
& $m_{\tilde {\chi^{\pm}}_{2}}$\\
\hline 
 500 &{\bf universal} & 66.8 & 128.3 & 305.9 & 330.2 & 126.8 & 329.5 \\
\hline
 500 &{\bf 24} & 37.6 & 209.8 & 312.2 & 323 & 210.8 & 328.3 \\
\hline
 500 &{\bf 75} & 276 & 294 & 371.1 & 474.1 & 276.4 & 474.2 \\
\hline
 500 &{\bf 200} & 232.73 & 303.83 & 365.66 & 729.01 & 235.36 & 369.37  \\
\hline
500 &{\bf 54(i)} & 68.13 & 276.34 & 312.36 & 378.34 & 280.24 & 379.78  \\
\hline
500 &{\bf 54(ii)} & 72.79 & 210.45 & 311.88 & 323.44 & 211.43 & 328.44  \\
\hline
1000 &{\bf universal} & 140.4 & 243.1 & 304.5 & 373.6 & 238.2 & 372.3  \\
\hline
1000 &{\bf 24} & 75.5 & 291.2 & 309.3 & 474.8 & 294.4 & 475.0  \\
\hline
1000 &{\bf 75} & 294.1 & 300.3 & 751.7 & 927.7 & 294.4 & 927.7  \\
\hline
1000 &{\bf 200} & 285.47 & 302.46 & 631.27 & 1509.18 & 288.47 & 631.47  \\
\hline
1000 &{\bf 54(i)} & 142.82 & 296.65 & 312.53 & 723.07 & 299.76 & 723.11  \\
\hline
1000 &{\bf 54(ii)} & 148.42 & 292.45 & 308.76 & 475.10 & 294.41 & 475.30  \\
\hline
1500 &{\bf universal} & 211.42 & 293.78 & 303.64 & 491.75 & 278.76 & 491.42  \\
\hline
1500 &{\bf 24} & 114.81 & 298.36 & 307.56 & 718.07 & 299.73 & 718.10  \\
\hline
1500 &{\bf 75} & 296.86 & 300.48 & 1155.07 & 1413.08 & 297.09 & 1413.08  \\
\hline
1500 &{\bf 200} & 292.48 & 301.79 & 951.77 & 2322.19 & 294.64 & 951.81  \\
\hline
1500 &{\bf 54(i)} & 216.22 & 298.63 & 318.30 & 1098.88 & 300.77 & 1098.89  \\
\hline
1500 &{\bf 54(ii)} & 224.27 & 302.87 & 306.93 & 710.06 & 299.67 & 710.08  \\
\hline
\end {tabular}\\

\noindent

\end {center}

\noindent

\begin{center}

\begin{tabular}{|c|c|c|c|c|c|c|c|}

\multicolumn{8}{c}{Table A2: Neutralino and Chargino spectra (GeV) for SU(5) and SO(10) {\bf pMSSM}}\\
\multicolumn{8}{c}{$m_{\tilde f}=~1000$ GeV,
$\mu$= 300 GeV, $\tan \beta$= 5} \\
\multicolumn{8}{c}{(Figures 3 and 12)}\\
\hline
 $m_{\tilde g}$ & Model & $m_{\tilde {\chi^{0}}_{1}}$ & 
$m_{\tilde {\chi^{0}}_{2}}$ &$m_{\tilde {\chi^{0}}_{3}}$& 
$m_{\tilde {\chi^{0}}_{4}}$ & $m_{\tilde {\chi^{\pm}}_{1}}$ 
& $m_{\tilde {\chi^{\pm}}_{2}}$\\
\hline 
 500 &{\bf universal} & 68.8 & 131.6 & 305.7 & 330.4 & 130.0 & 329.6 \\
\hline
 500 &{\bf 24} & 38.6 & 213.9 & 312.0 & 323.5 & 215.0 & 328.6 \\
\hline

 500 &{\bf 75} & 277.4 & 295.1 & 379.2 & 481.8 & 277.8 & 482.0 \\
\hline
 500 &{\bf 200} & 235.87 & 303.71 & 368.42 & 746.27 & 238.55 & 371.84  \\
\hline
500 &{\bf 54(i)} & 69.95 & 278.83 & 312.21 & 385.02 & 282.81 & 386.24  \\
\hline
500 &{\bf 54(ii)} & 74.84 & 213.99 & 311.70 & 323.81 & 215.00 & 328.63  \\
\hline
1000 &{\bf universal} & 142.3 & 245.1 & 304.4 & 374.9 & 240.1 & 373.5  \\
\hline
1000 &{\bf 24} & 76.5 & 291.7 & 309.2 & 479.3 & 294.8 & 479.5  \\
\hline
1000 &{\bf 75} & 294.2 & 300.3 & 760.1 & 935.9 & 294.5 & 935.9  \\
\hline
1000 &{\bf 200} & 285.83 & 302.39 & 636.12 & 1524.92 & 288.76 & 636.30  \\
\hline
1000 &{\bf 54(i)} & 144.38 & 296.80 & 312.51 & 729.18 & 299.86 & 729.22  \\
\hline
1000 &{\bf 54(ii)} & 150.16 & 292.92 & 308.64 & 478.87 & 294.81 & 479.05  \\
\hline
1500 &{\bf universal} & 212.88  & 294.61 & 303.56 & 494.02 & 279.26 & 493.70  \\
\hline
1500 &{\bf 24} & 115.53 & 298.38 & 307.51 & 713.30 & 299.75 & 713.33  \\
\hline
1500 &{\bf 75} & 296.92 & 300.46 & 1161.27 & 1419.11 & 297.13 & 1419.11  \\
\hline
1500 &{\bf 200} & 292.6 & 301.75 & 955.63 & 233.44 & 294.72 & 955.67  \\
\hline
1500 &{\bf 54(i)} & 217.29 & 298.68 & 318.41 & 1103.54 & 300.79 & 1103.55  \\
\hline
1500 &{\bf 54(ii)} & 225.64 & 303.05 & 306.85 & 713.38 & 299.75 & 713.40  \\
\hline
\end {tabular}\\

\noindent

\end {center}

\noindent

\begin{center}

\begin{tabular}{|c|c|c|c|c|c|c|c|}

\multicolumn{8}{c}{Table A3: Neutralino and Chargino spectra (GeV) for SU(5) and SO(10) {\bf pMSSM} }\\
\multicolumn{8}{c}{$m_{\tilde f}=~500$ GeV,
$\mu$= 1000 GeV, $\tan \beta$= 5} \\
\multicolumn{8}{c}{(Figures 5 and 14)}\\
\hline
 $m_{\tilde g}$ & Model & $m_{\tilde {\chi^{0}}_{1}}$ & 
$m_{\tilde {\chi^{0}}_{2}}$ &$m_{\tilde {\chi^{0}}_{3}}$& 
$m_{\tilde {\chi^{0}}_{4}}$ & $m_{\tilde {\chi^{\pm}}_{1}}$ 
& $m_{\tilde {\chi^{\pm}}_{2}}$\\
\hline 
 500 &{\bf universal} & 73.0 & 148.0 & 1002.2 & 1006.6 & 147.9 & 1006.9 \\
\hline
 500 &{\bf 24} & 37.7 & 228.6 & 1003.1 & 1004.9 & 228.6 & 1006.1 \\
\hline
 500 &{\bf 75} & 371.0 & 449.0 & 1002.0 & 1009.1 & 449.0 & 1009.5 \\
\hline
 500 &{\bf 200} & 299.19 & 738.18 & 1001.81 & 1011.24 & 299.21 & 1007.88  \\
\hline
500 &{\bf 54(i)} & 73.25 & 354.58 & 1003.53 & 1004.71 & 354.59 & 1006.29  \\
\hline
500 &{\bf 54(ii)} & 74.62 & 228.62 & 1003.12 & 1004.85 & 228.62 & 1006.15  \\
\hline
1000 &{\bf universal} & 149.6 & 302.7 & 1002.0 & 1007.9 & 302.7 & 1007.9  \\
\hline
1000 &{\bf 24} & NA & NA & NA & NA & NA & NA  \\
\hline
1000 &{\bf 75} & NA & NA & NA & NA & NA & NA  \\
\hline
1000 &{\bf 200} & NA & NA & NA & NA & NA & NA  \\
\hline
1000 &{\bf 54(i)} & 150.47 & 716.90 & 1004.14 & 1007.33 & 716.94 & 1009.40  \\
\hline
1000 &{\bf 54(ii)} & 151.57 & 462.04 & 1004.19 & 1004.27 & 462.05 & 1006.65  \\
\hline
1500 &{\bf universal} & 228.81  & 461.31 & 1001.78 & 1009.98 & 461.26 & 1009.63  \\
\hline
1500 &{\bf 24} & 116.03 & 700.85 & 1003.8 & 1007.11 & 700.89 & 1009.06  \\
\hline
1500 &{\bf 75} & NA & NA & NA & NA & NA & NA  \\
\hline
1500 &{\bf 200} & NA & NA & NA & NA & NA & NA  \\
\hline
1500 &{\bf 54(i)} & 230.56 & 983.92 & 1003.83 & 1117.94 & 985.38 & 1118.16  \\
\hline
1500 &{\bf 54(ii)} & 231.35 & 700.91 & 1003.69 & 1007.21 & 700.95 & 1009.06  \\
\hline
\end {tabular}\\

\noindent

\end {center}

\noindent

\begin{center}

\begin{tabular}{|c|c|c|c|c|c|c|c|}

\multicolumn{8}{c}{Table A4 :  Neutralino and Chargino spectra (GeV) for SU(5) and SO(10) {\bf pMSSM}}\\
\multicolumn{8}{c}{$m_{\tilde f}=~1000$ GeV,
$\mu$= 1000 GeV, $\tan \beta$= 5} \\
\multicolumn{8}{c}{(Figures 7 and 16)}\\
\hline
 $m_{\tilde g}$ & Model & $m_{\tilde {\chi^{0}}_{1}}$ & 
$m_{\tilde {\chi^{0}}_{2}}$ &$m_{\tilde {\chi^{0}}_{3}}$& 
$m_{\tilde {\chi^{0}}_{4}}$ & $m_{\tilde {\chi^{\pm}}_{1}}$ 
& $m_{\tilde {\chi^{\pm}}_{2}}$\\
\hline 
 500 &{\bf universal} & 74.1 & 149.8 & 1002.2 & 1006.6 & 149.8 & 1006.8 \\
\hline
 500 &{\bf 24} & 38.2 & 231.4 & 1003.0 & 1004.8 & 465.5 & 1006.5 \\
\hline
 500 &{\bf 75} & 376.4 & 454.4 & 1002.2 & 1009.1 & 454.4 & 1009.4 \\
\hline
 500 &{\bf 200} & 302.60 & 748.21 & 1001.77 & 1011.38 & 302.62 & 1007.80  \\
\hline
500 &{\bf 54(i)} & 74.33 & 358.80 & 1003.48 & 1004.67 & 358.81 & 1006.20  \\
\hline
500 &{\bf 54(ii)} & 75.76 & 231.46 & 1003.26 & 1004.80 & 231.46 & 1006.55  \\
\hline
1000 &{\bf universal} & 150.9 & 305.0 & 1001.9 & 1007.9 & 305.0 & 1007.8  \\
\hline
1000 &{\bf 24} & 76.8 & 465.5 & 1004.1 & 1004.2 & 465.5 & 1006.5  \\
\hline
1000 &{\bf 75} & 763.2 & 894.0 & 1003.3 & 1042.3 & 894.0 & 1042.5  \\
\hline
1000 &{\bf 200} & 614.03 & 1001.39 & 1009.07 & 1536.64 & 614.09 & 1012.64  \\
\hline
1000 &{\bf 54(i)} & 151.60 & 721.34 & 1004.10 & 1007.29 & 721.39 & 1009.33  \\
\hline
1000 &{\bf 54(ii)} & 152.91 & 465.37 & 1004.15 & 1004.21 & 465.38 & 1006.55  \\
\hline
1500 &{\bf universal} & 230.06  & 463.39 & 1001.75 & 1009.92 & 463.34 & 1009.54  \\
\hline
1500 &{\bf 24} & 116.59 & 703.66 & 1003.76 & 1007.03 & 703.69 & 1008.95  \\
\hline
1500 &{\bf 75} & NA & NA & NA & NA & NA & NA  \\
\hline
1500 &{\bf 200} & 904.54 & 1001.14 & 1045.98 & 2344.38 & 905.13 & 1047.34  \\
\hline
1500 &{\bf 54(i)} & 231.64 & 984.71 & 1003.80 & 1121.25 & 986.17 & 1121.45  \\
\hline
1500 &{\bf 54(ii)} & 233.01 & 704.02 & 1003.65 & 1007.13 & 704.06 & 1008.96  \\
\hline
\end {tabular}\\

\noindent

\end {center}

\noindent

\begin{center}

\begin{tabular}{|c|c|c|c|c|c|c|c|}

\multicolumn{8}{c}{Table A5 :  Neutralino and Chargino spectra (GeV) for SU(5) and SO(10) {\bf pMSSM}}\\
\multicolumn{8}{c}{$m_{\tilde f}=~500$ GeV,
$\mu$= 300 GeV, $\tan \beta$= 40} \\
\multicolumn{8}{c}{(Figures 2 and 11)}\\
\hline
 $m_{\tilde g}$ & Model & $m_{\tilde {\chi^{0}}_{1}}$ & 
$m_{\tilde {\chi^{0}}_{2}}$ &$m_{\tilde {\chi^{0}}_{3}}$& 
$m_{\tilde {\chi^{0}}_{4}}$ & $m_{\tilde {\chi^{\pm}}_{1}}$ 
& $m_{\tilde {\chi^{\pm}}_{2}}$\\
\hline 
 500 &{\bf universal} & 69.3 & 134.5 & 309.2 & 323.7 & 134.2 & 326.2 \\
\hline
 500 &{\bf 24} & 35.3 & 198.8 & 309.3 & 332.2 & 199.2 & 334.8 \\
\hline
 500 &{\bf 75} & 281.3 & 291.2 & 373.6 & 468.1 & 283.5 & 468.5 \\
\hline
 500 &{\bf 200} & 241.96 & 306.0 & 356.96 & 722.48 & 245.05 & 361.29  \\
\hline
500 &{\bf 54(i)} & 69.89 & 266.23 & 309.42 & 386.17 & 268.02 & 387.19  \\
\hline
500 &{\bf 54(ii)} & 70.27 & 198.94 & 309.06 & 332.50 & 199.13 & 334.79  \\
\hline
1000 &{\bf universal} & 143.3 & 250.3 & 307.0 & 364.7 & 248.6 & 364.8  \\
\hline
1000 &{\bf 24} & 73.1 & 286.6 & 307.1 & 478.5 & 286.9 & 478.7  \\
\hline
1000 &{\bf 75} & 295.5 & 300.5 & 750.8 & 925.5 & 297.1 & 925.5  \\
\hline
1000 &{\bf 200} & 288.96 & 303.85 & 626.87 & 1501.91 & 292.89 & 627.07  \\
\hline
1000 &{\bf 54(i)} & 144.95 & 294.75 & 309.58 & 722.85 & 296.05 & 722.89  \\
\hline
1000 &{\bf 54(ii)} & 145.09 & 288.49 & 306.64 & 478.55 & 286.93 & 478.66  \\
\hline
1500 &{\bf universal} & 216.13 & 294.81 & 305.68 & 486.87 & 285.51 & 486.81  \\
\hline
1500 &{\bf 24} & 112.51 & 297.31 & 305.75 & 710.55 & 295.87 & 710.58  \\
\hline
1500 &{\bf 75} & 297.63 & 300.75 & 1153.72 & 1411.23 & 298.74 & 1411.23  \\
\hline
1500 &{\bf 200} & 294.32 & 302.81 & 948.35 & 2314.19 & 297.22 & 948.39  \\
\hline
1500 &{\bf 54(i)} & 219.55 & 297.85 & 314.36 & 1097.87 & 298.58 & 1097.88  \\
\hline
1500 &{\bf 54(ii)} & 219.94 & 304.27 & 305.244 & 710.57 & 295.87 & 710.58  \\
\hline
\end {tabular}\\

\noindent

\end {center}

\noindent

\begin{center}

\begin{tabular}{|c|c|c|c|c|c|c|c|}

\multicolumn{8}{c}{Table A6 :  Neutralino and Chargino spectra (GeV) for SU(5) and SO(10) {\bf pMSSM}}\\
\multicolumn{8}{c}{$m_{\tilde f}=~1000$ GeV,
$\mu$= 300 GeV, $\tan \beta$= 40} \\
\multicolumn{8}{c}{(Figures 4 and 13)}\\
\hline
 $m_{\tilde g}$ & Model & $m_{\tilde {\chi^{0}}_{1}}$ & 
$m_{\tilde {\chi^{0}}_{2}}$ &$m_{\tilde {\chi^{0}}_{3}}$& 
$m_{\tilde {\chi^{0}}_{4}}$ & $m_{\tilde {\chi^{\pm}}_{1}}$ 
& $m_{\tilde {\chi^{\pm}}_{2}}$\\
\hline 
 500 &{\bf universal} & 71.3 & 137.8 & 309.0 & 323.9 & 137.5 & 326.2 \\
\hline
 500 &{\bf 24} & 36.3 & 202.9 & 309.1 & 332.9 & 203.3 & 335.4 \\
\hline
 500 &{\bf 75} & 282.5 & 292.5 & 381.7 & 476.3 & 284.7 & 476.6 \\
\hline
 500 &{\bf 200} & 254.4 & 305.84 & 360.0 & 741.6 & 248.60 & 364.0  \\
\hline
500 &{\bf 54(i)} & 71.72 & 268.66 & 309.30 & 390.94 & 270.45 & 391.84  \\
\hline
500 &{\bf 54(ii)} & 72.23 & 203.11 & 308.88 & 333.26 & 203.30 & 335.39  \\
\hline
1000 &{\bf universal} & 144.7 & 251.8 & 306.9 & 365.4 & 250.0 & 365.4  \\
\hline
1000 &{\bf 24} & 73.9 & 287.2 & 307.0 & 482.1 & 287.5 & 482.3  \\
\hline
1000 &{\bf 75} & 295.7 & 300.7 & 758.3 & 932.9 & 297.2 & 932.9  \\
\hline
1000 &{\bf 200} & 289.26 & 302.18 & 1279.9 & 3148.39 & 298.5 & 1279.9  \\
\hline
1000 &{\bf 54(i)} & 146.47 & 294.93 & 309.54 & 728.67 & 296.19 & 728.71  \\
\hline
1000 &{\bf 54(ii)} & 146.76 & 289.12 & 306.53 & 482.16 & 287.46 & 482.26  \\
\hline
1500 &{\bf universal} & 217.50  & 295.43 & 305.6 & 488.9 & 285.91 & 488.83  \\
\hline
1500 &{\bf 24} & 112.73 & 297.44 & 305.67 & 713.34 & 295.97 & 713.37  \\
\hline
1500 &{\bf 75} & 297.54 & 301.07 & 1416.72 & 1577.18 & 298.76 & 1416.72  \\
\hline
1500 &{\bf 200} & 294.43 & 302.77 & 952.30 & 2326.64 & 297.28 & 952.34  \\
\hline
1500 &{\bf 54(i)} & 220.66 & 297.91 & 314.41 & 1102.44 & 298.62 & 1102.45  \\
\hline
1500 &{\bf 54(ii)} & 220.49 & 304.43 & 305.17 & 713.75 & 295.98 & 713.76  \\
\hline
\end {tabular}\\

\noindent

\end {center}

\noindent

\begin{center}

\begin{tabular}{|c|c|c|c|c|c|c|c|}

\multicolumn{8}{c}{Table A7 :  Neutralino and Chargino spectra (GeV) for SU(5) and SO(10) {\bf pMSSM}}\\
\multicolumn{8}{c}{$m_{\tilde f}=~500$ GeV,
$\mu$= 1000 GeV, $\tan \beta$= 40} \\
\multicolumn{8}{c}{(Figures 6 and 15)}\\
\hline
 $m_{\tilde g}$ & Model & $m_{\tilde {\chi^{0}}_{1}}$ & 
$m_{\tilde {\chi^{0}}_{2}}$ &$m_{\tilde {\chi^{0}}_{3}}$& 
$m_{\tilde {\chi^{0}}_{4}}$ & $m_{\tilde {\chi^{\pm}}_{1}}$ 
& $m_{\tilde {\chi^{\pm}}_{2}}$\\
\hline 
 500 &{\bf universal} & 73.0 & 149.0 & 1003.5 & 1005.0 & 149.0 & 1006.5 \\
\hline
 500 &{\bf 24} & NA & NA & NA & NA & NA & NA \\
\hline
 500 &{\bf 75} & NA & NA & NA & NA & NA & NA \\
\hline
 500 &{\bf 200} & 299.23 & 732.41 & 1002.83 & 1008.35 & 299.25 & 1007.09  \\
\hline
500 &{\bf 54(i)} & 73.26 & 349.6 & 1003.53 & 1005.47 & 349.61 & 1007.11  \\
\hline
500 &{\bf 54(ii)} & 73.30 & 224.59 & 1003.64 & 1004.85 & 224.59 & 1006.62  \\
\hline
1000 &{\bf universal} & 149.4 & 303.7 & 1003.1 & 1005.9 & 303.7 & 1007.1  \\
\hline
1000 &{\bf 24} & NA & NA & NA & NA & NA & NA  \\
\hline
1000 &{\bf 75} & NA & NA & NA & NA & NA & NA  \\
\hline
1000 &{\bf 200} & NA & NA & NA & NA & NA & NA  \\
\hline
1000 &{\bf 54(i)} & 150.40 & 709.85 & 1003.10 & 1011.17 & 709.88 & 1012.32  \\
\hline
1000 &{\bf 54(ii)} & 150.09 & 457.27 & 1003.14 & 1006.60 & 457.27 & 1007.85  \\
\hline
1500 &{\bf universal} & 228.57  & 462.53 & 1002.78 & 1007.48 & 462.52 & 1008.27  \\
\hline
1500 &{\bf 24} & NA & NA & NA & NA & NA & NA  \\
\hline
1500 &{\bf 75} & NA & NA & NA & NA & NA & NA  \\
\hline
1500 &{\bf 200} & NA & NA & NA & NA & NA & NA  \\
\hline
1500 &{\bf 54(i)} & NA & NA & NA & NA & NA & NA  \\
\hline
1500 &{\bf 54(ii)} & NA & NA & NA & NA & NA & NA  \\
\hline
\end {tabular}\\

\noindent

\end {center}

\noindent

\begin{center}

\begin{tabular}{|c|c|c|c|c|c|c|c|}

\multicolumn{8}{c}{Table A8 :  Neutralino and Chargino spectra (GeV) for SU(5) and SO(10) {\bf pMSSM}}\\
\multicolumn{8}{c}{$m_{\tilde f}=~1000$ GeV,
$\mu$= 1000 GeV, $\tan \beta$= 40} \\
\multicolumn{8}{c}{(Figures 8 and 17)}\\
\hline
 $m_{\tilde g}$ & Model & $m_{\tilde {\chi^{0}}_{1}}$ & 
$m_{\tilde {\chi^{0}}_{2}}$ &$m_{\tilde {\chi^{0}}_{3}}$& 
$m_{\tilde {\chi^{0}}_{4}}$ & $m_{\tilde {\chi^{\pm}}_{1}}$ 
& $m_{\tilde {\chi^{\pm}}_{2}}$\\
\hline 
 500 &{\bf universal} & 74.2 & 151.1 & 1003.4 & 1005.9 & 151.1 & 1006.5 \\
\hline
 500 &{\bf 24} & 37.3 & 227.4 & 1003.6 & 1004.8 & 227.4 & 1006.6 \\
\hline
 500 &{\bf 75} & 327.7 & 454.2 & 1003.5 & 1006.8 & 454.2 & 1008.1 \\
\hline
 500 &{\bf 200} & 303.39 & 744.63 & 1002.79 & 1008.48 & 303.4 & 1007.02  \\
\hline
500 &{\bf 54(i)} & 74.39 & 354.00 & 1003.5 & 1005.45 & 354.01 & 1007.07  \\
\hline
500 &{\bf 54(ii)} & 74.52 & 227.66 & 1003.60 & 1004.82 & 227.67 & 1006.56  \\
\hline
1000 &{\bf universal} & 150.8 & 306.1 & 1003.0 & 1005.9 & 306.1 & 1007.0  \\
\hline
1000 &{\bf 24} & 75.7 & 460.1 & 1003.2 & 1006.5 & 460.6 & 1007.8  \\
\hline
1000 &{\bf 75} & 758.4 & 899.7 & 1005.2 & 1033.5 & 900.0 & 1034.3  \\
\hline
1000 &{\bf 200} & 615.17 & 1002.19 & 1006.76 & 1528.79 & 615.21 & 1010.41  \\
\hline
1000 &{\bf 54(i)} & 151.65 & 714.62 & 1003.08 & 1011.24 & 714.65 & 1012.37  \\
\hline
1000 &{\bf 54(ii)} & 151.43 & 460.59 & 1003.11 & 1006.58 & 460.60 & 1007.80  \\
\hline
1500 &{\bf universal} & 229.98  & 464.89 & 1002.76 & 1007.45 & 464.89 & 1008.22  \\
\hline
1500 &{\bf 24} & 115.55 & 697.57 & 1002.82 & 1010.87 & 697.59 & 1011.80  \\
\hline
1500 &{\bf 75} & NA & NA & NA & NA & NA & NA  \\
\hline
1500 &{\bf 200} & NA & NA & NA & NA & NA & NA  \\
\hline
1500 &{\bf 54(i)} & 231.57 & 976.77 & 1002.85 & 1126.06 & 977.49 & 1126.21  \\
\hline
1500 &{\bf 54(ii)} & 230.95 & 697.53 & 1002.74 & 1011.01 & 697.54 & 1011.80  \\
\hline
\end {tabular}\\

\noindent

\end {center}

\noindent

\begin{center}

\begin{tabular}{|c|c|c|c|c|c|c|c|}

\multicolumn{8}{c}{Table A9 : Neutralino and Chargino spectra (GeV) for SU(5) and SO(10) {\bf SUGRA}}\\
\multicolumn{8}{c}{$m_{\tilde f}=~506$ GeV at $M_{GUT}$,
 $\tan \beta$= 5} \\
\multicolumn{8}{c}{(Figures 9 and 18)}\\
\hline
 $m_{\tilde g}$ & Model & $m_{\tilde {\chi^{0}}_{1}}$ & 
$m_{\tilde {\chi^{0}}_{2}}$ &$m_{\tilde {\chi^{0}}_{3}}$& 
$m_{\tilde {\chi^{0}}_{4}}$ & $m_{\tilde {\chi^{\pm}}_{1}}$ 
& $m_{\tilde {\chi^{\pm}}_{2}}$\\
\hline 
 500 &{\bf universal} & 70.74 & 129.16 & 289.03 & 316.94 & 127.91 & 314.65 \\
\hline
 500 &{\bf 24} & 42.54 & 199.39 & 252.60 & 288.42 & 200.04 & 289.92 \\
\hline
 500 &{\bf 75} & 136.36 & 147.69 & 400.26 & 470.33 & 138.90 & 467.43 \\
\hline
 500 &{\bf 200} & 202.80 & 249.30 & 348.56 & 792.82 & 207.99 & 348.32  \\
\hline
500 &{\bf 54(i)} & 66.97 & 169.16 & 196.78 & 376.84 & 169.85 & 372.59  \\
\hline
500 &{\bf 54(ii)} & 80.32 & 199.26 & 251.2 & 288.21 & 199.52 & 289.38  \\
\hline
1000 &{\bf universal} & 171.20 & 321.40 & 555.60 & 574.93 & 321.55 & 573.38 \\
\hline
1000 &{\bf 24} & 92.52 & 420.53 & 445.10 & 545.27 & 413.17 & 538.06  \\
\hline
1000 &{\bf 75} & NA & NA & NA & NA & NA & NA  \\
\hline
1000 &{\bf 200} & 414.84 & 433.79 & 686.93 & 1767.96 & 421.96 & 680.06  \\
\hline
1000 &{\bf 54(i)} & 158.55 & 251.39 & 271.43 & 795.42 & 245.44 & 785.1  \\
\hline
1000 &{\bf 54(ii)} & 179.6 & 419.19 & 442.6 & 544.84 & 411.23 & 537.62  \\
\hline
1500 &{\bf universal} & 275.57 & 519.73 & 819.60 & 834.40 & 520.22 & 833.47 \\
\hline
1500 &{\bf 24} & 145.87 & 624.14 & 638.06 & 831.61 & 608.20 & 818.44  \\
\hline
1500 &{\bf 75} & NA & NA & NA & NA & NA & NA  \\
\hline
1500 &{\bf 200} & 592.68 & 603.33 & 1059.96 & 2804.17 & 602.40 & 1048.06  \\
\hline
1500 &{\bf 54(i)} & 244.91 & 291.76 & 321.48 & 1234.77 & 281.58 & 1220.69  \\
\hline
1500 &{\bf 54(ii)} & 285.17 & 620.81 & 633.83 & 831.08 & 604.33 & 817.89  \\
\hline
\end {tabular}\\

\noindent

\end {center}

\newpage
\noindent
 {\large {\bf APPENDIX B}}

\noindent
In this appendix we tabulate the cross-sections in each channel 
for all representations of SU(5) and SO(10) in the region of parameter space
studied and depicted in figures 1-18. The cross-sections are named as follows:
 $\sigma_{1}$ for OSD, $\sigma_{2}$ for SSD, $\sigma_{3}$ for $(1\ell+jets)$, 
$\sigma_{4}$ for $jets$ and $\sigma_{5}$ for $(3\ell+jets)$.
The points for which we do not get 
consistent spectra are denoted by NA as earlier and the points which 
give null result (for $(3\ell+jets)$ 
channel only) is written as NULL. Bold faced 
entries correspond to cross-sections which are less than $2\sigma$ 
above the background for an integrated luminosity of 300 fb$^{-1}$.\\

\noindent

\begin{center}

\begin{tabular}{|c|c|c|c|c|c|c|}

\multicolumn{7}{c}{Table B1 : Cross-sections (pb) for SU(5) and SO(10) 
{\bf pMSSM}}\\
\multicolumn{7}{c}{$m_{\tilde f}=~500$ GeV,
$\mu$= 300 GeV, $\tan \beta$= 5} \\
\multicolumn{7}{c}{(Figures 1 and 10)}\\
\hline
 $m_{\tilde g}$ & Model & $\sigma_{1}$ & $\sigma_{2}$ &$\sigma_{3}$& 
$\sigma_{4}$ & $\sigma_{5}$ \\

\hline 
 500 &{\bf universal} & 0.3434 & 0.1157 & 0.0472 & 18.3140 & NULL \\
\hline
 500 &{\bf 24} & 0.4648 & 0.1223 & 0.0552 & 20.2893 & NULL \\
\hline
 500 &{\bf 75} & 0.0388 & 0.0185 & 0.0178 & 2.0555 & NULL  \\
\hline
 500 &{\bf 200} & 0.0576 & 0.0240 & 0.0133 & 5.5483 & NULL   \\
\hline
500 &{\bf 54(i)} & 0.3682 & 0.0970 & 0.0440 & 16.8908 & NULL \\
\hline
500 &{\bf 54(ii)} & 0.4456 & 0.1291 & 0.0483 & 18.3423 & NULL   \\
\hline
1000 &{\bf universal} & 0.1086 & 0.0261 & 0.0152 & 3.4062 & NULL \\
\hline
1000 &{\bf 24} & 0.0808 & 0.0340 & 0.0133 & 4.0154 & NULL   \\
\hline
1000 &{\bf 75} & 0.0089 & 0.0063 & 0.0054 & 1.3613 & NULL  \\
\hline
1000 &{\bf 200} & 0.0090 & 0.0072 & 0.0048 & 1.4017 & NULL \\
\hline
1000 &{\bf 54(i)} & 0.0446 & 0.0180 & 0.0114 & 2.8733 & NULL  \\
\hline
1000 &{\bf 54(ii)} & 0.0745 & 0.0316 & 0.0103 & 3.2941 & NULL  \\
\hline
1500 &{\bf universal} & 0.0346 & 0.0845 & 0.0512 & 0.7688  & NULL  \\
\hline
1500 &{\bf 24} & 0.0265 & 0.0096 & 0.0040 & 1.2308 & NULL  \\
\hline
1500 &{\bf 75} & 0.0037 & {\bf 0.0010} & 0.0020 & 0.2852 & NULL  \\
\hline
1500 &{\bf 200} & 0.0034 & 0.0019 & 0.0026 & 0.3110 & NULL \\
\hline
1500 &{\bf 54(i)} & 0.0167 & 0.0060 & 0.0033 & 0.6066 & NULL \\
\hline
1500 &{\bf 54(ii)} & 0.0239 & 0.0057 & 0.0036 & 0.6256  & NULL \\
\hline
\end {tabular}\\

\noindent

\end {center}

\noindent

\begin{center}

\begin{tabular}{|c|c|c|c|c|c|c|}

\multicolumn{7}{c}{Table B2 : Cross-sections (pb) for SU(5) and SO(10) 
{\bf pMSSM}}\\
\multicolumn{7}{c}{$m_{\tilde f}=~1000$ GeV,
$\mu$= 300 GeV, $\tan \beta$= 5} \\
\multicolumn{7}{c}{(Figures 3 and 12)}\\
\hline
  $m_{\tilde g}$ & Model & $\sigma_{1}$ & $\sigma_{2}$ &$\sigma_{3}$& 
$\sigma_{4}$ & $\sigma_{5}$ \\
\hline 
 500 &{\bf universal} & 0.1400 & 0.0440 & 0.0230 & 8.3310 & NULL \\
\hline
 500 &{\bf 24} & 0.1317 & 0.0463 & 0.0207 & 8.7260 & NULL \\
\hline

 500 &{\bf 75} & 0.0108 & 0.0048 & 0.0064 & 3.3280 & NULL \\
\hline
 500 &{\bf 200} & 0.0137 & 0.0068 & 0.0079 & 4.5549 & NULL  \\
\hline
500 &{\bf 54(i)} & 0.0600 & 0.0154 & 0.0239 & 8.7907 & NULL  \\
\hline
500 &{\bf 54(ii)} & 0.1396 & 0.0479 & 0.0225 & 8.1194 & NULL  \\
\hline
1000 &{\bf universal} & 0.0310 & 0.0132 & 0.0033 & 0.8462 & $2.0\times 10^{-5}$  \\
\hline
1000 &{\bf 24} & 0.0350 & 0.0196 & 0.0034 & 0.9417 & $5.0\times 10^{-5}$  \\
\hline
1000 &{\bf 75} & 0.0197 & 0.0137 & 0.0040 & 0.8528 & NULL  \\
\hline
1000 &{\bf 200} & 0.0145 & 0.0091 & 0.0027 & 0.7410 & $3.0\times 10^{-5}$  \\
\hline
1000 &{\bf 54(i)} & 0.0371 & 0.0242 & 0.0044 & 1.0666 & $7.0\times 10^{-5}$  \\
\hline
1000 &{\bf 54(ii)} & 0.0397 & 0.0228 & 0.0038 & 1.0930 & 0.0001  \\
\hline
1500 &{\bf universal} & 0.0091 & 0.0032 & 0.0010 & 0.2788 & NULL  \\
\hline
1500 &{\bf 24} & 0.0089 & 0.0037 & 0.0015 & 0.3422 & NULL  \\
\hline
1500 &{\bf 75} & {\bf 0.0006} & {\bf 0.0003} & {\bf 0.0006} & 0.1023 & NULL  \\
\hline
1500 &{\bf 200} & {\bf 0.0016} & {\bf 0.0006} & 0.0007 & 0.1259 & NULL  \\
\hline
1500 &{\bf 54(i)} & 0.0037 & {\bf 0.0009} & 0.0008 & 0.1766 & NULL  \\
\hline
1500 &{\bf 54(ii)} & 0.0090 & 0.0029 & 0.0013 & 0.3120 & NULL  \\
\hline
\end {tabular}\\

\noindent

\end {center}

\noindent

\begin{center}

\begin{tabular}{|c|c|c|c|c|c|c|}

\multicolumn{7}{c}{Table B3 : Cross-sections (pb) for SU(5) and SO(10) 
{\bf pMSSM}}\\
\multicolumn{7}{c}{$m_{\tilde f}=~500$ GeV,
$\mu$= 1000 GeV, $\tan \beta$= 5} \\
\multicolumn{7}{c}{(Figures 5 and 14)}\\
\hline
  $m_{\tilde g}$ & Model & $\sigma_{1}$ & $\sigma_{2}$ &$\sigma_{3}$& 
$\sigma_{4}$ & $\sigma_{5}$ \\
\hline 
 500 &{\bf universal} & 0.7456 & 0.1483 & 0.0680 & 18.8841 & NULL  \\
\hline
 500 &{\bf 24} & 0.3510 & 0.1814 & 0.0537 & 19.1663 & NULL  \\
\hline
 500 &{\bf 75} & 0.0356 & {\bf 0.0013} & 0.0013 & 0.1100 & NULL  \\
\hline
 500 &{\bf 200} & 0.0125 & 0.0075 & 0.0106 & 0.9345 & NULL  \\
\hline
500 &{\bf 54(i)} & 0.2831 & 0.1015 & 0.0439 & 14.3062 & NULL  \\
\hline
500 &{\bf 54(ii)} & 0.2979 & 0.1694 & 0.0567 & 17.4439 & 0.0007  \\
\hline
1000 &{\bf universal} & 0.0453 & 0.0293 & 0.0124 & 3.6705 & NULL  \\
\hline
1000 &{\bf 24} & NA & NA & NA & NA & NA   \\
\hline
1000 &{\bf 75} & NA & NA & NA & NA & NA   \\
\hline
1000 &{\bf 200} & NA & NA & NA & NA & NA   \\
\hline
1000 &{\bf 54(i)} & 0.0102 & 0.0057 & 0.0107 & 1.9349 & NULL  \\
\hline
1000 &{\bf 54(ii)} & 0.0337 & 0.0096 & 0.0118 & 2.1440 & 0.0001  \\
\hline
1500 &{\bf universal} & 0.0090 & 0.0036 & 0.0049 & 0.5811 & NULL  \\
\hline
1500 &{\bf 24} & 0.0062 & 0.0031 & 0.0066 & 0.4968 & NULL  \\
\hline
1500 &{\bf 75} & NA & NA & NA & NA & NA   \\
\hline
1500 &{\bf 200} & NA & NA & NA & NA & NA   \\
\hline
1500 &{\bf 54(i)} & 0.0036 & 0.0020 & 0.0047 & 0.3528 & NULL  \\
\hline
1500 &{\bf 54(ii)} & 0.0045 & 0.0019 & 0.0048 & 0.4229 & NULL \\
\hline
\end {tabular}\\

\noindent

\end {center}

\noindent

\begin{center}

\begin{tabular}{|c|c|c|c|c|c|c|}

\multicolumn{7}{c}{Table B4 : Cross-sections (pb) for SU(5) and SO(10) 
{\bf pMSSM}}\\
\multicolumn{7}{c}{$m_{\tilde f}=~1000$ GeV,
$\mu$= 1000 GeV, $\tan \beta$= 5} \\
\multicolumn{7}{c}{(Figures 7 and 16)}\\
\hline
$m_{\tilde g}$ & Model & $\sigma_{1}$ & $\sigma_{2}$ &$\sigma_{3}$& 
$\sigma_{4}$ & $\sigma_{5}$ \\
\hline 
 500 &{\bf universal} & 0.1022 & 0.0503 & 0.0185 & 7.9664 & NULL \\
\hline
 500 &{\bf 24} & 0.0878 & 0.0449 & 0.0255 & 8.7054 & NULL \\
\hline
 500 &{\bf 75} & 0.0047 & {\bf 0.0009} & 0.0017 & 0.7335 & NULL \\
\hline
 500 &{\bf 200} & 0.0028 & 0.0019 & 0.0038 & 2.5958 & NULL  \\
\hline
500 &{\bf 54(i)} & 0.0302 & 0.0141 & 0.0245 & 8.8041 & NULL  \\
\hline
500 &{\bf 54(ii)} & 0.0879 & 0.0454 & 0.0203 & 8.1171 & NULL  \\
\hline
1000 &{\bf universal} & 0.0098 & 0.0062 & 0.0032 & 1.0422 & NULL  \\
\hline
1000 &{\bf 24} & 0.0119 & 0.0037 & 0.0032 & 1.0220 & NULL  \\
\hline
1000 &{\bf 75} & 0.0044 & {\bf 0.0005} & {\bf 0.0004} & 0.1052 & NULL  \\
\hline
1000 &{\bf 200} & {\bf 0.0002} & {\bf 0.0001} & 0.0006 & 0.2546 & NULL  \\
\hline
1000 &{\bf 54(i)} & 0.0119 & 0.0035 & 0.0027 & 0.9998 & $1.0\times 10^{-5}$  \\
\hline
1000 &{\bf 54(ii)} & 0.0125 & 0.0043 & 0.0032 & 1.0401 & $1.5\times 10^{-5}$  \\
\hline
1500 &{\bf universal} & 0.0026 & 0.0017 & 0.0009 & 0.2781 & $0.5\times 10^{-5}$  \\
\hline
1500 &{\bf 24} & 0.0036 & {\bf 0.0012} & 0.0010 & 0.3007 & NULL  \\
\hline
1500 &{\bf 75} & NA & NA & NA & NA & NA   \\
\hline
1500 &{\bf 200} & {\bf $6.0\times 10^{-5}$} & {\bf $6.0\times 10^{-5}$} & 
{\bf 0.0001} & 0.0172 & NULL  \\
\hline
1500 &{\bf 54(i)} & {\bf 0.0006} & {\bf 0.0003} & 0.0006 & 0.1023 & NULL \\
\hline
1500 &{\bf 54(ii)} & 0.0034 & {\bf 0.0012} & 0.0011 & 0.2959 & NULL  \\
\hline
\end {tabular}\\

\noindent

\end {center}

\noindent

\begin{center}

\begin{tabular}{|c|c|c|c|c|c|c|}

\multicolumn{7}{c}{Table B5 : Cross-sections (pb) for SU(5) and SO(10) 
{\bf pMSSM}}\\
\multicolumn{7}{c}{$m_{\tilde f}=~500$ GeV,
$\mu$= 300 GeV, $\tan \beta$= 40} \\
\multicolumn{7}{c}{(Figures 2 and 11)}\\
\hline
 $m_{\tilde g}$ & Model & $\sigma_{1}$ & $\sigma_{2}$ &$\sigma_{3}$& 
$\sigma_{4}$ & $\sigma_{5}$ \\
\hline 
 500 &{\bf universal} & 0.5220 & 0.2281 & 0.0729 & 16.0476 & 0.0008 \\
\hline
 500 &{\bf 24} & 0.6831 & 0.3310 & 0.0581 & 18.3674 & NULL \\
\hline
 500 &{\bf 75} & 0.0393 & 0.0203 & 0.0215 & 1.0915 & NULL \\
\hline
 500 &{\bf 200} & 0.0983 & 0.0393 & 0.0222 & 2.4881 & NULL  \\
\hline
500 &{\bf 54(i)} & 0.5987 & 0.2071 & 0.0632 & 14.6693 & NULL  \\
\hline
500 &{\bf 54(ii)} & 0.5995 & 0.2563 & 0.0617 & 15.1236 & NULL  \\
\hline
1000 &{\bf universal} & 0.1033 & 0.0244 & 0.0142 & 3.3959 & 0.0003  \\
\hline
1000 &{\bf 24} & 0.0800 & 0.0300 & 0.0156 & 4.2959 & NULL  \\
\hline
1000 &{\bf 75} & 0.0089 & 0.0041 & 0.0060 & 1.3822 & NULL  \\
\hline
1000 &{\bf 200} & 0.0083 & 0.0032 & 0.0048 & 1.3759 & NULL  \\
\hline
1000 &{\bf 54(i)} & 0.0569 & 0.0161 & 0.0146 & 3.0397 & NULL  \\
\hline
1000 &{\bf 54(ii)} & 0.0653 & 0.0254 & 0.0131 & 3.5114 & NULL  \\
\hline
1500 &{\bf universal} & 0.0374 & 0.0071 & 0.0058 & 0.7133 & NULL  \\
\hline
1500 &{\bf 24} & 0.0306 & 0.0101 & 0.0056 & 1.4278 & 0.0001  \\
\hline
1500 &{\bf 75} & 0.0027 & {\bf 0.0009} & 0.0030 & 0.2765 & NULL  \\
\hline
1500 &{\bf 200} & 0.0023 & {\bf 0.0012} & 0.0027 & 0.3046 & NULL  \\
\hline
1500 &{\bf 54(i)} & 0.0273 & 0.0048 & 0.0056 & 0.5729 & NULL  \\
\hline
1500 &{\bf 54(ii)} & 0.0266 & 0.0055 & 0.0054 & 0.6889 & NULL  \\
\hline
\end {tabular}\\

\noindent

\end {center}

\noindent

\begin{center}

\begin{tabular}{|c|c|c|c|c|c|c|}

\multicolumn{7}{c}{Table B6 : Cross-sections (pb) for SU(5) and SO(10) 
{\bf pMSSM}}\\
\multicolumn{7}{c}{$m_{\tilde f}=~1000$ GeV,
$\mu$= 300 GeV, $\tan \beta$= 40} \\
\multicolumn{7}{c}{(Figures 4 and 13)}\\
\hline
 $m_{\tilde g}$ & Model & $\sigma_{1}$ & $\sigma_{2}$ &$\sigma_{3}$& 
$\sigma_{4}$ & $\sigma_{5}$ \\ 
\hline 
 500 &{\bf universal} & 0.1602 & 0.0059 & 0.0019 & 8.1530 & NULL \\
\hline
 500 &{\bf 24} & 0.1714 & 0.0745 & 0.0236 & 8.4541 & NULL \\
\hline
 500 &{\bf 75} & 0.0312 & 0.0234 & 0.0085 & 2.8467 & NULL \\
\hline
 500 &{\bf 200} & 0.0258 & 0.0139 & 0.0097 & 3.9270 & NULL  \\
\hline
500 &{\bf 54(i)} & 0.1264 & 0.0400 & 0.0233 & 8.3948 & NULL  \\
\hline
500 &{\bf 54(ii)} & 0.1706 & 0.0864 & 0.0193 & 7.7990 & 0.0002  \\
\hline
1000 &{\bf universal} & 0.0214 & 0.0069 & 0.0030 & 0.8446 & $6.0\times 10^{-5}$  \\
\hline
1000 &{\bf 24} & 0.0343 & 0.0175 & 0.0037 & 1.0486 & 0.0001  \\
\hline
1000 &{\bf 75} & 0.0182 & 0.0106 & 0.0043 & 0.8455 & $3.0\times 10^{-5}$  \\
\hline
1000 &{\bf 200} & 0.0120 & 0.0063 & 0.0027 & 0.7075 & NULL  \\
\hline
1000 &{\bf 54(i)} & 0.0368 & 0.0194 & 0.0040 & 1.0739 & $2.0\times 10^{-5}$  \\
\hline
1000 &{\bf 54(ii)} & 0.0359 & 0.0183 & 0.0038 & 1.0810 & $4.0\times 10^{-5}$  \\
\hline
1500 &{\bf universal} & 0.0088  & 0.0031 & 0.0009 & 0.2799 & $2.0\times 10^{-5}$  \\
\hline
1500 &{\bf 24} & 0.0082 & 0.0038 & 0.0013 & 0.3403 & NULL  \\
\hline
1500 &{\bf 75} & {\bf 0.0004} & {\bf 0.0003} & {\bf 0.0006} & 0.0981 & NULL \\
\hline
1500 &{\bf 200} & {\bf 0.0015} & {\bf 0.0006} & 0.0008 & 0.1272 & NULL \\
\hline
1500 &{\bf 54(i)} & 0.0039 & {\bf 0.0009} & 0.0008 & 0.1700 & NULL  \\
\hline
1500 &{\bf 54(ii)} & 0.0090 & 0.0027 & 0.0011 & 0.3157 & NULL  \\
\hline
\end {tabular}\\

\noindent

\end {center}

\noindent

\begin{center}

\begin{tabular}{|c|c|c|c|c|c|c|}

\multicolumn{7}{c}{Table B7 : Cross-sections (pb) for SU(5) and SO(10) 
{\bf pMSSM}}\\
\multicolumn{7}{c}{$m_{\tilde f}=~500$ GeV,
$\mu$= 1000 GeV, $\tan \beta$= 40} \\
\multicolumn{7}{c}{(Figures 6 and 15)}\\
\hline
$m_{\tilde g}$ & Model & $\sigma_{1}$ & $\sigma_{2}$ &$\sigma_{3}$& 
$\sigma_{4}$ & $\sigma_{5}$ \\
\hline 
 500 &{\bf universal} & 0.9410 & 0.3260 & 0.0715 & 17.3778 & NULL \\
\hline
 500 &{\bf 24} & NA & NA & NA & NA & NA  \\
\hline
 500 &{\bf 75} & NA & NA & NA & NA & NA  \\
\hline
 500 &{\bf 200} & 0.0283 & 0.0225 & 0.0193 & 1.6686 & NULL  \\
\hline
500 &{\bf 54(i)} & 0.2766 & 0.1517 & 0.1182 & 19.2174 & 0.0007  \\
\hline
500 &{\bf 54(ii)} & 0.6048 & 0.3087 & 0.0725 & 19.3995 & NULL  \\
\hline
1000 &{\bf universal} & 0.0467 & 0.0245 & 0.0130 & 3.6043 & NULL  \\
\hline
1000 &{\bf 24} & NA & NA & NA & NA & NA   \\
\hline
1000 &{\bf 75} & NA & NA & NA & NA & NA   \\
\hline
1000 &{\bf 200} & NA & NA & NA & NA & NA   \\
\hline
1000 &{\bf 54(i)} & 0.0217 & 0.0092 & 0.0098 & 2.0479 & NULL  \\
\hline
1000 &{\bf 54(ii)} & 0.0308 & 0.0183 & 0.0156 & 2.1147 & NULL  \\
\hline
1500 &{\bf universal} & 0.0100  & 0.0044 & 0.0055 & 5.5373 & NULL  \\
\hline
1500 &{\bf 24} & NA & NA & NA & NA & NA   \\
\hline
1500 &{\bf 75} & NA & NA & NA & NA & NA   \\
\hline
1500 &{\bf 200} & NA & NA & NA & NA & NA   \\
\hline
1500 &{\bf 54(i)} & NA & NA & NA & NA & NA   \\
\hline
1500 &{\bf 54(ii)} & NA & NA & NA & NA & NA   \\
\hline
\end {tabular}\\

\noindent

\end {center}

\noindent

\begin{center}

\begin{tabular}{|c|c|c|c|c|c|c|}

\multicolumn{7}{c}{Table B8 : Cross-sections (pb) for SU(5) and SO(10) 
{\bf pMSSM}}\\
\multicolumn{7}{c}{$m_{\tilde f}=~1000$ GeV,
$\mu$= 1000 GeV, $\tan \beta$= 40} \\
\multicolumn{7}{c}{(Figures 8 and 17)}\\
\hline
$m_{\tilde g}$ & Model & $\sigma_{1}$ & $\sigma_{2}$ &$\sigma_{3}$& 
$\sigma_{4}$ & $\sigma_{5}$ \\
\hline 
 500 &{\bf universal} & 0.1027 & 0.0440 & 0.0233 & 7.9509 & NULL \\
\hline
 500 &{\bf 24} & 0.1123 & 0.0373 & 0.0243 & 8.7752 & 0.0002 \\
\hline
 500 &{\bf 75} & 0.0059 & 0.0019 & 0.0010 & 0.8054 & NULL \\
\hline
 500 &{\bf 200} & 0.0023 & 0.0018 & 0.0053 & 2.5620 & NULL  \\
\hline
500 &{\bf 54(i)} & 0.0379 & 0.0138 & 0.0274 & 8.7898 & NULL  \\
\hline
500 &{\bf 54(ii)} & 0.1161 & 0.0394 & 0.0203 & 8.1907 & NULL  \\
\hline
1000 &{\bf universal} & 0.0204 & 0.0130 & 0.0039 & 0.9343 & $3.0\times 10^{-5}$  \\
\hline
1000 &{\bf 24} & 0.0209 & 0.0124 & 0.0043 & 0.9690 & $1.5\times 10^{-5}$  \\
\hline
1000 &{\bf 75} & 0.0314 & {\bf 0.0001} & {\bf 0.0004}  & 0.0771 & NULL  \\
\hline
1000 &{\bf 200} & {\bf 0.0018} & 0.0016 & 0.0010 & 0.1095 & NULL  \\
\hline
1000 &{\bf 54(i)} & 0.0182 & 0.0929 & 0.0037 & 0.7876 & $1.5\times 10^{-5}$  \\
\hline
1000 &{\bf 54(ii)} & 0.0216 & 0.0130 & 0.0038 & 0.9677 & NULL  \\
\hline
1500 &{\bf universal} & 0.0028  & 0.0016 & 0.0010 & 0.2775 & $0.6\times 10^{-5}$ \\
\hline
1500 &{\bf 24} & 0.0030 & 0.0014 & 0.0013 & 0.3063 & NULL  \\
\hline
1500 &{\bf 75} & NA & NA & NA & NA & NA   \\
\hline
1500 &{\bf 200} & NA & NA & NA & NA & NA   \\
\hline
1500 &{\bf 54(i)} & {\bf 0.0007} & {\bf 0.0003} & 0.0007 & 0.1044 & NULL \\
\hline
1500 &{\bf 54(ii)} & 0.0024 & 0.0007 & 0.0009 & 0.2052 & NULL  \\
\hline
\end {tabular}\\

\noindent

\end {center}

\noindent

\begin{center}

\begin{tabular}{|c|c|c|c|c|c|c|}

\multicolumn{7}{c}{Table B9 : Cross-sections (pb) for SU(5) and SO(10) 
{\bf SUGRA}}\\
\multicolumn{7}{c}{$m_{\tilde f}=~506$ GeV 
 at $M_{GUT}$, $\tan \beta= 5$} \\
\multicolumn{7}{c}{(Figures 9 and 18)}\\
\hline
$m_{\tilde g}$ & Model & $\sigma_{1}$ & $\sigma_{2}$ &$\sigma_{3}$& 
$\sigma_{4}$ & $\sigma_{5}$ \\
\hline 
 500 &{\bf universal} & 0.2818 & 0.1411 & 0.0445 & 16.5239 & NULL  \\
\hline
 500 &{\bf 24} & 0.3807 & 0.1900 & 0.0390 & 16.1696 & 0.0007  \\
\hline
 500 &{\bf 75} & 0.3685 & 0.3382 & 0.0282 & 10.1572 & 0.0003 \\
\hline
 500 &{\bf 200} & 0.0912 & 0.0667 & 0.0194 & 9.0323 & NULL   \\
\hline
500 &{\bf 54(i)} & 0.6041 & 0.4034 & 0.0293 & 13.1929 & 0.0004   \\
\hline
500 &{\bf 54(ii)} & 0.4153 & 0.2023 & 0.0313 & 14.7555 & 0.0004   \\
\hline
1000 &{\bf universal} & 0.0397 & 0.0266 & 0.0032 & 1.0060 & NULL  \\
\hline
1000 &{\bf 24} & 0.0315 & 0.0191 & 0.0029 & 0.8035 & 0.0001   \\
\hline
1000 &{\bf 75} & NA & NA & NA & NA & NA   \\
\hline
1000 &{\bf 200} & 0.0137 & 0.0105 & 0.0013 & 0.4504 & $2.0\times 10^{-5}$   \\
\hline
1000 &{\bf 54(i)} & 0.0312 & 0.0196 & 0.0032 & 0.7153 & NULL   \\
\hline
1000 &{\bf 54(ii)} & 0.0321 & 0.0195 & 0.0029 & 0.7662 & $1.0\times 10^{-5}$   \\
\hline
1500 &{\bf universal} & {\bf 0.0019} & {\bf 0.0011} & {\bf 0.0003} & 0.0735 & 
$3.0\times 10^{-5}$  \\
\hline
1500 &{\bf 24} & 0.0022 & {\bf 0.0012} & {\bf 0.0004} & 0.0750 & $2.0\times 10^{-5}$   \\
\hline
1500 &{\bf 75} & NA & NA & NA & NA & NA   \\
\hline
1500 &{\bf 200} & {\bf 0.0012} & {\bf 0.0009} & {\bf 0.0002} & 0.0368 & 
$2.0\times 10^{-5}$  \\
\hline
1500 &{\bf 54(i)} & {\bf 0.0017} & {\bf 0.0009} & {\bf 0.0004} & 0.0561 & NULL   \\
\hline
1500 &{\bf 54(ii)} & 0.0215 & {\bf 0.0013} & {\bf 0.0004} & 0.0728 & $4.0\times 10^{-5}$\\
\hline
\end {tabular}\\

\noindent

\end {center}

\end{document}